\documentclass[useAMS,usenatbib]{mn2e}

\usepackage{aas_macros}

\usepackage{graphicx}
\title[S0 galaxy structure in STAGES]{The environmental dependence of the structure of galactic discs in
STAGES S0 galaxies: implications for S0 formation}

\author[D.~T.~Maltby et al.]
{David~T.~Maltby,$^{1}$\thanks{E-mail: dtmaltby@gmail.com}
 Alfonso~Arag{\'o}n-Salamanca,$^{1}$ Meghan~E.~Gray,$^{1}$ Carlos~Hoyos,$^{2}$
\newauthor Christian~Wolf,$^{3}$ Shardha~Jogee$^{4}$ and Asmus~B{\"o}hm$^{5}$\\
$^{1}$School of Physics and Astronomy, The University of Nottingham, University Park, Nottingham NG7 2RD\\
$^{2}$Instituto de Astronomia, Geof\'{i}sica e Ci\^{e}ncias Atmosf\'{e}ricas, Rua do Mat\~{a}o 1226, Cidade Universit\'{a}ria S\~{a}o Paulo-SP, Brazil\\
$^{3}$Research School of Astronomy and Astrophysics, Australian National University, Canberra, ACT 2611, Australia\\
$^{4}$Department of Astronomy, University of Texas at Austin, 1 University Station, C1400 Austin, TX 78712-0259, USA\\
$^{5}$Institute for Astro- and Particle Physics, University of Innsbruck, Technikerstr. 25/8, 6020 Innsbruck, Austria}

\begin{document}

\date{Accepted Year Month Date. Received Year Month Date; in original form Year Month Date}

\pagerange{\pageref{firstpage}--\pageref{lastpage}} \pubyear{0000}

\maketitle

\label{firstpage}


\begin{abstract}
We present an analysis of $V$-band radial surface brightness $\mu(r)$ profiles for S0 galaxies in different
environments using {\em Hubble Space Telescope}/Advanced Camera for Surveys imaging and data from the Space
Telescope A901/2 Galaxy Evolution Survey (STAGES). Using a large sample of $\sim280$ field and cluster S0s,
we find that in both environments, $\sim25$ per cent have a pure exponential disc (Type~I) and $\sim50$ per
cent exhibit an up-bending disc break ({\em antitruncation}, Type~III). However, we find hardly any ($<5$ per
cent) down-bending disc breaks ({\em truncations}, Type~II) in our S0s and many cases ($\sim20$ per cent)
where no discernible exponential component was observed (i.e.\ general curvature). We also find no evidence
for an environmental dependence on the disc scalelength~$h$ or break strength~$T$ (outer-to-inner scalelength
ratio), implying that the galaxy environment does not affect the stellar distribution in S0 stellar discs.
Comparing disc structure (e.g.\ $h$, $T$) between these S0s and the spiral galaxies from our previous
studies, we find: i) no evidence for the Type~I scalelength~$h$ being dependent on morphology; and ii) some
evidence to suggest that the Type~II/III break strength~$T$ is smaller (weaker) in S0s compared to spiral
galaxies. Taken together, these results suggest that the stellar distribution in S0s is not drastically
affected by the galaxy environment. However, some process inherent to the morphological transformation of
spiral galaxies into S0s does affect stellar disc structure causing a weakening of $\mu(r)$ breaks and may
even eliminate truncations from S0 galaxies. In further tests, we perform analytical bulge--disc
decompositions on our S0s and compare the results to those for spiral galaxies from our previous studies. For
Type~III galaxies, we find that bulge light can account for the excess light at large radii in up to $\sim50$
per cent of S0s but in only $\sim15$ per cent of spirals. We propose that this result is consistent with a
fading stellar disc (evolving bulge-to-disc ratio) being an inherent process in the transformation of spiral
galaxies into S0s.
\end{abstract}

\begin{keywords}
galaxies: clusters: individual: A901/2 ---
galaxies: elliptical and lenticular, cD ---
galaxies: evolution ---
galaxies: spiral ---
galaxies: structure.
\vspace{-0.35cm}
\end{keywords}

\section{Introduction}

\label{Introduction}

It is now well established that correlations exist between the properties of galaxies, e.g.\ morphology,
colour and star-formation rate, and their local environment \citep{Dressler:1980,Weinmann_etal:2006}.
However, the exact mechanisms driving these correlations remain elusive. Certain physical processes inherent
to galaxy evolution and related to the galaxy environment may contribute, e.g.\ ram-pressure stripping of the
interstellar medium, mergers and harassment \citep[e.g.][] {Gunn_Gott:1972,Icke:1985,Moore_etal:1996}.
However, the relative importance of each of these processes in galaxy evolution remains uncertain.

An important aspect in studying how the environment could affect the formation and evolution of disc
galaxies is the structure of galactic discs. Their fragile outer regions are more easily affected by
interactions with other galaxies, and therefore their structural characteristics must be closely related to
their evolutionary history. Consequently, exploring the effect of the environment on the light distribution
(surface brightness $\mu$ profile) of disc galaxies should aid in our understanding of the physical processes
of galaxy evolution occurring in different environments.

The light profiles of disc galaxies are comprised of two main structural components: an inner bulge-dominated
component; and an outer exponentially declining disc with some minor deviations related to substructure
\citep{deVaucouleurs:1959b, Freeman:1970}. However, this `classical' picture fails for the majority of
disc galaxies in the Universe since the exponential component is often truncated (sharply cut off)
after several scalelengths \citep{vanderKruit:1979}. In fact, most disc $\mu$ profiles are
actually best described by a two slope model (broken exponential), characterised by an inner and outer
exponential scalelength separated by a relatively well-defined break radius $r_{\rm brk}$
\citep{Pohlen_etal:2002}. Many studies have reported (mainly using surface photometry) the existence of
broken exponential discs in both the local \citep{Pohlen_etal:2002,
Pohlen_etal:2007, Pohlen_Trujillo:2006, Bakos_etal:2008, Erwin_etal:2008, Erwin_etal:2012,
Gutierrez_etal:2011, Maltby_etal:2012a} and distant $z < 1$ Universe \citep{Perez:2004, Trujillo_Pohlen:2005,
Azzollini_etal:2008}. Broken exponential discs have also been reported through the use of resolved star
counts on some nearby galaxies \citep{Ibata_etal:2005, Ferguson_etal:2007}.

As a direct result of these studies, a comprehensive classification scheme for disc galaxies has emerged
based on break features in the outer disc component of their radial $\mu(r)$ profiles \citep[see e.g.][]
{Pohlen_Trujillo:2006, Erwin_etal:2008}. This classification scheme consists of three broad types (Type~I,
II and III): Type~I (no break) -- the galaxy has a simple exponential profile extending out to several
scalelengths \citep[e.g.][]{BlandHawthorn_etal:2005}; Type~II (down-bending break, {\em truncation}) -- a
broken exponential with a shallow inner and steeper outer region separated by a relatively well-defined
break radius $r_{\rm brk}$ \citep{vanderKruit:1979, Pohlen_etal:2002}; Type~III (up-bending break,
{\em antitruncation}) -- a broken exponential with the opposite behaviour to a Type~II profile \cite[a
shallower region beyond $r_{\rm brk}$;][]{Erwin_etal:2005}. In each case the classification refers to the
outer disc component of the galaxy $\mu(r)$ profile and does not consider the inner bulge component even if
the bulge is near exponential in nature.

At present, the physical origins of the different profile types are not well understood. Some models suggest
that Type~II profiles (truncations) could be the consequence of a radial star formation threshold
\citep[e.g.][]{Kennicutt:1989, Elmegreen_Parravano:1994, Schaye:2004}. Others suggest that Type~II profiles
are caused by a resonance phenomenon and a redistribution of angular momentum \citep{Debattista_etal:2006}.
However, most current theories incorporate both these ideas, suggesting that the inner disc forms as a
consequence of a star formation threshold while the outer disc forms by the outward migration of stars from
the inner disc to regions beyond the star formation threshold (i.e. break radius $r_{\rm brk}$). This
proposed migration could be due to resonant scattering with spiral arms \citep{Roskar_etal:2008a,
Roskar_etal:2008b} or clump disruptions \citep{Bournaud_etal:2007}. For Type~III profiles, their discovery is
still very recent \citep{Erwin_etal:2005} and therefore much less effort has been afforded to their origin.
\cite{Erwin_etal:2005} suggest that in some cases the excess light beyond the break radius $r_{\rm brk}$
could actually be attributed to light from the spheroidal bulge or halo extending beyond the end of the disc;
however, these cases seem to be quite rare \citep{Maltby_etal:2012b}. In general, it appears that Type~III
profiles are the consequence of a disturbed system and that recent minor mergers could produce up-bending
stellar profiles in the remnant galaxy \citep{Younger_etal:2007, SilChenko_etal:2011}.

Investigating the frequency of profile types for different morphologies and in regions of different galaxy
density, is a useful tool for exploring galaxy evolution and the role of the galaxy environment. However,
presently there have only been a few systematic searches for broken exponentials in stellar discs
\citep[e.g.][]{Pohlen_Trujillo:2006,Azzollini_etal:2008,Erwin_etal:2008, Erwin_etal:2012,Gutierrez_etal:2011,
Maltby_etal:2012a} and these rarely span the full range of disc morphologies (S0--Sdm).
\cite{Pohlen_Trujillo:2006} use a local sample of $\sim90$ late-type spirals (Sb--Sdm) and find that the
distribution of profile types I:II:III is approximately $10$:$60$:$30$ per cent. However,
\cite{Erwin_etal:2008} use a local sample of $66$ barred early-type disc galaxies (S0--Sb) and find a
distribution of approximately $30$:$40$:$25$ per cent (the remaining $\sim5$ per cent contained both Type~II
and Type~III features). The differences in the profile-type fractions between these two authors can easily
be attributed to the morphological range of their respective samples. This is because the shape of disc
galaxy $\mu(r)$ profiles is dependent on morphology \citep{Pohlen_Trujillo:2006, Gutierrez_etal:2011,
Maltby_etal:2012a}. \cite{Gutierrez_etal:2011} explored this issue by combining their local sample of $47$
unbarred early-type galaxies with those of \cite{Pohlen_Trujillo:2006} and \cite{Erwin_etal:2008} and
examined the profile-type distribution across the entire S0--Sdm range. They find that towards later Hubble
types (S0 $\rightarrow$ Sdm) the fraction of Type~I profiles decreases ($\sim30 \rightarrow 10$ per cent)
and the fraction of Type~II profiles increases ($\sim25 \rightarrow 80$ per cent). Consequently, morphology
is an important factor when comparing the profile-type distributions from these different studies.

Presently, the effect of the galaxy environment on the frequency of profile types has only been explored by
a few authors. For spiral galaxies, \cite{Maltby_etal:2012a} use a sample of $\sim300$ field and cluster
spirals (Sa--Sdm) and find no environmental dependence on the distribution of profile types in the outer
regions of the stellar disc ($\mu > 24\rm\,mag\,arcsec^{-2}$). We note that due to the limited surface
brightness range studied by \cite{Maltby_etal:2012a}, their profile type distributions cannot be directly
compared to those of previous works. However, an inspection of their $\mu(r)$ profiles yields a profile type
distribution I:II:III of approximately $10$:$50$:$40$ per cent in both the field and cluster environment,
which is in line with previous works. In contrast, for S0 galaxies \cite{Erwin_etal:2012} recently discovered
an intriguing environmental dependence among the shapes of their $\mu(r)$ profiles. Using a sample of
$\sim70$ field and cluster S0s they find that in the field the distribution is $25$:$25$:$50$ per cent while
in the cluster the distribution is $50$:$0$:$50$ per cent. Thus their cluster S0s show a complete lack of
Type~II profiles.

In this work, we expand on these environmental studies by performing a systematic search for broken
exponential discs in field and cluster S0s using the Space Telescope A901/2 Galaxy Evolution Survey
\citep[STAGES;][]{Gray_etal:2009}. This work builds on previous studies by using larger and more
statistically viable field and cluster samples and by being one of only a few studies to probe the
high-density environments \cite[see][for another example]{Erwin_etal:2012}. This work is also similar to
our companion study \citep{Maltby_etal:2012a}, which explores the effect of the galaxy environment on the
disc structure of spiral galaxies in STAGES\footnote{Note: unlike \cite{Maltby_etal:2012a}, in this work our
break classification is based on the entire disc and {\em not}  just the outer disc
($\mu > 24\rm\,mag\,arcsec^{-2}$) (see Section~\ref{Profile classification}).}. Since S0s are expected to
have evolved from spiral galaxies, by comparing the results of these two works we aim to provide some insight
into the potential evolutionary mechanisms involved in their morphological transformation.

The structure of this paper is as follows. In Section~\ref{Description of the data}, we give a brief
description of the STAGES data set relevant to this work and outline our sample selection in
Section~\ref{Sample selection}. In Section~\ref{Profile fitting}, we describe the method used to obtain our
radial surface brightness $\mu(r)$ profiles from the STAGES $V$-band imaging and explain our profile
classification scheme in Section~\ref{Profile classification}. We present our results for S0 galaxies in
Section~\ref{Results}, and then compare these results with those for STAGES spiral galaxies from
\cite{Maltby_etal:2012a} in Section~\ref{The structure of galactic discs in spiral and S0 galaxies}. In
Section~\ref{Antitruncated surface brightness profiles: bulge or disc related?}, we complement this work with
an examination of the impact of a \cite{deVaucouleurs:1948} bulge profile on the outer regions of our S0
$\mu(r)$ profiles, and compare our results with those from a similar study using spiral galaxies
\citep{Maltby_etal:2012b}. Finally, we draw our conclusions in Section~\ref{Conclusions}. Throughout this
paper, we adopt a cosmology of $H_0 = 70\rm\,km\,s^{-1}\,Mpc^{-1}$, $\Omega_\Lambda = 0.7$ and
$\Omega_{\rm m} = 0.3$, and use AB magnitudes unless stated otherwise.

\vspace{-0.3cm}
\section{Description of the data}

\label{Description of the data}

This work is entirely based on the STAGES data published by \cite{Gray_etal:2009}. STAGES is an extensive
multiwavelength survey targeting the Abell(A) 901/902 multicluster system ($z\sim0.167$) and covering a wide
range of galaxy environments. These environments span from the general field to the intermediate densities of
the A901/2 clusters \citep{Heiderman_etal:2009}. {\em Hubble Space Telescope} ({\em HST})/Advanced Camera for
Surveys (ACS) $V$-band (F606W) imaging covering the full $0.5^{\circ}\times0.5^{\circ}$
($\sim5\times5\rm\,Mpc^2$) of the multicluster system is available and complemented by extensive
multiwavelength observations. These include high-precision photometric redshifts and observed-/rest-frame
spectral energy distributions (SEDs) from the $17$-band COMBO-17 photometric redshift survey
\citep{Wolf_etal:2003}. These photometric redshifts are accurate to $1$ per cent in $\delta{z}/(1+z)$ at
$R < 21$. However, photo-$z$ quality degrades for progressively fainter galaxies reaching accuracies of $2$
per cent for galaxies with $R\sim22$ and $10$ per cent for galaxies with $R > 24$ \citep{Wolf_etal:2004,
Wolf_etal:2008}. Stellar mass estimates derived from the SED fitting of the COMBO-17 photometry are also
available \citep{Borch_etal:2006, Gray_etal:2009}.

\cite{Gray_etal:2009} have also performed S{\'e}rsic profile fitting using the {\sc galfit} code
\citep{Peng_etal:2002} on all {\em HST}/ACS images and conducted simulations to quantify the completeness of
the survey, all of which are publicly available\footnote{http://www.nottingham.ac.uk/astronomy/stages}.
Additionally, all galaxies with $R < 23.5$ and $z_{\rm phot} < 0.4$ ($5090$ galaxies) were visually
classified by seven members of the STAGES team into Hubble-type morphologies (E, S0, Sa, Sb, Sc, Sd, Irr) and
their intermediate classes (Gray et al., in preparation). This classification ignored bars and degrees of
asymmetry and defined S0s to be disc galaxies with a visible bulge but no spiral arms (smooth disc).

\vspace{-0.3cm}
\subsection{Sample selection}

\label{Sample selection}

Our sample of field and cluster S0 galaxies is drawn from \cite{Maltby_etal:2010}. This consists of a large,
mass-limited ($M_* > 10^9\rm\,M_\odot$), visually classified sample of $276$ S0s from both the field and
cluster environments in STAGES ($60$ field and $216$ cluster). In the following, we give a brief summary of
the relevant field and cluster sample selection presented in \cite{Maltby_etal:2010}.

The cluster S0 sample is selected from a parent sample of cluster galaxies suggested for STAGES by
\cite{Gray_etal:2009}. This parent sample of cluster galaxies is defined solely from photometric redshifts.
The photo-$z$ distribution of cluster galaxies is assumed to be Gaussian while the distribution of field
galaxies is assumed to be consistent with the average galaxy counts $N(z, R)$ outside the cluster and to vary
smoothly with redshift and magnitude. Cluster galaxies are then defined simply via a redshift interval around
the known spectroscopic redshift of the cluster, \mbox{$z_{\rm phot} = [0.17-\Delta{z},0.17+\Delta{z}]$}, the
width of which varies with $R$-magnitude. The half-width $\Delta{z}$ as a function of $R$-magnitude is
\begin{equation}
\Delta{z}(R) = \sqrt{0.015^2 + 0.0096525^2(1 + 10^{0.6[R_{\rm tot} - 20.5]})}.
\end{equation}

This cluster selection adopts a narrow redshift range for bright $R$-magnitudes due to the high precision of
the COMBO-17 photometric redshifts; however, the interval increases in width towards fainter $R$-magnitudes
to accommodate for the increase in the photo-$z$ error. \cite{Gray_etal:2009} calculate the completeness and
contamination of this cluster selection as a function of $R$-magnitude by using the counts of their smooth
models. In these calculations, they compromise $\Delta{z}$ so that the completeness of the cluster selection
is $>90$ per cent at all magnitudes. The completeness of this selection converges to nearly $100$ per cent
for bright galaxies (see \citealt{Gray_etal:2009}, for further details). This cluster sample is then limited
by stellar mass (${\rm log}\,M_*/{\rm M_\odot} > 9$) and morphology (visually classified S0s) in order to
create our final sample of $216$ cluster S0 galaxies.

The field sample is selected from STAGES galaxies by applying a redshift interval either side of the cluster
redshift ($z_{\rm cl} = 0.167$) that avoids the cluster selection. We use a lower redshift interval at
$z = [0.05, 0.14]$ and an upper redshift interval at $z = [0.22, 0.30]$, based on a similar sample selection
used by \cite{Wolf_etal:2009}. This field sample is then limited by stellar mass
(${\rm log}\,M_*/{\rm M_\odot} > 9$) and morphology (visually classified S0s) in order to create our final
sample of $60$ field S0 galaxies. For full details of the field and cluster sample selection used in this
work, see \cite{Maltby_etal:2010}.

In the data catalogue published by \cite{Gray_etal:2009}, there are two sets of derived values for galaxy
properties such as magnitude and stellar mass: one value based on the \mbox{photo-$z$} estimate and another
assuming the galaxy is located at the known spectroscopic redshift of the cluster \mbox{($z_{\rm cl} = 0.167$)}.
In this work, we use the original photo-$z$ estimates for our field sample and the fixed redshift values for our
cluster sample. This practice prevents the propagation of photo-$z$ errors into the physical values of our
cluster galaxies.

\begin{figure}
\includegraphics[width=0.41\textwidth]{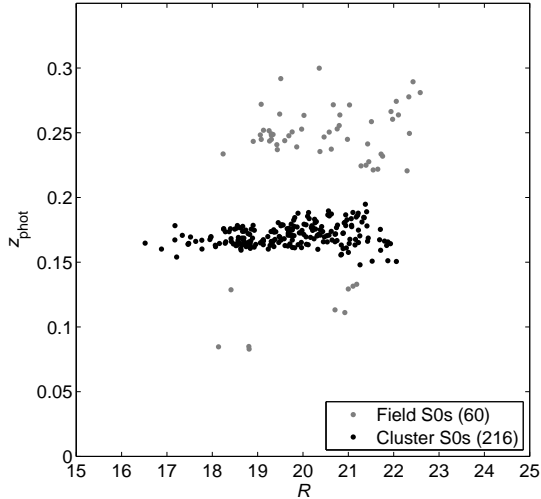}
\centering
\caption{\label{S0s:R_vs_z} The photometric redshift $z_{\rm phot}$ versus total $R$-band magnitude (Vega)
for the field (grey points) and cluster (black points) S0 galaxy samples. The field sample reaches
$R\sim23$ and the cluster sample reaches $R\sim22$. Respective sample sizes are shown in the legend.
\vspace{-0.3cm}}
\end{figure}

\begin{table}
\begin{minipage}{80mm}
\centering
\caption{\label{Sample properties}{Properties of the field and cluster S0 samples.}}
\begin{tabular}{lccc}
\hline
Property			& Field			& Cluster		\\
\hline
$N_{\rm gal}$			& $60$			& $216$			\\
${\rm Completeness}$		& $>70$ per cent	& $>90$ per cent	\\
${\rm Contamination}$		& $-$			& $<25$ per cent	\\
$R_{\rm mean}$			& $20.44$		& $19.71$		\\
$M_{B\rm (min)}$		& $-16.54$		& $-16.45$		\\
$M_{B\rm (max)}$		& $-21.19$		& $-21.78$		\\
$z_{\rm phot,mean}$		& $0.230$		& $0.171$		\\
$z_{\rm phot,min}$		& $0.083$		& $0.148$		\\
$z_{\rm phot,max}$		& $0.300$		& $0.195$		\\
${\rm log} M_{\rm *,mean}$	& $10.41$		& $10.45$		\\
${\rm Bar\,fraction}$		& $7$ per cent		& $14$ per cent		\\
\hline
\end{tabular}
\end{minipage}
\end{table}

The completeness of STAGES is $>90$ per cent for $R < 23.5$ \citep{Gray_etal:2009} which is true for both our
field and cluster S0 samples (see Fig.~\ref{S0s:R_vs_z}). However, based on previous COMBO-17 experience
\cite{Wolf_etal:2009} estimate that at low stellar masses $M_* < 10^{9.5}\rm\,M_\odot$, the field sample
could have an additional $20$ per cent incompleteness. Consequently, our field sample is essentially $>70$
per cent complete. For our cluster sample, completeness is $>90$ per cent and contamination by the field is
$<25$ per cent based on the $R$-magnitude the cluster sample reaches (see Fig.~\ref{S0s:R_vs_z} and
\citealt{Gray_etal:2009}: fig.~14)\footnote{Note: using spectroscopy, \cite{Bosch_etal:2013a} report that
the contamination of this cluster sample may actually be $<10$ per cent.}. No further incompleteness is
introduced by selecting only visually classified S0 galaxies \citep{Maltby_etal:2010}. The properties of the
field and cluster S0 samples are shown in Table~\ref{Sample properties}.

\subsection{Galaxy inclination}

\label{Galaxy inclination}

\begin{figure}
\includegraphics[width=0.395\textwidth]{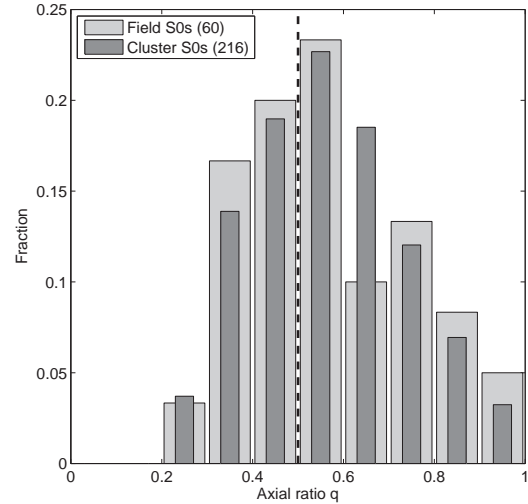}
\centering
\caption{\label{S0s:AR} The distribution of minor-to-major axial ratio $q$ for our field (light grey) and
cluster (dark grey) S0 galaxies. The $q$ cut used by previous works ($q > 0.5$, represented by a black
dashed line) is shown for reference. Errors in $q$ are $<3$ per cent. Respective sample sizes are shown in
the legend.\vspace{-0.3cm}}
\end{figure}

In the majority of studies that use surface photometry to explore broken exponential stellar discs, the disc
galaxy samples are limited by galaxy inclination $i$ to be face-on to intermediately inclined
\citep[e.g.][]{Pohlen_Trujillo:2006, Erwin_etal:2008, Gutierrez_etal:2011, Maltby_etal:2012a}. In general,
the minor-to-major axial ratio $q$ ($q = b/a = 1 - e$, where $a$ and $b$ are the semimajor and semiminor
axes, respectively, and $e$ is the ellipticity) is restricted to correspond to an inclination~$i$ of less
than $60^\circ$ ($q > 0.5$ or $e < 0.5$). The purpose of this inclination $i$ cut is to
\begin{enumerate}
\item minimise the influence of dust on the galaxy $\mu(r)$ profiles -- this is particularly important in
the case of spiral galaxies, but less of an issue with S0s;
\item allow for reliable information on disc sub-structure, e.g.\ bars, rings and spiral arms;
\item ensure observations/measurements of the disc component can actually probe the disc light outside of
the bulge-dominated region.
\end{enumerate}

The axial ratio $q$ for our S0 galaxies is determined from the STAGES {\sc galfit} models
\citep{Gray_etal:2009} and the axial ratio $q$ distributions for our field/cluster S0s are shown in
Fig.~\ref{S0s:AR}. The suggested inclination cut \mbox{($q > 0.5$)} would remove $\sim40$ per cent of our
field and cluster S0s. Unfortunately, this would have a drastic effect on the number of galaxies in our field
S0 sample and the quality of our field property distributions and subsequent results. Therefore, in order to
maintain our field sample size, we do not limit our S0 samples by galaxy inclination $i$ in this study.
However, we take account of the above considerations by performing parallel analysis on both the full S0
sample and a low-axis ratio ($q > 0.5$) S0 sub-sample (36 field and 137 cluster S0s). Reassuringly, we find
that the application of such an inclination $i$ cut ($i < 60^\circ$, $q > 0.5$) has no effect on the overall
significance of our results or our conclusions.

\section[]{Profile Fitting}

\label{Profile fitting}

For each galaxy in our field and cluster sample, we use the {\sc iraf} task ellipse\footnote{{\sc stsdas}
package - version 2.12.2} in order to obtain azimuthally-averaged radial surface brightness $\mu(r)$ profiles
from the STAGES {\em HST}/ACS $V$-band imaging. The ACS images used include the sky background and the
necessary sky subtraction is performed after profile fitting (see Section~\ref{Sky subtraction}).

We run ellipse using bad pixel masks that remove all sources of contamination from our isophotal fits, e.g.\
background/companion galaxies and foreground stars (everything not associated with the galaxy itself). In
this work, we use the bad pixel masks of \cite{Gray_etal:2009} but also apply some additional manual masking.
\cite{Gray_etal:2009} use the data pipeline `Galaxy Analysis over Large Areas: Parameter Assessment by
{\sc galfit}ting Objects from {\sc SExtractor}' \citep[{\sc galapagos};][]{Barden_etal:2012} to extract source
galaxies from the STAGES {\em HST}/ACS $V$-band imaging and fit \cite{Sersic:1968} $\mu(r)$ models to each
galaxy image using the {\sc galfit} code \citep{Peng_etal:2002}. Bad pixel masks are automatically generated
by {\sc galapagos} for each galaxy image and in most cases the companion galaxies are completely masked out.
However, occasionally in crowded regions {\sc galfit} performs multiobject fitting and therefore the
companion galaxies in these cases are not removed by the bad pixel masks as they are too close to the subject
galaxy. In these cases (three field and $27$ cluster, $\sim10$ per cent), we remove the companion galaxies
from the ACS image by the subtraction of their {\sc galfit} surface brightness model. The residuals of these
companion galaxies are not expected to have any significant effect on the azimuthally-averaged radial $\mu$
profile for the subject galaxy. However, we still modify the bad pixel masks of \cite{Gray_etal:2009} by
applying some additional manual masking in order to remove these residual features and also some low surface
brightness objects not detected by the {\sc galapagos} pipeline.

Using a similar procedure to previous works \citep{Pohlen_Trujillo:2006, Erwin_etal:2008, Maltby_etal:2012a},
we fit two different sets of ellipses to each galaxy image\footnote{Note: all our isophotal fits use
logarithmic radial sampling (steps -- 0.03 dex) and a fixed isophotal centre (galaxy centre) determined from
the {\sc galfit} S{\'e}rsic model \citep{Gray_etal:2009}.}.
The first is a free-parameter fit (fixed centre,
free ellipticity $e$ and position angle $\rm PA$) and tends to follow morphological features such as bars and
rings. Consequently, these free fits are not suitable for the characterisation of the underlying stellar disc
studied in this paper. However, their $e(r)$ and $\rm PA\it(r)$ radial profiles may be used to determine the
$e$ and $\rm PA$ of the outer stellar disc component. This was achieved using an estimate for the semimajor
axis of the end of the stellar disc $a_{\rm disc\,lim}$ (where the galaxy surface brightness enters the
background noise) obtained from a visual inspection of the ACS image. A fixed-parameter fit (fixed centre,
$e$ and $\rm PA$ using the $e$ and $\rm PA$ determined for the outer disc) is then used to produce our final
measured $\mu(r)$ profiles. During these fixed-parameter fits, four iterations of a $3\sigma$ rejection are
applied to deviant points below and above the average in order to smooth some of the bumps in the surface
brightness profiles that are due to non-axisymmetric features, i.e. not part of the disc (e.g.\ star-forming
regions and supernovae). The necessary sky subtraction is then performed using the sky level estimates of
\cite{Gray_etal:2009} generated by the {\sc galapagos} pipeline (see Section~\ref{Sky subtraction}).

All our S0 $\mu(r)$ profiles are then corrected for Galactic foreground extinction, individual galaxy
inclination $i$ and surface brightness dimming (the $\mu$ profiles of our field galaxies,
$0.05 < z_{\rm phot} < 0.30$, are corrected to the redshift of the cluster $z_{\rm cl} = 0.167$). Full
details of the fitting procedure (performed on a different sample of galaxies), subsequent photometric
calibration and an estimation of the error in the sky subtraction can be found in \cite{Maltby_etal:2012a}.

\vspace{-0.1cm}
\subsection[]{Sky subtraction}

\label{Sky subtraction}

During the {\sc galfit} S{\'e}rsic model fitting performed by the {\sc galapagos} pipeline
\citep{Gray_etal:2009}, the sky level is calculated individually for each source galaxy by evaluating a flux
growth curve and using the full science frame. In this paper, for each sample galaxy we use the sky level
determined by {\sc galapagos} ($sky_{\rm gal}$) for our sky subtraction. The $1\sigma$ error in this sky
subtraction is $\pm0.18$ counts \citep[see][]{Maltby_etal:2012a}.

For our $\mu(r)$ profiles, the error in the sky subtraction dominates over the individual errors produced by
ellipse in the fitting process. At $\mu < 25\rm\,mag\,arcsec^{-2}$, the fit error dominates over the error in
the sky subtraction but has a negligible effect on the $\mu(r)$ profile. However, at
$\mu > 25\rm\,mag\,arcsec^{-2}$ the sky subtraction error dominates the error in the $\mu(r)$ profile. The
sky subtraction error can have a significant effect on the $\mu(r)$ profile of the S0 galaxies, especially in
the outer regions where the surface brightness $\mu$ approaches that of the sky background. However, for
any particular galaxy the global sky subtraction error is approximately constant across the length of the
$\mu(r)$ profile. Therefore, we can specify the error in our $\mu(r)$ profiles by generating profiles for
when the sky background is oversubtracted and undersubtracted by $\pm1\sigma$.

The $\pm1\sigma$ error in the sky background corresponds to a critical surface brightness limit
$\mu_{\rm crit}$ below which the sky subtracted $\mu(r)$ profile of a galaxy becomes unreliable. This
critical surface brightness $\mu_{\rm crit}$ is approximately $27.7\rm\,mag\,arcsec^{-2}$. We also define a
limiting surface brightness $\mu_{\rm lim}$, corresponding to a $\pm3\sigma$ sky error, below which
identifying profile breaks becomes unreliable. The limiting surface brightness $\mu_{\rm lim}$ is
approximately $26.5\rm\,mag\,arcsec^{-2}$.

\vspace{-0.1cm}
\subsection[]{Reliability of the {\sc galapagos} sky background}

\label{Reliability of sky subtraction}

\begin{figure*}
\centering
\includegraphics[width=0.79\textwidth]{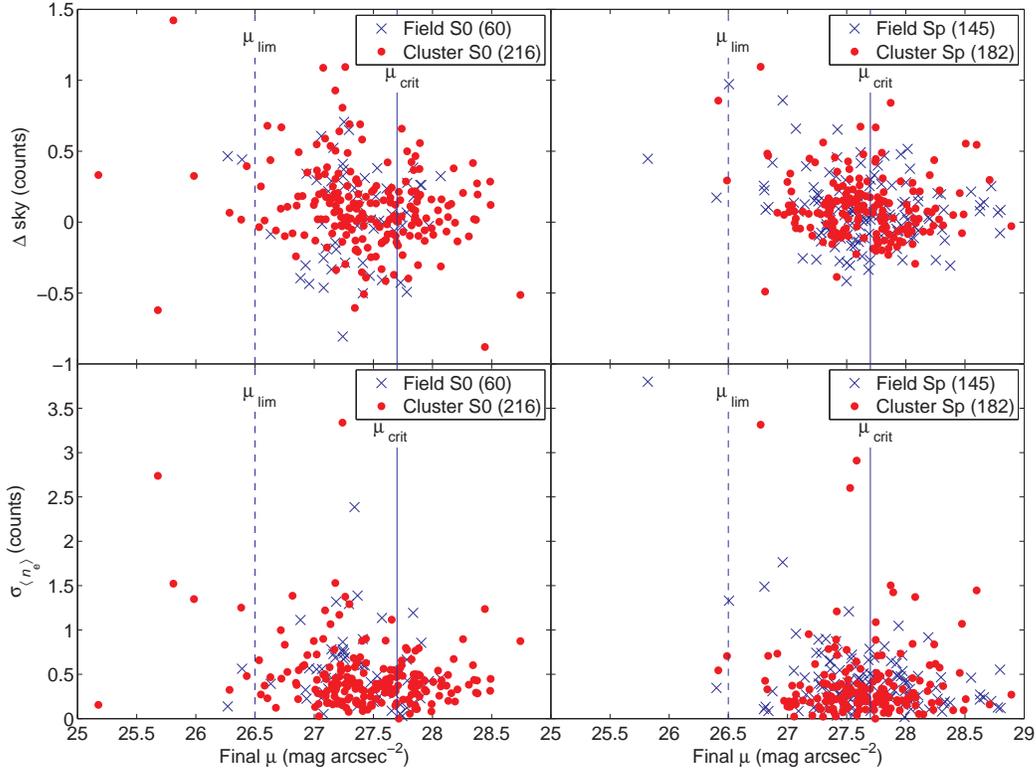}
\caption{\label{Sky comparisons} An evaluation of the sky subtraction for our S0 galaxies. Left-hand
panels: a plot of the final surface brightness reached by the isophotal fit $\mu(r\rightarrow\infty)$
against the difference in our two sky values $\Delta sky$ (top) and the corner-to-corner rms in the mean
sky value for each galaxy image $\sigma_{\langle{n_{\rm e}}\rangle}$ (bottom). Right-hand panels: a similar
evaluation for the spiral galaxies from \protect\cite{Maltby_etal:2012a}. Respective sample sizes are
shown in the legends.}
\end{figure*}

An initial inspection of our S0 $\mu(r)$ profiles revealed two key observations: i) many cases exhibiting
significant curvature throughout the $\mu(r)$ profile; and ii) a distinct lack of truncations (Type~II
features). We wanted to ensure that these observations were not just the manifestation of a sky subtraction
problem, causing the sky to be either oversubtracted or undersubtracted in our $\mu(r)$ profiles.
Therefore, we assess the reliability of our sky subtraction by the comparison of our {\sc galapagos} sky
values $sky_{\rm gal}$ with an additional rough estimate for the sky background $sky_{\rm est}$.

For each S0 galaxy, we obtain this sky estimate $sky_{\rm est}$ by using pixels obtained from the four
corners of the galaxy ACS image (postage stamp). The sizes of these ACS postage stamps are variable and were
designed to optimally contain the galaxy during the STAGES {\sc galfit} model fits (see
\citealt{Gray_etal:2009}, for full details)\footnote{Note: the size of the postage stamps are a multiple
($2.5\times$) of the \cite{Kron:1980} radius, and therefore by definition contain $\geq95$ per cent of the
subject galaxy's light \citep{Barden_etal:2012}.}. Consequently, in sampling the corners of the postage
stamp we have a reasonable expectation of probing the actual sky background. These corner pixels were
selected using quarter-circle wedges of side equal to $5$ per cent of the smallest image dimension. We then
apply our bad pixel masks to ensure only `dark' pixels are used and obtain the mean pixel value
$\langle n_{\rm e}\rangle$ in each wedge. The corner-to-corner rms in these mean pixel values
$\sigma_{\langle n_{\rm e}\rangle}$ is then calculated in order to determine if there is any large-scale
variation in the sky level across the galaxy image. In the vast majority of cases ($>90$ per cent),
$\sigma_{\langle{n_{\rm e}}\rangle}<1$ count (count $\equiv$ ACS pixel values). Finally, we obtain our sky
estimate $sky_{\rm est}$ by calculating the weighted mean of $\langle{n_{\rm e}}\rangle$ from the four
corners of the image
\begin{equation}
sky_{\rm est} = \sum_{i=1}^{4} w_i \langle{n_{\rm e}}\rangle_{i}.
\end{equation}
The weight factor $w_i$ is necessary due to the bad pixel masking and is given by
\begin{equation}
w_i = \frac{N_{i{\rm :total}} - N_{i{\rm :masked}}}{\sum_{i=1}^{4} N_{i{\rm :total}} - N_{i{\rm :masked}}},
\end{equation}
where $N_{i{\rm :masked}}$ is the number of flagged pixels and $N_{i{\rm :total}}$ is the total number of
pixels in the respective corner wedge.

In the vast majority of cases ($>90$ per cent), the agreement between the {\sc galapagos} sky level
$sky_{\rm gal}$ and our rough sky estimate $sky_{\rm est}$ was very good (\mbox{$|\Delta sky|<0.5$ counts},
where \mbox{$\Delta sky=sky_{\rm est}-sky_{\rm gal}$}). Furthermore, in $\sim50$ per cent of cases the
agreement was within the $\pm1\sigma$ error in $sky_{\rm gal}$ ($|\Delta sky| < 0.18$ counts; see
Section~\ref{Sky subtraction}). However, we need to ensure that any difference in these two sky values
($\Delta sky$), or indeed any variation in the sky level across the ACS image
($\sigma_{\langle n_{\rm e}\rangle}$), will not lead to a significant oversubtraction or undersubtraction of
the sky in our S0 $\mu(r)$ profiles and e.g.\ cause general curvature in the outer regions.

To address this issue, we compare $\Delta sky$ and $\sigma_{\langle n_{\rm e}\rangle}$ with the final mean
surface brightness reached by the galaxy $\mu(r)$ profile $\mu_{r\rightarrow\infty}$ (see
Fig.~\ref{Sky comparisons}). In general, $\mu_{r\rightarrow\infty}$ is taken to be the mean $\mu$ after the
oversubtracted $\mu(r)$ profile ($-1\sigma$ sky) drops below the $\mu_{\rm crit}$ level\footnote{Note: in
the $\sim5$ per cent of cases where the oversubtracted $\mu(r)$ does not drop below the $\mu_{\rm crit}$
level, we use the mean $\mu$ after the declining $\mu(r)$ profile levels off (i.e.\ enters the sky
background).}. The difference between $\mu_{r\rightarrow\infty}$ and $\mu_{\rm crit}$ ($1\sigma$ above the
sky) is a measure of the quality of the sky subtraction. Ideally $\mu_{r\rightarrow\infty}>\mu_{\rm crit}$,
but this is not always the case due to measurement errors in the sky background. In reality, the sky
subtraction is only suspect if $\mu_{r\rightarrow\infty}$ approaches the $\mu_{\rm lim}$ level ($3\sigma$
above the sky). In the few cases where $\mu_{r\rightarrow\infty}<\mu_{\rm lim}$, projection effects or nearby
stars are known to have affected the measured $\mu(r)$ and are flagged in our analysis. If a large potential
sky error (large $\Delta sky$) or a large sky variation (large $\sigma_{\langle n_{\rm e}\rangle}$) were
causing a significant error in our sky subtraction, one would expect a correlation between
$\mu_{r\rightarrow\infty}$ and either $\Delta sky$ or $\sigma_{\langle n_{\rm e}\rangle}$, respectively.
However, no such correlations are observed (see Fig.~\ref{Sky comparisons}). These results suggest that cases
where $\mu_{r\rightarrow\infty}$ approaches the $\mu_{\rm lim}$ level are not the consequence of a small
measurement error in our {\sc galapagos} sky background or small variations in the sky level across the ACS
image. Therefore, we conclude that our {\sc galapagos} sky values are robust and that any small sky errors
are not likely to effect the outcome of this study.

To further validate this result, we perform the same tests using the sample of spiral galaxies from
\cite{Maltby_etal:2012a}, where no sky subtraction problems were suspected. Reassuringly, the
distributions of $\mu_{r\rightarrow\infty}$ with $\Delta sky$ and $\sigma_{\langle n_{\rm e}\rangle}$ for
these spiral galaxies are essentially the same as for our S0s (see Fig~\ref{Sky comparisons}). Therefore,
we can conclude that the {\sc galapagos} sky values are adequate for this study, and that the general
curvature and lack of truncations observed in our S0 $\mu(r)$ profiles appear {\em not} to be a
manifestation of a sky subtraction error and are a real feature of our S0 galaxies.

\section[]{Profile Classification}

\label{Profile classification}

\subsection[]{Profile inspection}

\label{Profile inspection}

\begin{figure*}
\includegraphics[width=0.87\textwidth]{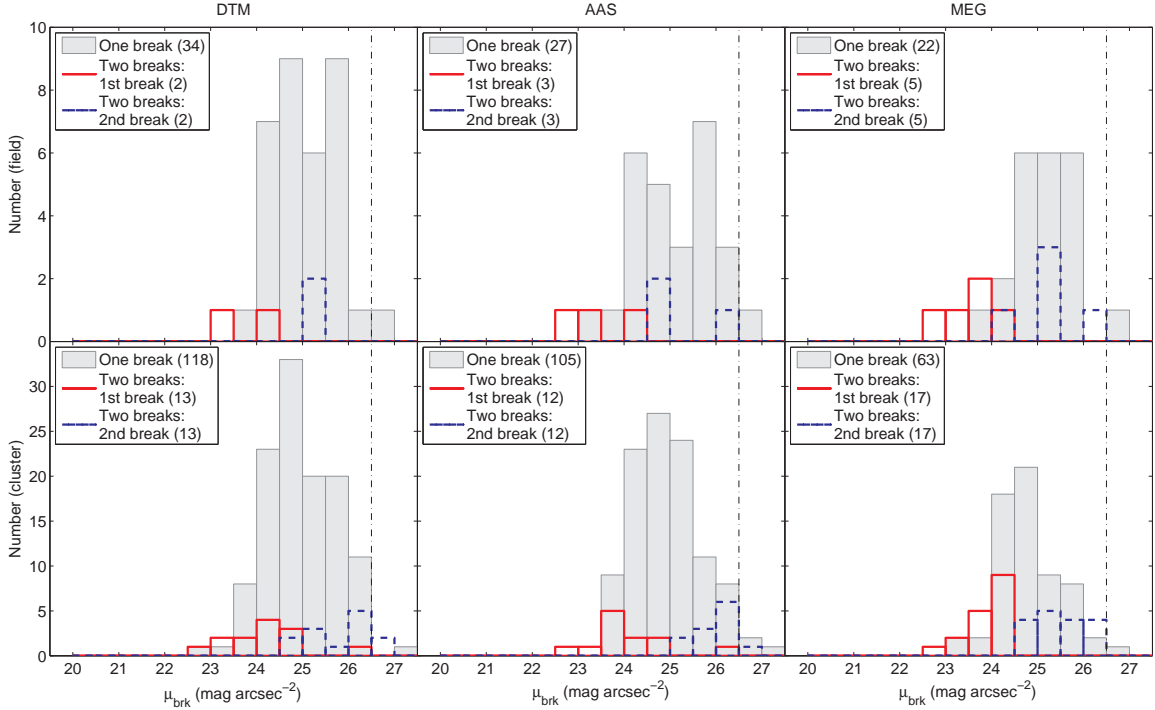}
\caption{\label{S0: break mu distribution} The distribution of break surface brightness $\mu_{\rm brk}$ for
our S0 galaxies. The surface brightness at the break radius $\mu_{\rm brk}$ for field (top row) and cluster
(bottom row) S0 galaxies as determined by DTM (left-hand column), AAS (centre column) and MEG (right-hand
column). The distributions show galaxies with one break (grey shaded area), and both the inner (red line) and
outer break (blue dashed line) of galaxies with two breaks. Respective sample sizes are shown in the legends.
Systematic errors in $\mu_{\rm brk}$ due to the error in the sky subtraction are
$<0.25\,{\rm mag\,arcsec}^{-2}$. Contamination of the cluster sample by the field is $<25$ per cent. Due to
the subjective nature of some galaxy profile classifications, the number of galaxies with either one or two
breaks varies subtly between the different assessors.}
\end{figure*}

For each S0 galaxy in our field and cluster sample, the azimuthally-averaged radial surface brightness
$\mu(r)$ profile was visually inspected in order to identify potential profile breaks (inflection points in
the exponential region of the $\mu$ profile). Due to the subjective nature of some profile classifications,
this inspection was carried out by three independent assessors (DTM, AAS, MEG). Four possible cases were
considered: i)~a simple exponential profile with no break; ii)~a single broken exponential either
down-bending or up-bending; iii)~cases with two profile breaks; and iv)~no discernible exponential component
(i.e.\ general curvature throughout the $\mu$ profile). In each case, break identification relates to the
outer disc component of the galaxy $\mu(r)$ profile and does not consider the inner varying bulge component.
We do not trust breaks with a break surface brightness $\mu_{\rm brk}$ fainter than the $\mu_{\rm lim}$ level
($26.5\,{\rm mag\,arcsec}^{-2}$, $3\sigma$ above the sky), as this is where break identification becomes
unreliable due to the deviation of the $\mu(r)$ profiles generated by oversubtracting and undersubtracting
the sky by $\pm1\sigma$. We therefore restrict our analysis to $\mu$ breaks that have
$\mu_{\rm brk} < 26.5\,{\rm mag\,arcsec}^{-2}$.

In this work, we follow the same procedure as \cite{Maltby_etal:2012a} for the identification of our $\mu$
profile breaks. Therefore, our break identification is based solely on the $\mu(r)$ profiles and without
direct inspection of the ACS images. As with \cite{Maltby_etal:2012a}, we have chosen not to relate the
$\mu(r)$ breaks to visually identified structural features because we wanted a break identification method
that treated all galaxies equally, in a self-consistent manner and avoided the prejudice that image inspection
could introduce. In addition to this, the aims of this work are to explore the effect of the galaxy
environment on the structure of S0 galactic discs, regardless of the origins of any identified structural
features. Note however, that although we use the same procedure as \cite{Maltby_etal:2012a} to identify breaks,
we do not adopt their classification scheme (based on $\mu_{\rm brk} > 24\rm\,mag\,arcsec^{-2}$) and use
the standard scheme based on the entire disc component (see Section~\ref{S0: profile types}).

If a $\mu$ profile break was identified, the radial limits of exponential regions either side of the break
radius $r_{\rm brk}$ were also estimated. For the inner exponential, the inner boundary is manually selected
to avoid the region dominated by the bulge component. For the outer exponential, the outer boundary is
generally taken to be where the $\mu(r)$ profile reaches the critical surface brightness $\mu_{\rm crit}$
($1\sigma$ above the sky) but may be at a higher $\mu$ depending on the nature of the profile. A small
manually selected transition region (non-exponential) is allowed between the exponential regions either side
of the break. The break radius $r_{\rm brk}$ is defined as the mean radius of the two radial limits for this
transition region.

The distributions of break surface brightness $\mu_{\rm brk}$ [$\mu(r_{\rm brk})$] for the breaks identified
by the three assessors are shown in Fig.~\ref{S0: break mu distribution}. The $\mu_{\rm brk}$
distributions for both one and two break cases are similar for each assessor. However, due to the subjective
nature of some galaxy profile classifications, the number of galaxies with either no, one, or two breaks,
varies subtly between the different assessors. To account for this, in what follows we perform parallel
analysis on the breaks identified by each assessor and compare the final results.

\subsection{Profile types}

\label{S0: profile types}

We classify our S0 galaxies into four main types: those classified to be Type~I, Type~II, or Type~III
depending on break features in their stellar disc; and Type~c -- cases with no significant exponential
component (general curvature). If the galaxy has a single exponential $\mu(r)$ profile with no break it is
classified as Type~I. If the $\mu(r)$ profile has a down-bending break then the galaxy has a stellar disc
truncation and is classified as Type~II. If the $\mu(r)$ profile has an up-bending break then the galaxy has
an antitruncation in the stellar disc and is classified as Type~III. However, if the $\mu(r)$ profile has no
discernible exponential component (general curvature throughout) then the galaxy is classified as Type~c.
Note that this classification scheme is based on the entire disc component and {\em not} on the outer disc
($\mu > 24\rm\,mag\,arcsec^{-2}$) as in \cite{Maltby_etal:2012a}.

\begin{figure*}
\includegraphics[width=0.300\textwidth]{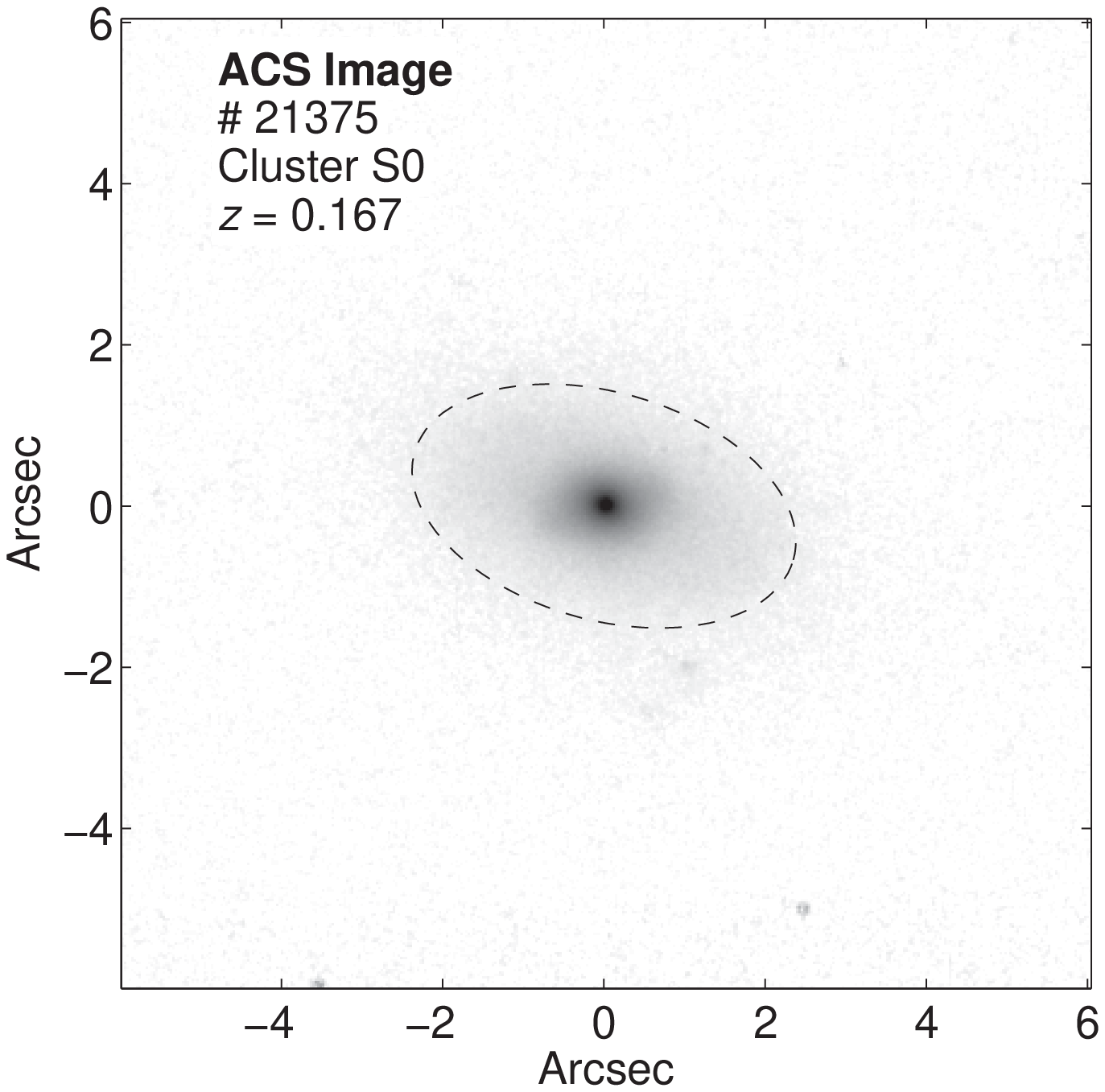}
\includegraphics[width=0.375\textwidth]{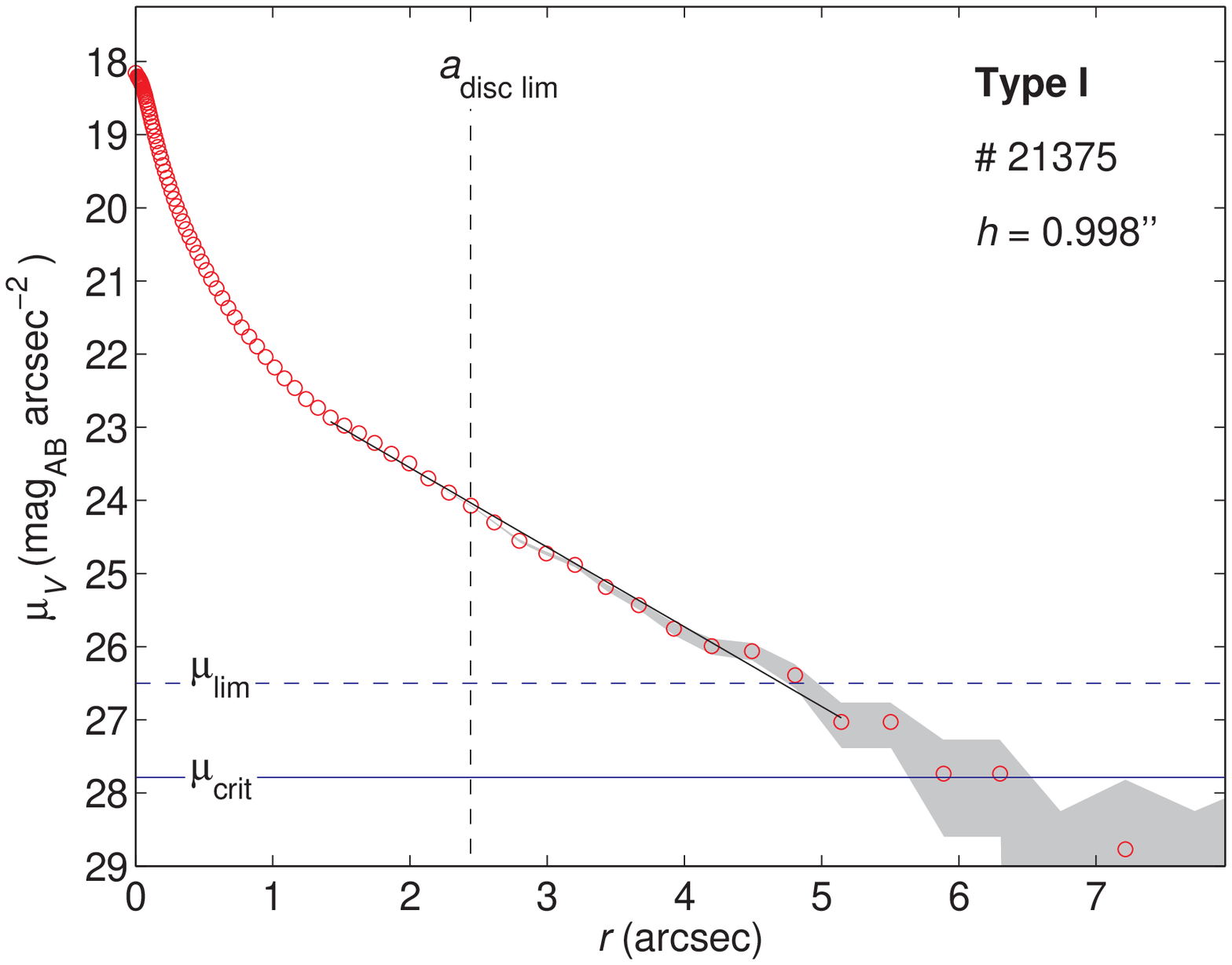}	\\
\includegraphics[width=0.300\textwidth]{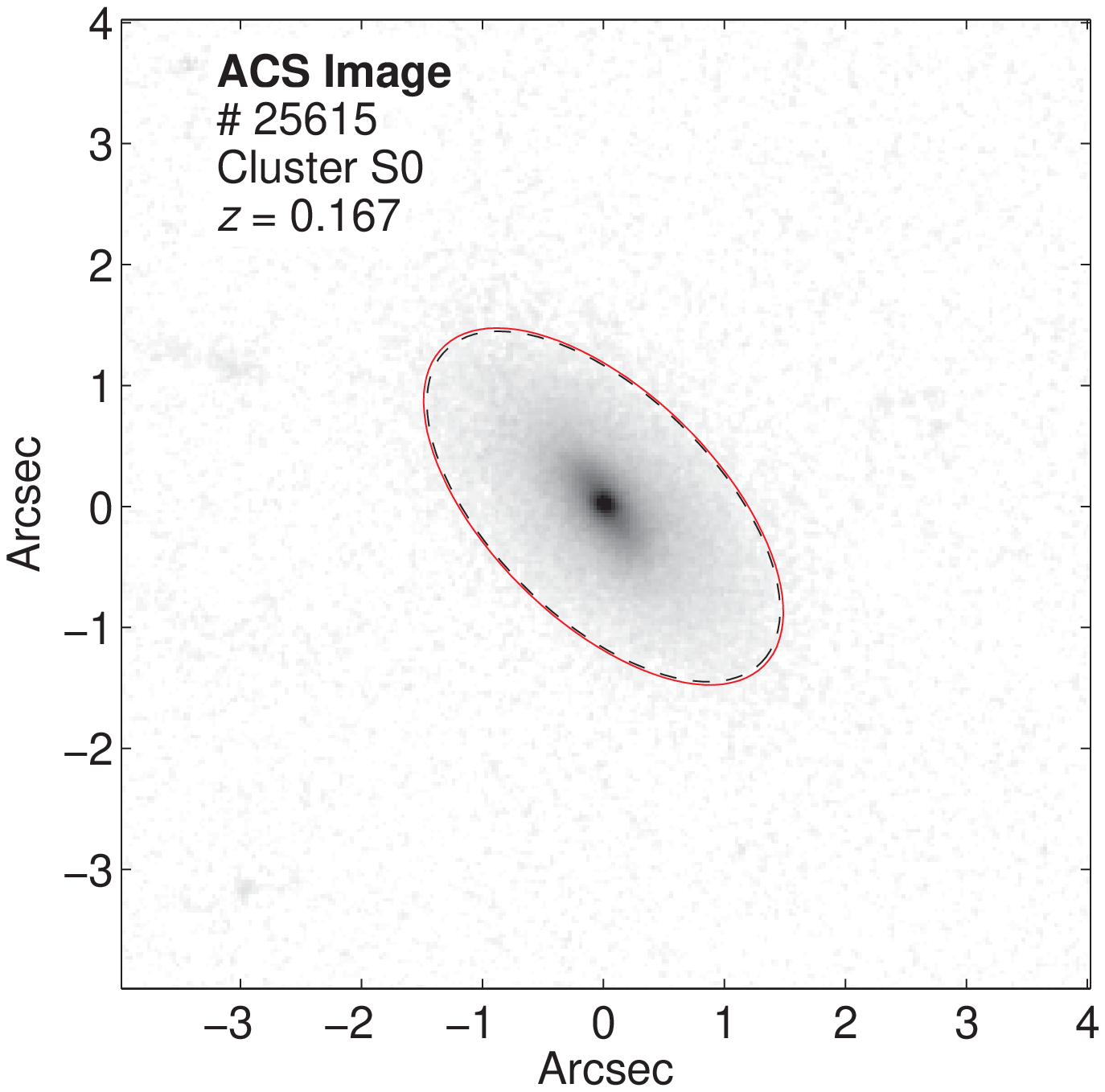}
\includegraphics[width=0.375\textwidth]{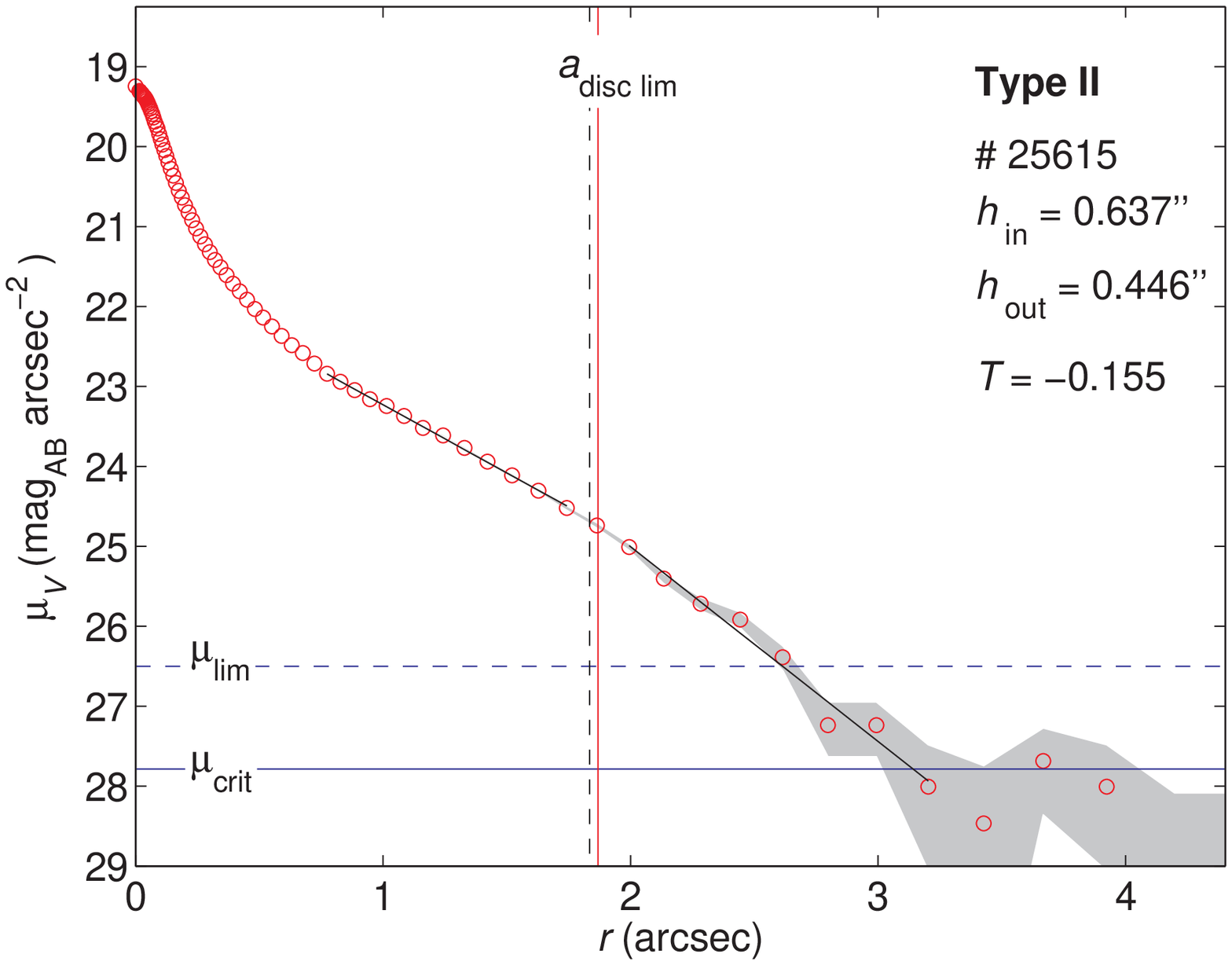}	\\
\includegraphics[width=0.300\textwidth]{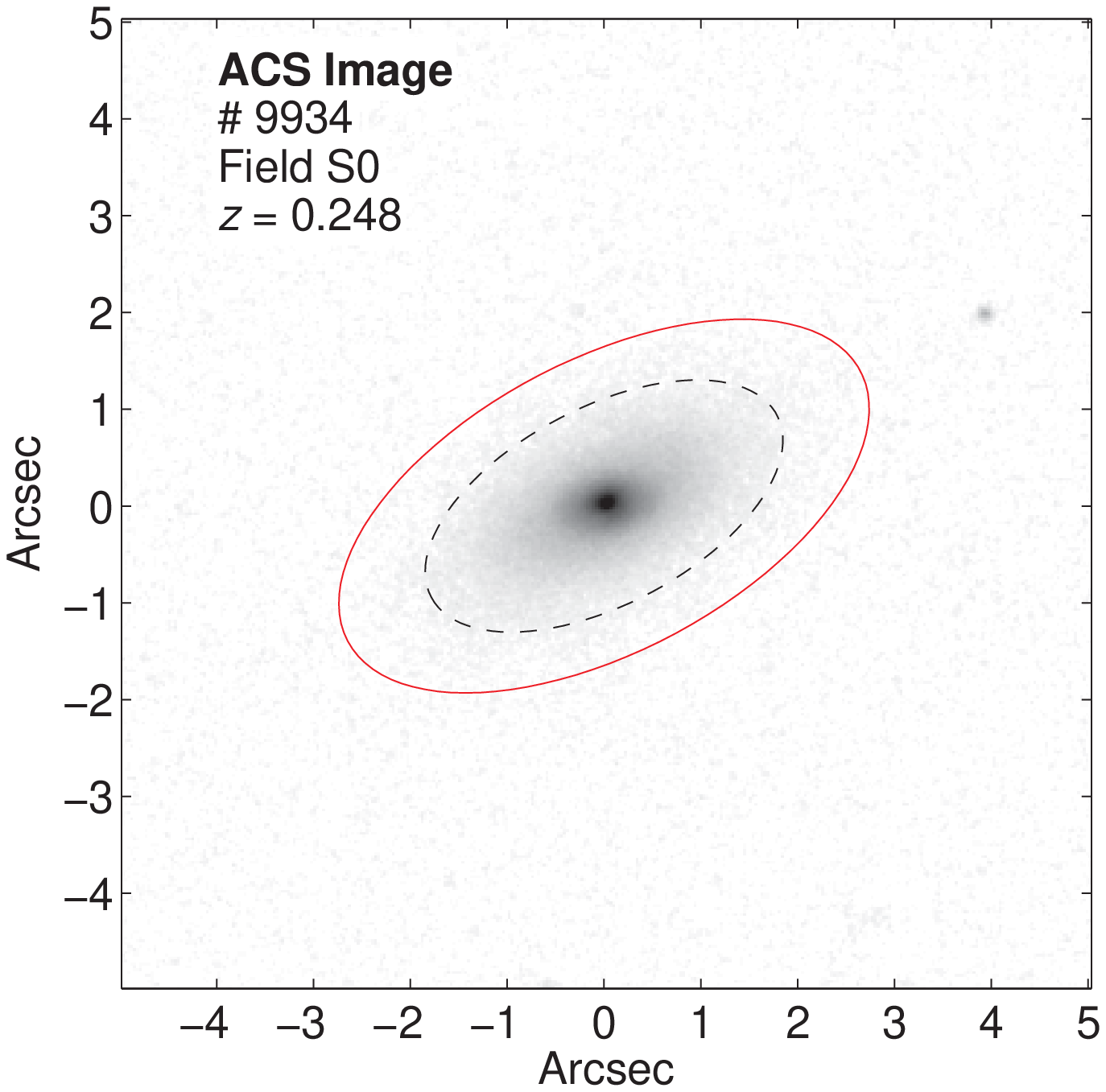}
\includegraphics[width=0.375\textwidth]{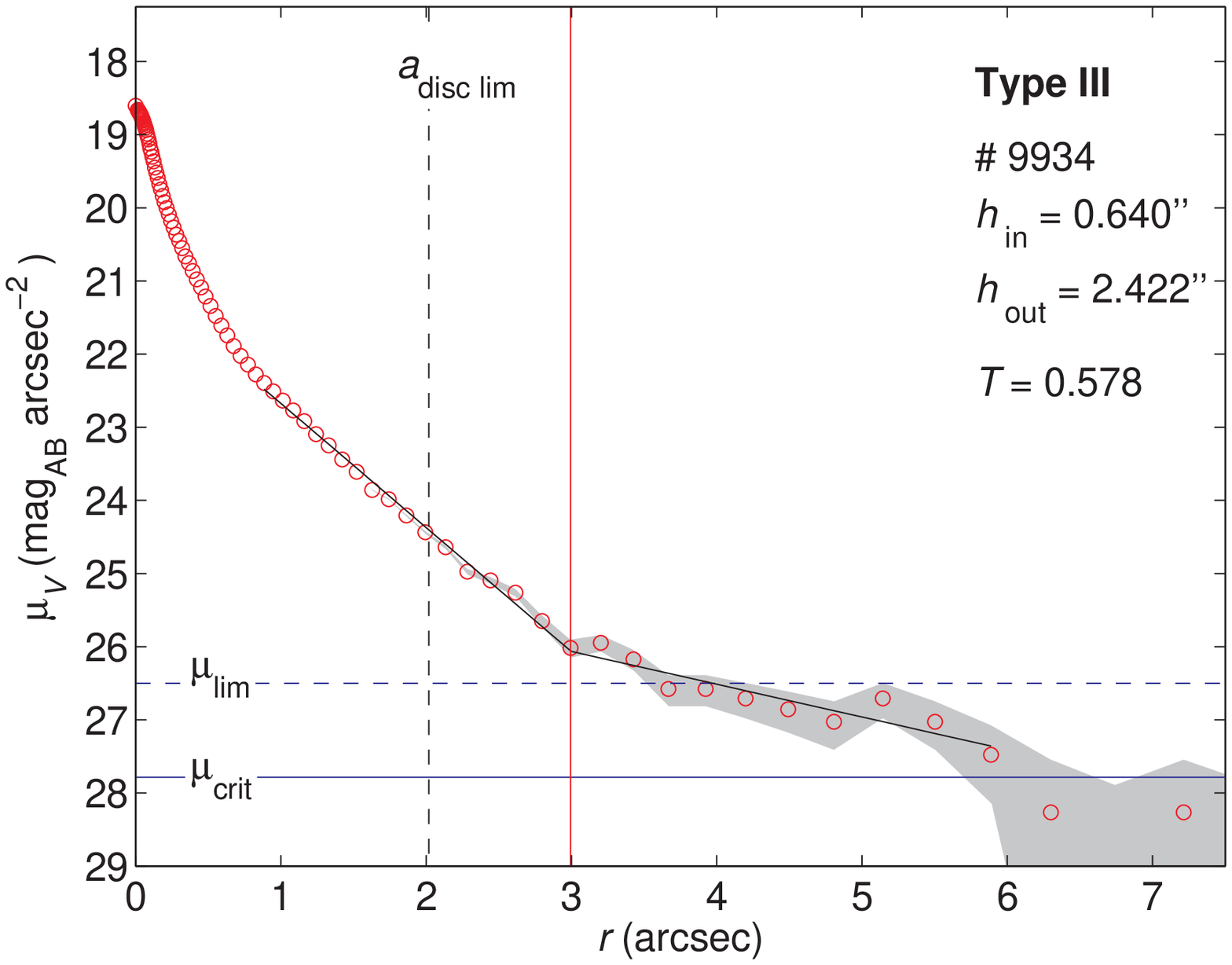}	\\
\includegraphics[width=0.300\textwidth]{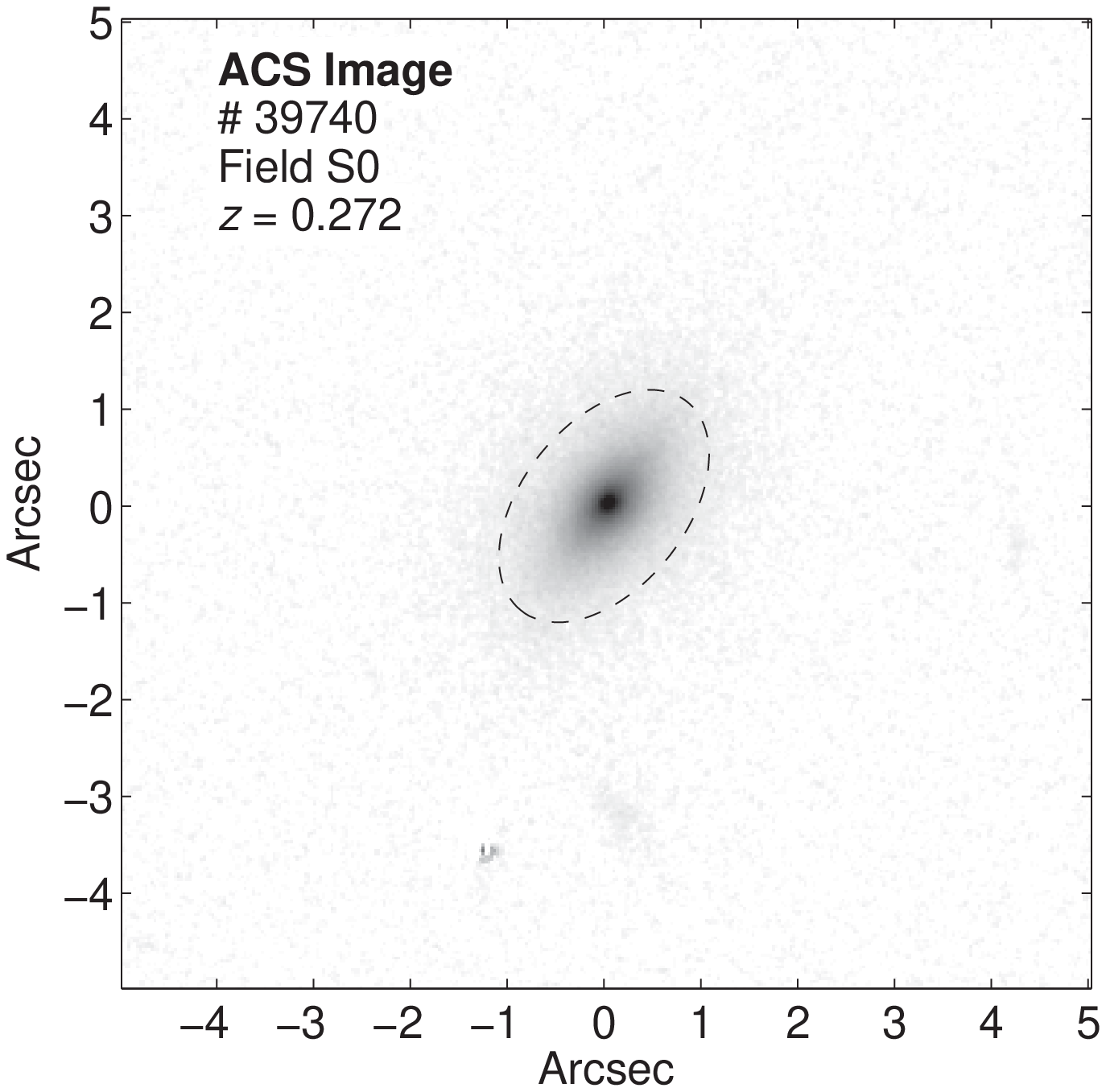}
\includegraphics[width=0.375\textwidth]{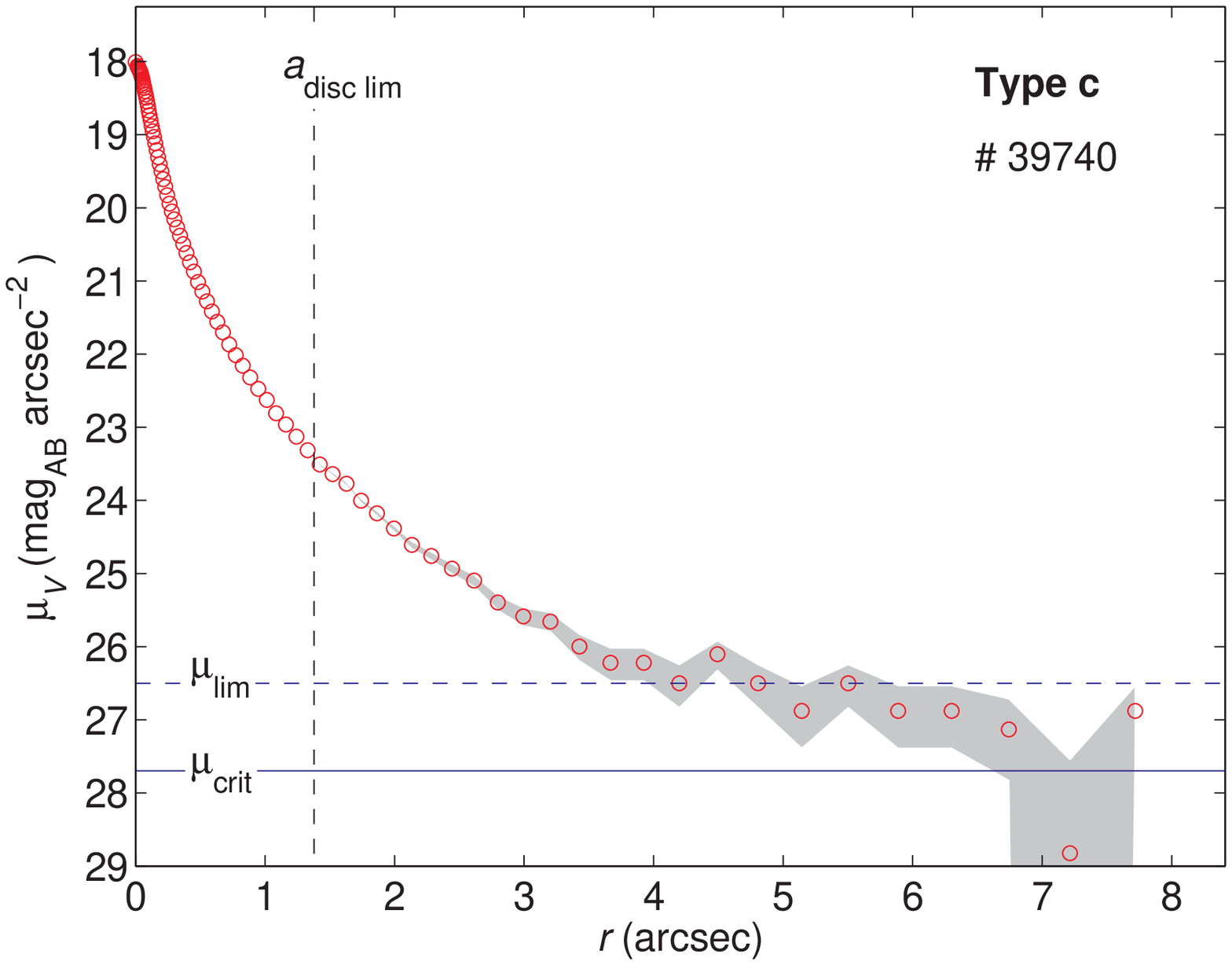}	\\
\caption{\label{Profile types} Examples of each class of S0 profile (DTM classification). Top to bottom:
Type~I, Type~II, Type~III and Type~c (no discernible exponential component, i.e.\ general curvature).
Left-hand panels: ACS $V$-band images. Right-hand panels: azimuthally-averaged $V$-band radial surface
brightness profiles. We overplot the break radii where applicable (red solid lines) and the stellar disc
limit $a_{\rm disc\,lim}$ (a visual estimate of the end of the stellar disc, black dashed line). The inner
and outer scalelength, $h_{\rm in}$ and $h_{\rm out}$, respectively, and the break strength $T$ are also
shown for reference. Errors in the $\mu(r)$ profiles are for an oversubtraction and an undersubtraction of
the sky by $1\sigma$. The $\mu_{\rm crit}$/$\mu_{\rm lim}$ levels represent $+1\sigma$/$+3\sigma$ above the
sky, respectively. The ACS images are in a logarithmic grey-scale. Note that in our samples, Type~II S0
profiles are very rare ($<5$ per cent).}
\end{figure*}

This classification assumes only one $\mu(r)$ break at most in the stellar disc. This is the case for
$\sim95$ per cent of our field and cluster S0 galaxies (see Fig.~\ref{S0: break mu distribution}). In this
study, we wish to consider the effect of the galaxy environment on the outer regions of S0 stellar discs.
Therefore, if two breaks are present the outer break is used for classification as any effect of the
environment should be stronger in the outer, fainter and more fragile break. Examples of each profile type
(Type I, II, III and Type~c) are shown in Fig.~\ref{Profile types} along with their ACS images showing the
break radius $r_{\rm brk}$ isophote.

\subsection{Measuring scalelength and break strength}

\label{Measuring scalelength and break strength}

In the following section, we obtain our exponential fits by using a linear least-squares fit to the original
$\mu(r)$ profile between the radial limits identified during the visual inspection (see
Section~\ref{Profile inspection}).

For S0 galaxies with no $\mu(r)$ profile break in their disc component (Type~I, pure exponentials), we obtain
the disc scalelength $h$ from a simple exponential fit across the length of the disc component
\begin{equation}
h = 1.086\times\Delta{r}/\Delta{\mu_{\rm fit}(r)},
\end{equation}
where $\mu_{\rm fit}$ is the surface brightness from the exponential fit. For these galaxies, the mean
random error in scalelength due to the exponential fitting routine is $<10$ per cent and the mean systematic
error in scalelength due to the error in the sky subtraction $\pm1\sigma$ (see Section~\ref{Sky subtraction})
is also $<10$ per cent.

For S0 galaxies where a $\mu(r)$ profile break was observed (Type~II/III), we obtain the inner/outer
scalelength $h$ from exponential fits to the stellar disc either side of the break radius $r_{\rm brk}$. The
inner exponential disc extends from a radius of $r_{\rm in,min}$ to $r_{\rm in,max}$ and has a scalelength
$h_{\rm in}$ given by
\begin{equation}
h_{\rm in} = 1.086\times\frac{r_{\rm in,max}-r_{\rm in,min}}{\mu_{\rm fit}(r_{\rm in,max})-\mu_{\rm fit}
(r_{\rm in,min})},
\end{equation}
Similarly, the outer exponential disc extends from a radius of $r_{\rm out,min}$ to $r_{\rm out,max}$ and has
a scalelength $h_{\rm out}$ given by
\begin{equation}
h_{\rm out} = 1.086\times\frac{r_{\rm out,max}-r_{\rm out,min}}{\mu_{\rm fit}(r_{\rm out,max})-\mu_{\rm fit}
(r_{\rm out,min})}.
\end{equation}
For these galaxies, the mean random error in scalelength due to the exponential fitting routine is $<10$ per
cent for $h_{\rm in}$ and $<20$ per cent for $h_{\rm out}$.

In the outer regions of the surface brightness profile, negative sky-subtracted flux can occur. As surface
brightness $\mu$ cannot be defined for a negative flux, these points are removed from our linear exponential
fits to the $\mu(r)$ profile. However, the removal of these negative fluxes is not expected to introduce any
bias on our scalelength $h$ measurements \citep{Maltby_etal:2012a}. Some example $\mu(r)$ profiles with
fitted exponential regions and overplotted break radii are shown in Fig.~\ref{Profile types}. The ACS images
and $\mu(r)$ profiles for all our S0 galaxies are presented in Appendix~A (on-line version only).

In order to measure the strength of our Type~II/III profile breaks, we define a break strength $T$ as the
logarithm of the outer-to-inner scalelength ratio,
\begin{equation}
T = {\rm log}_{10}\,h_{\rm out}/h_{\rm in}.
\end{equation}
A Type~I galaxy (pure exponential) has no break and therefore has a break strength of $T = 0$. A Type~II
galaxy (down-bending break, truncation) has a smaller outer scalelength $h_{\rm out}$ with respect to its
inner scalelength $h_{\rm in}$, and therefore has a negative break strength ($T < 0$). Similarly, a Type~III
galaxy (up-bending break, antitruncation) has a larger outer scalelength $h_{\rm out}$ with respect to its
inner scalelength $h_{\rm in}$, and therefore has a positive break strength ($T > 0$). For our S0 galaxy
samples, the mean random error in $T$ due to the exponential fitting routine is $\sim\pm0.1$ ($<20$ per
cent) and the mean systematic error in $T$ due to the sky subtraction error is also $\sim\pm0.1$.

In this study, the majority of our field S0s have a redshift \mbox{$z\sim0.23$} while our cluster S0s have
$z\sim0.167$. However, evolutionary effects are not expected to have a significant impact on our results. The
break strength $T$ of our field galaxies show no correlation with redshift and evolutionary effects on
the disc scalelength $h$ between the mean redshifts of our field and cluster samples are only $\sim5$ per cent
(based on the fits of \citealt{Buitrago_etal:2008}, for the expected size evolution of massive disc galaxies).

\vspace{-0.45cm}
\section[]{Results}

\label{Results}

The frequencies of the profile classifications by each assessor (DTM, AAS, MEG) for S0 galaxies in the field
and cluster environments are shown in Table~\ref{S0 profile frequencies}. Profile frequencies are also shown
for the low-axis-ratio ($q > 0.5$) S0 sub-sample (see Section~\ref{Galaxy inclination}). These profile
classifications are based on single disc breaks only and in multiple break cases the outer break is used for
classification. The $1\sigma$ uncertainty in the frequency/fraction of profile types
$\delta{f_{i}}$ ($f_{i} = N_{i}/N_{\rm tot}$) is determined using the \cite{Wilson:1927} binomial confidence
interval
\begin{equation}
\label{Equation: Frequency error}
f_{i} \pm \delta{f_{i}} = \frac {N_{i} + \kappa^2/{2}}{N_{\rm tot} + \kappa^2} \pm 
\frac{\kappa\sqrt{N_{\rm tot}}}{N_{\rm tot} + \kappa^2}\sqrt{f_{i}(1- f_{i}) + \frac{\kappa^2}{4N_{\rm tot}}},
\end{equation}
where $\kappa$ is the $100(1-\alpha/2)\rm{th}$ percentile of a standard normal distribution ($\alpha$ being
the error percentile corresponding to the $1\sigma$ level; see \citealt{Brown_etal:2001} for further details).

\begin{table*}
\begin{minipage}{145mm}
\centering
\caption{\label{S0 profile frequencies}{The frequency of profile types for S0 galaxies in the field and
cluster environments and for the three independent assessors (DTM, AAS, MEG). Profile frequencies are also
shown for the low-axis-ratio ($q > 0.5$) S0 sub-sample. Percentage errors are calculated using equation
\ref{Equation: Frequency error}.}}
\begin{tabular}{lcccccc}
\hline
{Assessor}				&\multicolumn{3}{c}{Disc profile types}								&{Curvature}			&{Unclassified}			\\
{}					&{Type~I}			&{Type~II}			&{Type~III}			&{Type~c}			&{}				\\[0.5ex]
\hline																									\\[-1.5ex]
{Field S0 galaxies}			&{}				&{}				&{}				&{}				&{}				\\
{DTM}					&{$18$ ($30\pm6\,\%$)}		&{$1$ ($2\,^{+3}_{-1}\,\%$)}	&{$34$ ($57\pm6\,\%$)}		&{$5$  ($8\,^{+4}_{-3}\,\%$)}	&{$2$ ($3\,^{+3}_{-2}\,\%$)}	\\[0.3ex]
{AAS}					&{$18$ ($30\pm6\,\%$)}		&{$0$ ($0+2\,\%$)}		&{$29$ ($48\pm6\,\%$)}		&{$13$ ($22\,^{+6}_{-5}\,\%$)}	&{$0$ ($0+2\,\%$)}		\\[0.3ex]
{MEG}					&{$17$ ($28\,^{+6}_{-5}\,\%$)}	&{$0$ ($0+2\,\%$)}		&{$26$ ($43\pm6\,\%$)}		&{$14$ ($23\,^{+6}_{-5}\,\%$)}	&{$3$ ($5\,^{+4}_{-2}\,\%$)}	\\[1.5ex]
{Cluster S0 galaxies}			&{}				&{}				&{}				&{}				&{}				\\
{DTM}					&{$43$ ($20\pm3\,\%$)}		&{$8$ ($4\,^{+2}_{-1}\,\%$)}	&{$122$ ($56\pm3\,\%$)}		&{$37$ ($17\,^{+3}_{-2}\,\%$)}	&{$6$ ($3\pm1\,\%$)}		\\[0.3ex]
{AAS}					&{$58$ ($27\pm3\,\%$)}		&{$5$ ($2\pm1\,\%$)}		&{$109$ ($50\pm3\,\%$)}		&{$41$ ($19\pm3\,\%$)}		&{$3$ ($1\pm1\,\%$)}		\\[0.3ex]
{MEG}					&{$56$ ($26\pm3\,\%$)}		&{$2$ ($1\pm1\,\%$)}		&{$77$ ($36\pm3\,\%$)}		&{$68$ ($31\pm3\,\%$)}		&{$13$ ($6\,^{+2}_{-1}\,\%$)}	\\[1.5ex]
{Field S0 galaxies ($q > 0.5$)}		&{}				&{}				&{}				&{}				&{}				\\
{DTM}					&{$11$ ($31\,^{+8}_{-7}\,\%$)}	&{$1$ ($3\,^{+4}_{-2}\,\%$)}	&{$20$ ($56\pm8\,\%$)}		&{$3$  ($8\,^{+6}_{-4}\,\%$)}	&{$1$ ($3\,^{+4}_{-2}\,\%$)}	\\[0.3ex]
{AAS}					&{$13$ ($36\,^{+8}_{-7}\,\%$)}	&{$0$ ($0+3\,\%$)}		&{$14$ ($39\pm8\,\%$)}		&{$9$ ($25\,^{+8}_{-6}\,\%$)}	&{$0$ ($0+3\,\%$)}		\\[0.3ex]
{MEG}					&{$14$ ($39\pm8\,\%$)}		&{$0$ ($0+3\,\%$)}		&{$13$ ($36\,^{+8}_{-7}\,\%$)}	&{$8$ ($22\,^{+8}_{-6}\,\%$)}	&{$1$ ($3\,^{+4}_{-2}\,\%$)}	\\[1.5ex]
{Cluster S0 galaxies ($q > 0.5$)}	&{}				&{}				&{}				&{}				&{}				\\
{DTM}					&{$37$ ($27\pm4\,\%$)}		&{$6$ ($4\,^{+2}_{-1}\,\%$)}	&{$71$ ($52\pm4\,\%$)}		&{$17$ ($12\pm3\,\%$)}		&{$6$ ($4\,^{+2}_{-1}\,\%$)}	\\[0.3ex]
{AAS}					&{$47$ ($34\pm4\,\%$)}		&{$3$ ($2\,^{+2}_{-1}\,\%$)}	&{$60$ ($44\pm4\,\%$)}		&{$24$ ($18\pm3\,\%$)}		&{$3$ ($2\,^{+2}_{-1}\,\%$)}	\\[0.3ex]
{MEG}					&{$45$ ($33\pm4\,\%$)}		&{$2$ ($1\pm1\,\%$)}		&{$40$ ($29\pm4\,\%$)}		&{$38$ ($28\pm4\,\%$)}		&{$12$ ($9\,^{+3}_{-2}\,\%$)}	\\[0.5ex]
\hline
\end{tabular}
\end{minipage}
\end{table*}

Due to the subjective nature of some profile classifications, the frequency obtained for each profile type
varies subtly between the different assessors. The agreement between the assessors is generally very
good, especially for Type~I and II profiles. However, the agreement is slightly weaker for the
frequencies of Type~III and Type~c (non-exponential) profiles due to the increased level of subjectivity
involved in differentiating between these two profile types.

The frequencies of the profile types (Type~I, II and III) are approximately the same in the field and cluster
environments. For both field and cluster S0s, $\sim25$ per cent have a simple exponential profile
(Type~I), $<5$ per cent exhibit a down-bending break (truncation, Type~II) and $\sim50$ per cent exhibit an
up-bending break (antitruncation, Type~III)\footnote{Note: in Section~\ref{Antitruncations in S0 galaxies},
we find that a significant fraction of our Type~III S0 profiles (up to $\sim50$ per cent) may actually be
related to light from an extended bulge component and not an antitruncated stellar disc.}. The frequency of
profiles with no discernible exponential component (i.e.\ general curvature, Type~c) is also approximately
the same in the field and cluster environments and is $\sim20$ per cent. Restricting our analysis to low
inclination systems ($i < 60^\circ$, $q > 0.5$) affects our profile fractions by $\sim5$ per cent but has no
effect on our conclusions (see Table~\ref{S0 profile frequencies}). These results suggest that the profile
type of S0 galaxies is not significantly affected by the galaxy environment from the general field to the
intermediate densities of the STAGES A901/2 clusters.

\vspace{-0.35cm}
\subsection{The absence of S0 Type~II profiles}

\label{The absence of S0 Type II profiles}

The distinct lack of disc truncations (Type~II profiles) in both our field and cluster S0s is of particular
interest. Type~II profiles are very common in spiral galaxies, occurring in approximately $40$--$60$ per cent
of cases \citep{Pohlen_Trujillo:2006, Erwin_etal:2008, Gutierrez_etal:2011}. Therefore, it seems whatever
process transforms spiral galaxies into S0s may well erase these truncations from their $\mu(r)$ profiles.
This result is in partial agreement with a similar result reported recently by \cite{Erwin_etal:2012}. Using
a sample of $\sim70$ field and cluster S0 galaxies, \cite{Erwin_etal:2012} find no Type~II S0s in the cluster
environment but a Type~II fraction of $\sim30$ per cent for their field S0s. Therefore, our Type~II S0
fractions are in perfect agreement with \cite{Erwin_etal:2012} for the cluster environment, but differ
significantly for the field. The origin of this disagreement is uncertain but may be related to the low
fraction of barred S0s in our field sample (see Table~\ref{Sample properties}).

In order to explore this issue, it is important to note that Type~II profiles can be further classified into
two main sub-types depending on their potential origin -- Type~II-CT and Type~II-OLR
\citep{Pohlen_Trujillo:2006, Erwin_etal:2008}. Type-II-CT profiles (classical truncations) are cases where
the truncation appears to be related to a radial star formation threshold, whereas Type~II-OLR profiles are
cases where the presence of a bar appears to have introduced a Type~II feature related to the outer Lindblad
resonance (OLR). In our field sample, the bar fraction is only $\sim10$ per cent (see
Table~\ref{Sample properties}). This contrasts with a field bar fraction of $>50$ per cent in
\cite{Erwin_etal:2012}. Consequently, the lower fraction of barred S0s (and hence Type~II-OLR profiles) in
our field sample could easily be the origin of the observed differences between our results and those of
\cite{Erwin_etal:2012}.

An explanation for the lack of Type~II profiles in our S0s can be hypothesised by considering their potential
origin. For classical truncations (Type~II-CT), current theories suggest that their formation is via a radial
star formation threshold and the outward scattering of inner disc stars to regions beyond this threshold
\citep[i.e.\ break radius $r_{\rm brk}$; e.g.][]{Debattista_etal:2006}. In these cases, the outer disc should
be populated by old stars as these are the ones that have had enough time to make the disc migration. The
discovery that the stellar mass surface density $\Sigma^{M_*}(r)$ profiles of late-type Type~II galaxies tend
to be purely exponential \citep{Bakos_etal:2008, Martinez-Serrano_etal:2009} supports this scenario and
suggests that Type~II $\mu(r)$ breaks are {\em not} related to the stellar mass distribution but due to a
radial change in the age of the stellar population. Assuming this formation scenario, and an inside-out growth
for the inner disc [i.e.\ negative age$(r)$ gradient], when star formation is suppressed throughout the galaxy
(e.g.\ via gas stripping) the age of the stellar population in the `break region' will gradually increase. As
a result, the relative difference in stellar population age between the break region and inner/outer discs
will decrease and the mass-to-light ratio ($M/L$) across the $\mu(r)$ break will converge. Consequently, the
$\mu$ break will get weaker and may even disappear. For Type~II-OLR profiles, the $\mu$ break is expected to
be related to a resonance phenomenon and therefore the above scenario does not hold. However,
\cite{Erwin_etal:2012} suggest that the depletion/removal of gas from a barred galaxy would cause a weakening
of the OLR effect and may weaken or remove the Type~II-OLR break from the $\mu(r)$ profile. Considering these
theories, and since spiral galaxies are thought to transform into S0s via the depletion/removal of gas and
the subsequent termination of star formation \citep[e.g.][]{Aragon-Salamanca_etal:2006}, it seems natural to
expect Type~II $\mu(r)$ breaks to be weaker/rarer in S0s compared to spiral galaxies.

\subsection[]{S0: pure exponential discs (Type~I)}

\label{S0: Type I}

\begin{figure*}
\includegraphics[width=0.84\textwidth]{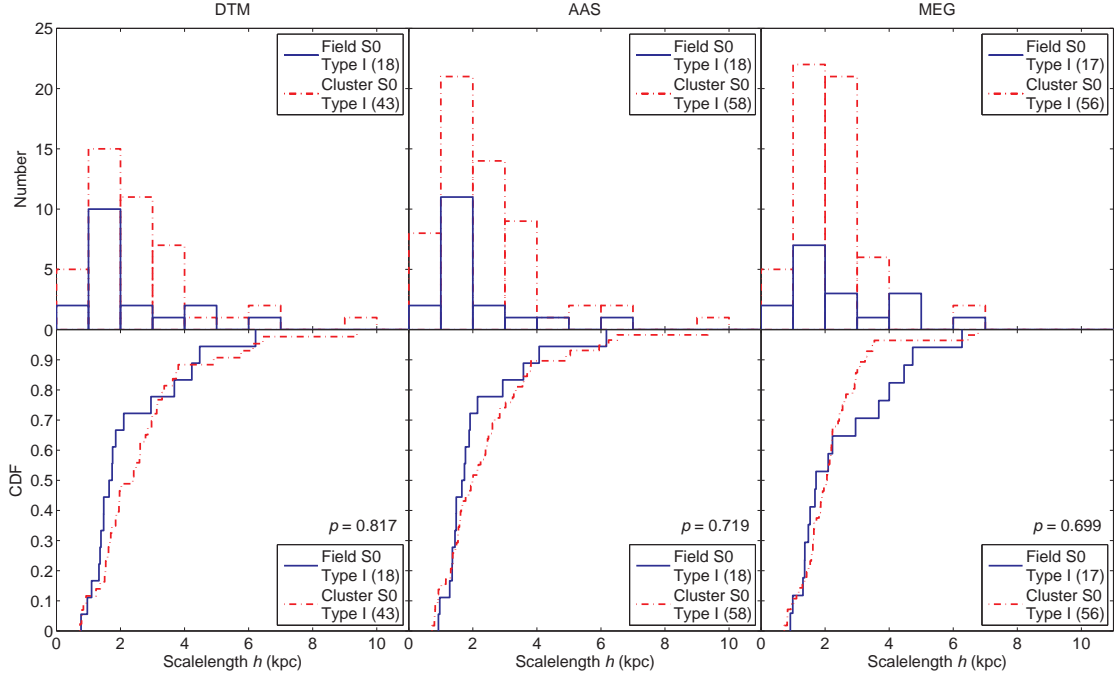}
\caption{\label{S0: Type I analysis} Comparing Type~I disc scalelength $h$ distributions in different
environments for S0 galaxies (full sample). Top row: scalelength $h$ distributions for Type I S0 galaxies in
the field (blue line) and cluster (red dashed line) environment as classified by DTM (left-hand panel), AAS
(centre panel) and MEG (right-hand panel). Bottom row: the corresponding scalelength $h$ CDFs showing the
probability $p$ that compared samples are {\em not} drawn from the same continuous $h$ distributions in the
bottom right of each plot. Respective sample sizes are shown in the legends. Random errors in scalelength
are typically $<10$ per cent. Systematic errors in scalelength due to the error in the sky subtraction are
also typically $<10$ per cent. Contamination of the cluster sample by the field is $<25$ per cent. We find
no significant difference between the CDFs in each environment and no evidence to suggest that the
scalelengths $h$ of our Type~I S0 galaxies are {\em not} drawn from the same continuous $h$ distributions.}
\end{figure*}

For S0s where no $\mu(r)$ break was identified (Type~I), we compare the scalelength $h$ distributions in the
field and cluster environments to see whether there is any evidence for an environmental dependence on the
scalelength $h$ (see Fig.~\ref{S0: Type I analysis}). In these comparisons, our scalelengths $h$ were
transformed into intrinsic linear scales using the fixed cluster redshift ($z_{\rm cl} = 0.167$) for our
cluster S0s and the COMBO-17 photo-$z$ estimate for our field S0s. Therefore, photo-$z$ errors only propagate
into the intrinsic scalelengths of our field galaxies and not our cluster galaxies. The mean error in $h$
associated with this photo-$z$ error is $<10$ per cent \citep[i.e.\ the error in the distance to the
galaxy;][]{Maltby_etal:2010}.

For our $\mu(r)$ profiles, the error in the sky background (see Section~\ref{Sky subtraction}) can have a
significant effect on both our scalelength $h$ and break strength $T$ measurements, especially at large
radii where the $\mu(r)$ profile approaches the critical surface brightness $\mu_{\rm crit}$
($27.7\rm\,mag\,arcsec^{-2}$, $1\sigma$ above the sky). However, for any particular galaxy the sky
subtraction error can be taken to be approximately constant across the length of the $\mu(r)$ profile.
Therefore, we can account for this error by performing parallel analyses for when the sky background is
oversubtracted and undersubtracted by $\pm1\sigma$ ($\pm0.18$ counts). For our Type~I S0s, the mean error in
$h$ due to the sky subtraction error is $\sim\pm0.3\rm\,kpc$ ($<10$ per cent). Random errors in $h$ due to
the exponential fitting routine are also typically $<10$ per cent (see
Section~\ref{Measuring scalelength and break strength}). We also perform parallel analyses on the Type~I
samples generated by the three assessors (DTM, AAS, MEG) in order to account for the subjective nature of the
profile classifications and compare the final results.

In all cases (all parallel analyses), we observe no clear difference between the distributions of scalelength
$h$ for our Type~I S0s in the field and cluster environments (see Fig.~\ref{S0: Type I analysis}). This is
also the case when considering only the low-axis-ratio ($q > 0.5$) S0 sub-sample.
%
In order to test the significance of these results, we construct scalelength $h$ cumulative distribution
functions (CDFs, see Fig.~\ref{S0: Type I analysis}) for our Type~I S0 samples and perform
Kolmogorov--Smirnov (K--S) tests between corresponding samples from the field and cluster environments. These
K--S tests are used in order to obtain the probability $p_{\rm(field/cluster)}$ that the field and cluster
Type~I S0 samples are {\em not} drawn from the same continuous $h$ distributions. The results of these K--S
tests, for both the full S0 sample and the low-axis-ratio ($q > 0.5$) S0 sub-sample, are shown in 
Table~\ref{S0: K-S results}.

In this study, we only consider an environmental effect on the Type~I scalelength $h$ to be significant if
K--S tests yield a $2\sigma$ level probability for $p_{\rm(field/cluster)}$. However,
$p_{\rm(field/cluster)}$ is below the $2\sigma$ level for each assessor and for when the sky background is
oversubtracted and undersubtracted by $\pm1\sigma$ (see Table~\ref{S0: K-S results}). This is also the case
for the low-axis-ratio S0 sub-sample. Therefore, we find no evidence to suggest that the disc scalelength $h$
of our Type~I S0s is dependent on the galaxy environment. This result is also robust to the error in the sky
subtraction and the subjective nature of the profile classifications.


\begin{table*}
\centering
\begin{minipage}{120mm}
\centering
\caption{\label{S0: K-S results} The K--S test results for Type~I and Type~III S0s as classified by DTM,
AAS and MEG. Results are also shown for the low-axis-ratio ($q > 0.5$) S0 sub-sample. \mbox{K--S} tests give
the probability $p_{\rm(field/cluster)}$ that the respective field and cluster samples are {\em not} drawn
from the same $h$ distributions for Type~I S0s and $\mu_{\rm brk}$/$T$ distributions for Type~III S0s.
Results are also shown for when the sky is oversubtracted and undersubtracted by $\pm1\sigma$. We find no
environmental dependence on the structural properties of the stellar disc ($h$, $\mu_{\rm brk}$, $T$) in our
S0 galaxies.}
\begin{tabular}{lccccccc}
\hline
{}						&\multicolumn{3}{c}{S0 sample}			&{}	&\multicolumn{3}{c}{S0 sub-sample ($q > 0.5$)}	\\
{}						&\multicolumn{3}{c}{$p_{\rm(field/cluster)}$}	&{}	&\multicolumn{3}{c}{$p_{\rm(field/cluster)}$}	\\ [1ex]
\cline{2-4}												\cline{6-8}					\\ [-2ex]
{Sky subtraction}				&\multicolumn{1}{c}{Under}	&\multicolumn{1}{c}{Nominal}	&\multicolumn{1}{c}{Over}&{}
						&\multicolumn{1}{c}{Under}	&\multicolumn{1}{c}{Nominal}	&\multicolumn{1}{c}{Over}		\\
{}						&\multicolumn{1}{c}{($-1\sigma$)}	&{}	&\multicolumn{1}{c}{($+1\sigma$)}	&{}
						&\multicolumn{1}{c}{($-1\sigma$)}	&{}	&\multicolumn{1}{c}{($+1\sigma$)}			\\
\hline
{S0: Type~I ($h$)}				&{}		&{}		&{}		&{}	&{}		&{}		&{}		\\
{DTM}						&{$0.811$}	&{$0.817$}	&{$0.822$}	&{}	&{$0.572$}	&{$0.572$}	&{$0.787$}	\\
{AAS}						&{$0.755$}	&{$0.719$}	&{$0.804$}	&{}	&{$0.370$}	&{$0.327$}	&{$0.344$}	\\
{MEG}						&{$0.589$}	&{$0.699$}	&{$0.621$}	&{}	&{$0.566$}	&{$0.497$}	&{$0.383$}	\\ [1ex]
{S0: Type~$\rm III$ ($\mu_{\rm brk}$)}		&{}		&{}		&{}		\\
{DTM}						&{$0.536$}	&{$0.504$}	&{$0.682$}	&{}	&{$0.589$}	&{$0.398$}	&{$0.352$}	\\
{AAS}						&{$0.099$}	&{$0.206$}	&{$0.121$}	&{}	&{$0.523$}	&{$0.477$}	&{$0.222$}	\\
{MEG}						&{$0.902$}	&{$0.812$}	&{$0.937$}	&{}	&{$0.983$}	&{$0.984$}	&{$0.981$}	\\ [1ex]
{S0: Type~$\rm III$ ($T$)}			&{}		&{}		&{}		\\
{DTM}						&{$0.732$}	&{$0.226$}	&{$0.357$}	&{}	&{$0.226$}	&{$0.062$}	&{$0.331$}	\\
{AAS}						&{$0.909$}	&{$0.864$}	&{$0.436$}	&{}	&{$0.523$}	&{$0.381$}	&{$0.551$}	\\
{MEG}						&{$0.925$}	&{$0.821$}	&{$0.154$}	&{}	&{$0.877$}	&{$0.761$}	&{$0.083$}	\\
\hline
\end{tabular}
\end{minipage}
\end{table*}

\subsection[]{S0: broken exponential discs (Type~II/III)}

\label{S0: Type II/III}

For S0 galaxies where a $\mu(r)$ break was identified\footnote{Note: if two $\mu(r)$ breaks are
identified in any one galaxy, the outer break is used in the analysis.} (Type~II/III), we compare the break
surface brightness $\mu_{\rm brk}$ and break strength $T$ distributions in the field and cluster environments.
These comparisons are presented in Section~\ref{S0: Break surface brightness} and
Section~\ref{S0: Break strength}, respectively.

\subsubsection{Break surface brightness $\mu_{\rm brk}$}

\label{S0: Break surface brightness}

Certain physical processes inherent to galaxy evolution and related to the galaxy environment could
potentially affect the position of $\mu(r)$ breaks. For example in the cluster environment,
tidal/ram-pressure stripping \citep{Gunn_Gott:1972, Faber:1973} could remove gas from the stellar disc,
causing star formation to cease in the outer regions as the gas density drops below the star-formation
threshold. This would cause the outer regions to gradually fade as the stellar population ages. For classical
truncations (Type~II-CT), gas (and hence star formation) is not expected to occur beyond the break radius
$r_{\rm brk}$ (see Section~\ref{The absence of S0 Type II profiles}); and therefore such gas processes should
only act on the inner disc. This may result in $\mu_{\rm brk}$ (i.e. the end of the starforming inner disc)
evolving to a slightly brighter $\mu$ in the cluster environment. However, for Type~III profiles the outer
disc is thought to form via the displacement of disc stars in a minor merger
\citep[e.g.][]{Younger_etal:2007}; and therefore gas (and star formation) may still be prevalent beyond
$r_{\rm brk}$. In these cases, cluster processes (e.g. ram-pressure stripping) may result in a general fading
across the break and cause $\mu_{\rm brk}$ to evolve to a fainter $\mu$ in the cluster environment (although
the exact effect will be dependent on the relative fading of the inner and outer discs). These are just two
possible scenarios (there may be many others), and illustrate the potential for an effect of the environment
on break surface brightness $\mu_{\rm brk}$. Consequently, the comparison of $\mu_{\rm brk}$ distributions
for Type~II/III galaxies could provide some insight into the effect of the galaxy environment on galactic
discs.

\begin{figure*}
\includegraphics[width=0.85\textwidth]{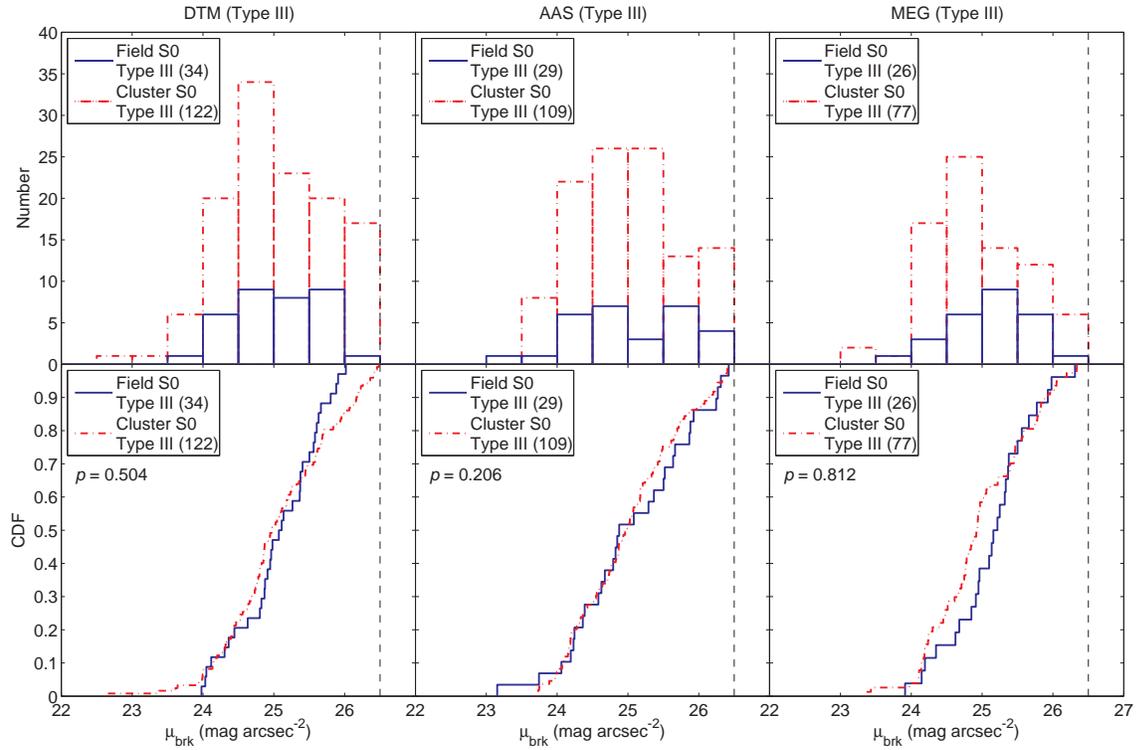}
\centering
\caption{\label{S0: Type III mu brk analysis} Comparing break surface brightness $\mu_{\rm brk}$
distributions in different environments for Type~III S0 galaxies (full sample). Format as in
Fig.~\ref{S0: Type I analysis}. Respective sample sizes are shown in the legends. Contamination of the
cluster sample by the field is $<25$ per cent. We find no significant difference between the CDFs in each
environment and no evidence to suggest that the $\mu_{\rm brk}$ of our Type~III S0s are {\em not} drawn from
the same continuous $\mu_{\rm brk}$ distributions.}
\end{figure*}

Therefore, for S0 galaxies where a Type~II/III $\mu(r)$ break was identified, we compare the break surface
brightness $\mu_{\rm brk}$ distributions in the field and cluster environments in order to see if there is
any evidence for an environmental dependence on $\mu_{\rm brk}$. However, in such comparisons, it is
important to remember that Type~II breaks generally occur at a brighter $\mu_{\rm brk}$ than Type~III breaks
and that environmental processes may affect these breaks in different ways. Therefore, in order to ensure our
S0 $\mu_{\rm brk}$ comparisons are fair, we only compare the $\mu_{\rm brk}$ distributions for our
field/cluster S0 Type~III galaxies (see Fig.~\ref{S0: Type III mu brk analysis}). Note that we cannot
compare the $\mu_{\rm brk}$ distributions for Type~II profiles since they are  very rare in our S0 galaxies
(see Table~\ref{S0 profile frequencies}). Analogous parallel analyses and statistical tests are performed
as in our Type~I profile analysis (see Section~\ref{S0: Type I}) and the results of the \mbox{K--S} tests,
for both the full S0 sample and the low-axis-ratio ($q > 0.5$) S0 sub-sample, are shown in
Table~\ref{S0: K-S results}.

For our Type~III S0 galaxies, in all cases (all assessor samples and sky versions), we observe no significant
difference between the $\mu_{\rm brk}$ distributions in the field and cluster environments. The probability
$p_{\rm(field/cluster)}$ is below the $2\sigma$ level in each case. For our S0 ($q > 0.5$) sub-sample, in
a few cases the probability $p_{\rm(field/cluster)}$ is above the $2\sigma$ level (MEG results, see
Table~\ref{S0: K-S results}). However, this result is not consistent between the different assessors (i.e.\
robust to the subjective nature of the profile classifications). Therefore, we conclude that there is no
evidence to suggest that the break surface brightness $\mu_{\rm brk}$ of our S0s is dependent on the galaxy
environment. However, we stress that this result may suffer from low number statistics and a larger field
sample may yield different results. We also note that for our Type~III S0 galaxies, a significant fraction
(up to $\sim50$ per cent) may actually be related to light from an extended bulge component and not due to an
antitruncated stellar disc (see Section~\ref{Antitruncations in S0 galaxies}). In such cases, the suppression
of star-formation in the disc may cause $\mu_{\rm brk}$ to evolve to a brighter $\mu$. This is due to the
inner star-forming disc fading more quickly than the old extended bulge component ($r>r_{\rm brk}$), once
star formation has ceased. This could potentially mask an environmental effect in our Type~III
$\mu_{\rm brk}$ comparisons. We shall return to this issue in Section~\ref{SpS0: break surface brightness}.

For our Type~III S0s, we also perform an analogous comparison using the break radius $r_{\rm brk}$
distributions in the field and cluster environments  ($r_{\rm brk}$ in units of the effective radius of the
{\sc galfit} S{\'e}rsic model). As with our $\mu_{\rm brk}$ analysis, we find no significant difference
between the $r_{\rm brk}$ distributions of our Type~III S0s in the field and cluster environments. However,
we note that the S{\'e}rsic effective radius is not an ideal unit to measure the break radius $r_{\rm brk}$
and a non-parameterized effective radius may yield more robust results.

\subsubsection{Break strength $T$}

\label{S0: Break strength}

\begin{figure*}
\includegraphics[width=0.85\textwidth]{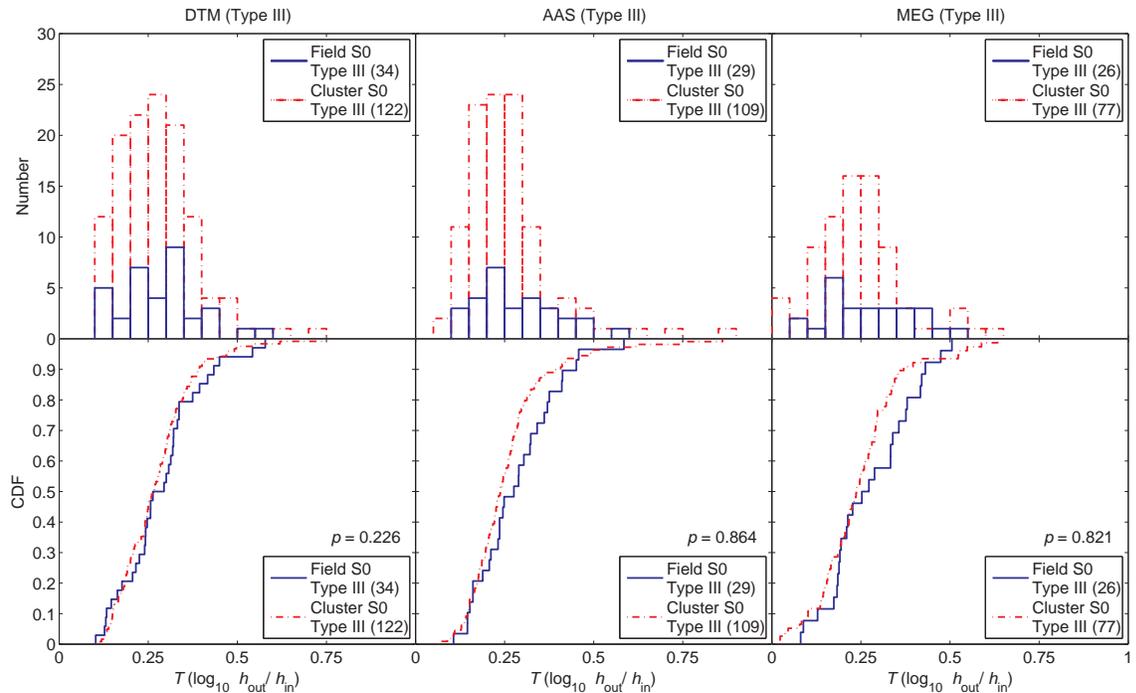}
\centering
\caption{\label{S0: Type II/III analysis} Comparing break strength $T$
(${\rm log}_{10}\,h_{\rm out}/{h}_{\rm in}$) distributions in different environments for Type~III S0 galaxies
(full sample). Format as in Fig.~\ref{S0: Type I analysis}. Respective sample sizes are shown in the legends.
Random errors in $T$ are typically $<0.1$. Systematic errors in $T$ due to the error in the sky subtraction
are also $\sim\pm0.1$. Contamination of the cluster sample by the field is $<25$ per cent. We find no
significant difference between the CDFs in each environment and no evidence to suggest that the break
strengths $T$ of our Type~III S0s are {\em not} drawn from the same continuous $T$ distributions.}
\end{figure*}

For our Type~III S0s, we also compare the break strength $T$ distributions in the field and cluster
environments in order to see if there is any evidence for an environmental dependence on break strength $T$
(see Fig.~\ref{S0: Type II/III analysis}). Note that we only compare the $T$ distributions for our Type~III
S0s since i) the effect of environment may be different for Type~II/III profiles; and ii) Type~II profiles
are very rare in our S0 galaxies. Similar parallel analyses and statistical tests are also carried out as in
our S0 Type~I profile analysis (Section~\ref{S0: Type I}). The mean error in $T$ due to the sky subtraction
error is $\pm0.1$ and random errors in $T$ due to exponential fitting routine are also typically $\pm0.1$
(see Section~\ref{Measuring scalelength and break strength}). The results of the K--S tests, for both the
full S0 sample and the low-axis-ratio ($q > 0.5$) S0 sub-sample, are shown in Table~\ref{S0: K-S results}.

In all cases (all assessor samples and sky versions), we observe no significant difference between the break
strength $T$ distributions in the field and cluster environments. The probability $p_{\rm(field/cluster)}$
is below the $2\sigma$ level in each case. Therefore, we find no evidence to suggest that the break strength
$T$ of our Type~III S0 galaxies is dependent on the galaxy environment. This result is also robust to the
error in the sky subtraction and the subjective nature of the profile classifications.


\section[]{The structure of galactic discs in spiral and S0 galaxies}

\label{The structure of galactic discs in spiral and S0 galaxies}

In this study, we have so far explored the effect of the galaxy environment on the structure of galactic
discs in STAGES S0 galaxies. This work is analogous to our companion study \citep{Maltby_etal:2012a}, which
explores the effect of the galaxy environment on the structure of galactic discs in STAGES {\em spiral}
galaxies. For spiral galaxies, \cite{Maltby_etal:2012a} found no evidence to suggest that their $\mu(r)$
profiles were affected by the galaxy environment. Both the scalelength~$h$ and break strength $T$ of their
spiral galaxies showed no evidence for an environmental dependence from the general field to the intermediate
galaxy densities probed by the STAGES survey. Therefore, our conclusion that there is no evidence for an
effect of the galaxy environment on the structure of S0 galactic discs is in qualitative agreement with the
conclusions presented in \cite{Maltby_etal:2012a} for spiral galaxies.

In this section, we compare our results for S0s with those for spiral galaxies from \cite{Maltby_etal:2012a}
in order to assess the effect of galaxy morphology on the structure of galactic discs. Such comparisons of
disc structure (e.g.\ profile type, scalelength $h$, break strength $T$, break surface brightness
$\mu_{\rm brk}$) between different Hubble-type morphologies are a useful tool in exploring the potential
evolutionary link between spiral and S0 galaxies.

However, in these comparisons it is important to note that the break classification scheme used by
\cite{Maltby_etal:2012a} differs from the one used in this work. In \cite{Maltby_etal:2012a}, break
classification is based on the outer stellar disc ($\mu > 24\rm\,mag\,arcsec^{-2}$), while in this work the
entire disc component is used. Fortunately, \cite{Maltby_etal:2012a} identify breaks across the entire disc
component prior to their classification and also identify Type~I profiles based on the entire disc.
Therefore, their Type~I scalelength $h$ and $\mu_{\rm brk}$ distributions are not limited by surface
brightness and can be directly compared with the results of this work (see Sections~\ref{SpS0: Type I} and
\ref{SpS0: break surface brightness}). However, for break strength $T$ the situation is different because
\cite{Maltby_etal:2012a} only performed these measurements on Type~II/III profiles in the outer stellar disc
($\mu_{\rm brk} > 24\rm\,mag\,arcsec^{-2}$), cases which they classify as Type~$\rm II_o/III_o$ profiles
(o -- outer). Consequently, when we compare our S0 break strength $T$ distributions with those for spiral
galaxies from \cite{Maltby_etal:2012a}, we also limit our S0 breaks by $\mu_{\rm brk}$ to allow for a fair
comparison (see Section~\ref{SpS0: break strength}).

\subsection[]{Spiral/S0: galaxy samples}

\label{SpS0: galaxy samples}

Both this work and \cite{Maltby_etal:2012a} use the same parent sample of morphologically classified galaxies
in STAGES from which to draw their samples \citep[see][]{Maltby_etal:2010}. Both works also perform analogous
$\mu(r)$ profile fitting and break identification (see Section~\ref{Profile inspection}).
\cite{Maltby_etal:2012a} use a large, mass-limited ($M_* > 10^9\rm\,M_{\odot}$), visually classified
(Sa--Sdm) sample of $327$ face-on to intermediately inclined ($i<60^\circ$) spiral galaxies from both the
field and cluster environments ($145$ field and $182$ cluster spirals). In this work, our S0 sample selection
is analogous to this spiral selection except for the lack of an inclination $i$ cut (see
Section~\ref{Sample selection}). Therefore, in order to allow for a fair comparison of these spiral and S0
galaxies, we use our low-axis-ratio ($i<60^\circ$, $q >0.5$) S0 sub-sample of $173$ S0s ($36$ field and $137$
cluster) in all our spiral/S0 comparisons (see Section~\ref{Galaxy inclination}).

\subsection[]{Profile type (Type~I, II and III)}

\label{SpS0: profile type}

Previous works \citep[e.g.][]{Pohlen_Trujillo:2006, Erwin_etal:2008, Gutierrez_etal:2011} have found that for
spiral galaxies the distribution of profile types I:II:III is approximately $20$:$50$:$30$ $\pm10$ per
cent. This is also true for \cite{Maltby_etal:2012a} if their spiral $\mu(r)$ profiles are re-classified
based on the entire disc (see Section~\ref{Introduction}). However, in this work we find that for S0s,
$\sim25$ per cent are Type~I, $<5$ per cent are Type~II, $\sim50$ per cent are Type~III and $\sim20$ per cent
have no discernible exponential component (see Section~\ref{Results}, Table~\ref{S0 profile frequencies}). In
comparing the profile types (disc structure) of these spiral/S0 galaxies, the most striking difference is the
lack of truncations (Type~II profiles) in our S0s compared to their abundance in spiral galaxies. Therefore,
it seems whatever mechanism transforms spiral galaxies into S0s may well erase these Type~II features from
the galaxy $\mu(r)$ profiles. In the case of classical truncations (Type~II-CT), recent studies suggest that
the Type~II profile is related to a radial change in the age of the stellar population throughout the disc,
with the outer disc being populated by old stars and an inside-out growth
for the inner disc \citep{Debattista_etal:2006,Bakos_etal:2008, Martinez-Serrano_etal:2009}. Consequently,
the absence of Type~II-CT profiles in our S0s may actually be a natural consequence of an ageing stellar
population in the `break region' as the spirals transform into S0s (see Section~\ref{Results}, for a full
explanation). For bar-related truncations (Type~II-OLR), the depletion/removal of galactic gas should weaken
the OLR effect and may also result in the weakening/removal of the Type~II feature as spirals transform into
S0s \citep{Erwin_etal:2012}.

Another observation is that Type~III profiles seem to be significantly more frequent in S0s than in spiral
galaxies. This suggests a continuation of the observed trend for spiral galaxies where Type~III profiles
become more frequent with progressively earlier Hubble types \citep[e.g.][]{Pohlen_Trujillo:2006,
Maltby_etal:2012a}. Since Type~III profiles are thought to form via minor mergers, this trend is consistent
with a minor merger scenario for the formation of S0 galaxies. However, in
Section~\ref{Antitruncations in S0 galaxies} we find that for our Type~III S0s, a significant fraction
(potentially up to $\sim50$ per cent) may actually be related to light from an extended bulge component and
not due to an antitruncated stellar disc. This is in contrast to spiral galaxies where the vast majority of
Type~III profiles appear to be genuine antitruncated discs \citep{Maltby_etal:2012b}. Consequently, the
fraction of genuine antitruncated discs may actually be lower in S0s than in spiral galaxies. This could be
due to the fading inner/outer disc causing $\mu_{\rm brk}$ to drop below the level of the bulge $\mu$ profile
(or sky background) as some spirals evolve into S0s.

\subsection[]{Spiral/S0: pure exponential discs (Type~I)}

\label{SpS0: Type I}

\begin{figure*}
\includegraphics[width=0.85\textwidth]{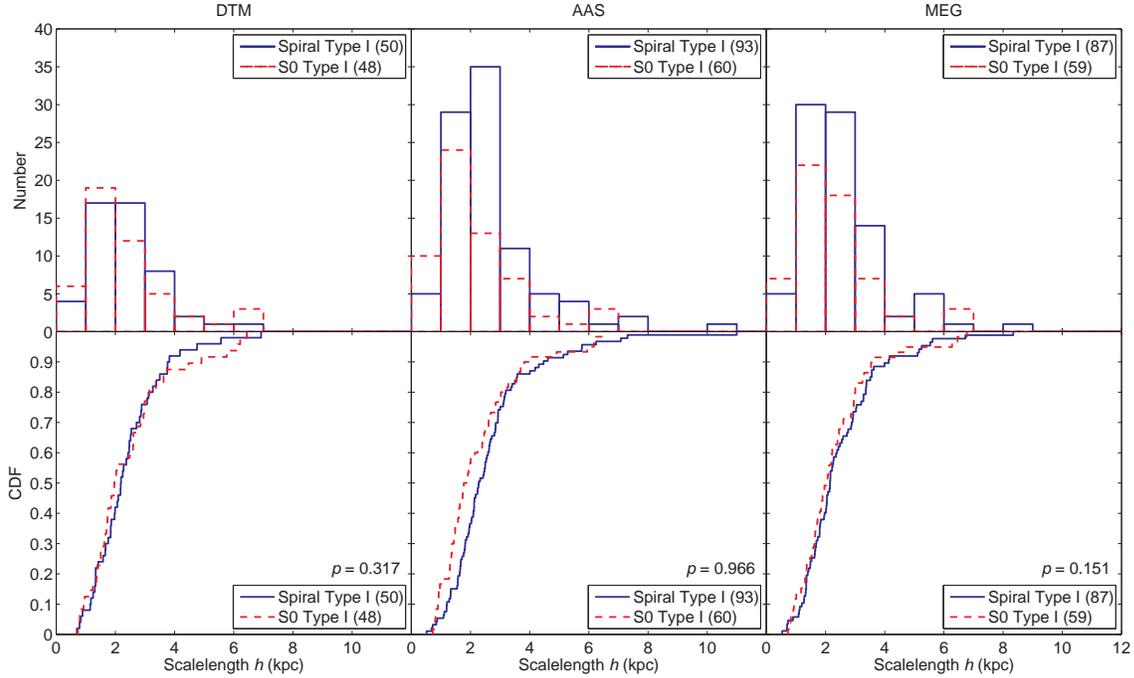}
\caption{\label{SpS0: Type I analysis} Comparing the disc scalelength $h$ distributions for different
Hubble-type morphologies. Top row: scalelength $h$ distributions for spiral (blue line) and S0 (red dashed
line) Type~I galaxies as classified by DTM (left-hand panel), AAS (centre panel), and MEG (right-hand
panel). Bottom row: the corresponding scalelength $h$ CDFs showing the probability $p$ that compared
samples are {\em not} drawn from the same continuous $h$ distributions in the bottom right of each plot.
Respective sample sizes are shown in the legends. Random errors in scalelength are typically $<10$ per
cent. Systematic errors in scalelength $h$ due to the error in the sky subtraction are also typically $<10$
per cent. We find no significant difference between the CDFs for each morphology and no evidence to suggest
that the disc scalelength $h$ of our Type~I galaxies are {\em not} drawn from the same continuous $h$
distributions.}
\end{figure*}

For spiral/S0 galaxies where no disc $\mu(r)$ break was identified (Type~I), we compare the scalelength $h$
distributions for spiral and S0 morphologies (see Fig.~\ref{SpS0: Type I analysis}). These comparisons are
independent of the galaxy environment (our field and cluster samples are combined) and allow for an
assessment of whether the scalelength $h$ of the stellar disc is affected by the Hubble-type morphology.
Similar parallel analyses are also carried out as in previous tests.

In most cases (most parallel analyses), we observe no clear difference between the scalelength $h$
distributions for our Type~I spiral and S0 galaxies (see Fig.~\ref{SpS0: Type I analysis}).
In order to test the significance of these results we construct scalelength $h$ CDFs for our Type~I spiral/S0
samples and perform K--S tests in order to obtain the probability $p_{\rm(spiral/S0)}$ that they are
{\em not} drawn from the same continuous $h$ distributions. The results of these K--S tests are presented in
Table~\ref{SpS0: K-S results tbl}.

In most cases (most assessor samples and sky versions), the probability $p_{\rm(spiral/S0)}$ is below the
$2\sigma$ level. However, in a few cases the probability $p_{\rm(spiral/S0)}$ is above the $2\sigma$ level
(AAS results, see Table~\ref{SpS0: K-S results tbl}). As these high significance results are not robust to
the subjective nature of the profile classifications, we conclude that there is no evidence to suggest that
the scalelength $h$ of our Type~I galaxies is significantly affected by the Hubble-type morphology.

\subsection{Spiral/S0: broken exponential discs (Type~II/III)}

\label{SpS0: Type II/III}

For spiral/S0 galaxies where a $\mu(r)$ break was identified\footnote{Note: if two $\mu(r)$ breaks are
identified in any one galaxy, the outer break is used in the analysis.} (Type~II/III), we compare both the
break surface brightness $\mu_{\rm brk}$ and break strength $T$ distributions for our spiral and S0 galaxies.
These comparisons are presented in Section~\ref{SpS0: break surface brightness} and
Section~\ref{SpS0: break strength}, respectively. In these comparisons, it is important to note that for our
spiral galaxies (taken from \citealt{Maltby_etal:2012a}), the $\mu_{\rm brk}$ distribution is for breaks
identified across the entire disc, whereas the $T$ distribution is for breaks from the outer disc only
($\mu > 24\rm\,mag\,arcsec^{-2}$).

\subsubsection{Break surface brightness $\mu_{\rm brk}$}

\label{SpS0: break surface brightness}

The physical processes that drive the morphology--density relation \citep{Dressler:1980} and the
transformation of spiral galaxies into S0s are not well understood. Certain mechanisms inherent to the
cluster environment could be responsible, e.g.\ ram-pressure stripping of the interstellar medium, mergers
and harassment \citep{Gunn_Gott:1972, Icke:1985,Moore_etal:1996}. However, the existence of S0s in the
general field implies that either cluster processes are not ultimately responsible, or that S0s can form via
alternative processes in different environments. For example, it could be the case that cluster S0s form from
in-falling spirals via the tidal/ram-pressure stripping of interstellar gas, while field S0s are simply the
faded remnants of field spirals.

For these mechanisms, the processes causing star formation to be quenched and the subsequent morphological
transformation could have a distinct effect on the $\mu_{\rm brk}$ of any $\mu(r)$ break that was present
(see Section~\ref{S0: Break surface brightness}). For example, in a Type~III profile the suppression of star
formation could cause $\mu_{\rm brk}$ to evolve to a fainter $\mu$ as spirals evolve into S0s. Consequently,
the comparison of $\mu_{\rm brk}$ distributions for spiral and S0 Type~II/III galaxies could provide some
insight into the potential evolutionary links between them.

\begin{figure*}
\includegraphics[width=0.85\textwidth]{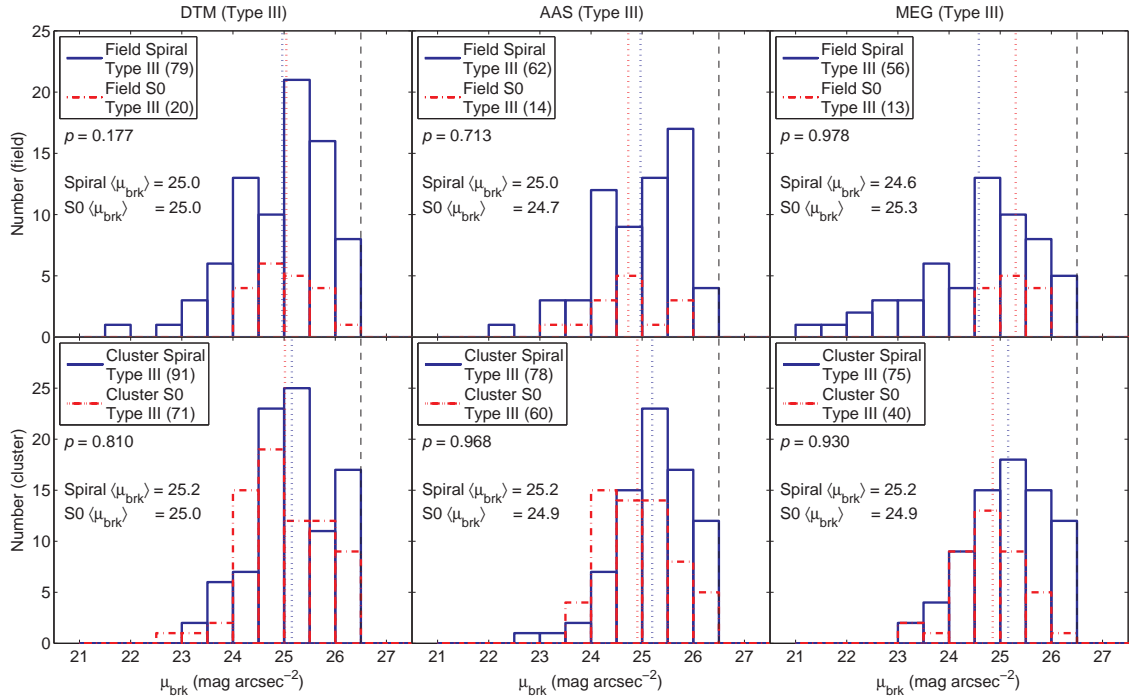}
\caption{\label{SpS0: Type III mu brk analysis} The break surface brightness $\mu_{\rm brk}$ distributions
for Type~III spiral and S0 galaxies in different environments. The $\mu_{\rm brk}$ distributions for field
(top row) and cluster (bottom row) galaxies as determined by DTM (left-hand column), AAS (centre column)
and MEG (right-hand column). The distributions show spiral galaxies (blue line), and S0 galaxies (red
dashed line). The mean $\mu_{\rm brk}$ for field/cluster spirals (blue dotted line) and S0s (red dotted
line) are also shown for reference. Respective sample sizes are shown in the legends. Contamination of the
cluster sample by the field is $< 25$ per cent.\vspace{-0.1cm}}
\end{figure*}

Therefore, for spiral/S0 galaxies where a $\mu(r)$ break was identified (Type~II/III), we compare the
$\mu_{\rm brk}$ distributions between field/cluster spiral and S0 galaxies. In such comparisons it is
important to remember that Type~II breaks generally occur at a brighter $\mu_{\rm brk}$ than Type~III breaks.
For our Type~II/III spirals, $\sim50$ per cent are Type~III (classification based on the entire stellar
disc), while our S0s are almost exclusively ($>95$ per cent) Type~III. Therefore, to ensure our spiral/S0
$\mu_{\rm brk}$ comparisons are fair, we only compare the $\mu_{\rm brk}$ distributions for our
field/cluster spiral and S0 Type~III galaxies~(see Fig~\ref{SpS0: Type III mu brk analysis}). Similar
parallel analyses and statistical tests (between corresponding spiral and S0 samples) are carried out as in
previous tests.

For our cluster spiral/S0 comparison, in most cases (most assessor samples) we observe no significant
difference between the $\mu_{\rm brk}$ distributions for spiral and S0 galaxies. The probability
$p_{\rm(spiral/S0)}$ is below the $2\sigma$ level in most cases. However, for one assessor (AAS) the
significance of a morphological dependence is above the $2\sigma$ level. We also note that for all assessors,
the mean $\mu_{\rm brk}$ for S0s ($24.9\rm\,mag\,arcsec^{-2}$) is brighter than that of spiral galaxies
($25.2\rm\,mag\,arcsec^{-2}$). However, since this result is inconclusive we deduce that there is no
evidence to suggest that $\mu_{\rm brk}$ in the cluster environment is dependent on the galaxy morphology.

For our field spiral/S0 comparison, we also observe no significant difference between the spiral/S0
$\mu_{\rm brk}$ distributions in the majority of cases (i.e.\ $p_{\rm(spiral/S0)} < 2\sigma$). For one
assessor (MEG), the significance of a morphological dependence is above the $2\sigma$ level, but this may be
due to low number statistics. It is interesting that for all assessors, the spiral $\mu(r)$ breaks reach to
a brighter $\mu_{\rm brk}$ than the S0 $\mu(r)$ breaks. This would be consistent with the hypothesis that
field S0s are the faded remnants of field spirals. However, since this result is based on low number
statistics we conclude that there is no evidence to suggest that $\mu_{\rm brk}$ in the field is dependent on
the galaxy morphology. Further investigation with a larger sample of field S0s is needed to suitably address
this issue and may yield more conclusive evidence.

In these $\mu_{\rm brk}$ comparisons, it is also important to note that for our Type~III S0s a significant
fraction (possibly up to $50$ per cent) may actually be related to light from an extended bulge component and
not due to an antitruncated stellar disc (see Section~\ref{Antitruncations in S0 galaxies}). In such cases,
the suppression of star-formation in the disc may cause $\mu_{\rm brk}$ to evolve to a brighter $\mu$. This is
due to the inner star-forming disc fading more quickly than the old extended bulge component ($r>r_{\rm brk}$),
once star formation has ceased. In contrast, for spiral galaxies the vast majority of our Type~III profiles 
appear to be genuine antitruncated discs \citep{Maltby_etal:2012b}. Consequently, a significant fraction of
our Type~III S0s may not have evolved from a Type~III spiral galaxy, potentially leading to a masking of any
morphological trends in our $\mu_{\rm brk}$ comparisons. A robust method for differentiating between genuine
antitruncated discs and those caused by an extended bulge is needed to address this problem (see
Section~\ref{Antitruncations in S0 galaxies}).

Similar analyses were also performed using the break radius $r_{\rm brk}$ distributions ($r_{\rm brk}$ in
units of the {\sc galfit} S{\'e}rsic model effective radius). As with our $\mu_{\rm brk}$ analysis, no
significant differences or trends were observed for $r_{\rm brk}$ in field/cluster spiral and S0 galaxies.
However, we note that the S{\'e}rsic effective radius is not an ideal unit to measure the break radius
$r_{\rm brk}$ and a non-parameterized effective radius may yield more robust results.


\begin{table}
\centering
\begin{minipage}{70mm}
\centering
\caption{\label{SpS0: K-S results tbl} The K--S test results for Type~I and Type~$\rm II_o/III_o$ galaxies
as classified by DTM, AAS and MEG. K--S tests give the probability $p_{\rm(spiral/S0)}$ that the respective
spiral and S0 samples are {\em not} drawn from the same continuous $h$ distributions for Type~I galaxies,
and $T$ distributions for Type~$\rm II_o/III_o$ galaxies. Results are also shown for when the sky is
oversubtracted and undersubtracted by $\pm1\sigma$.}
\begin{tabular}{lrrr}
\hline
{}				&\multicolumn{3}{c}{$p_{\rm(spiral/S0)}$}	\\ [1ex]
\cline{2-4}									\\ [-2ex]
{Sky subtraction}		&\multicolumn{1}{c}{Under}	&\multicolumn{1}{c}{Nominal}	&\multicolumn{1}{c}{Over}\\
{}				&\multicolumn{1}{c}{($-1\sigma$)}&{}
&\multicolumn{1}{c}{($+1\sigma$)}	\\
\hline
{Type~I ($h$)}			&{}		&{}		&{}		\\
{DTM}				&{$0.293$}	&{$0.317$}	&{$0.477$}	\\
{AAS}				&{$0.966$}	&{$0.966$}	&{$0.981$}	\\
{MEG}				&{$0.056$}	&{$0.151$}	&{$0.346$}	\\ [1ex]
{Type~$\rm II/III$ ($T$)}	&{}		&{}		&{}		\\
{DTM}				&{$0.996$}	&{$0.988$}	&{$0.989$}	\\
{AAS}				&{$0.996$}	&{$0.969$}	&{$0.997$}	\\
{MEG}				&{$0.983$}	&{$0.957$}	&{$0.972$}	\\ [1ex]
{Type~$\rm III$ ($T$)}		&{}		&{}		&{}		\\
{DTM}				&{$0.9998$}	&{$0.9994$}	&{$0.9999$}	\\
{AAS}				&{$0.99999$}	&{$0.99997$}	&{$0.995$}	\\
{MEG}				&{$0.99992$}	&{$0.9997$}	&{$0.975$}	\\
\hline
\end{tabular}
\end{minipage}
\end{table}

\subsubsection{Break strength $T$}

\label{SpS0: break strength}

In \cite{Maltby_etal:2012a}, break strength $T$ measurements were only performed on $\mu(r)$ breaks in the
outer regions of the stellar disc $\mu_{\rm brk} > 24\rm\,mag\,arcsec^{-2}$ (their criteria for selecting
intrinsically similar outer breaks in spiral galaxies). They refer to these `outer disc' breaks as
Type~{$\rm II_o$} and Type~{$\rm III_o$} profiles. Therefore, in order to allow for a fair comparison
between the break strength $T$ distributions of our S0s and the spiral galaxies of \cite{Maltby_etal:2012a},
we also need to limit our S0 break strength $T$ distributions by break surface brightness
($\mu_{\rm brk} > 24\rm\,mag\,arcsec^{-2}$). However, since the vast majority ($\sim95$ per cent) of our S0
Type~II/III profiles have $\mu_{\rm brk} > 24\rm\,mag\,arcsec^{-2}$ anyway, this has little effect on our S0
break strength $T$ distributions (see Fig.~\ref{S0: Type III mu brk analysis}).

\begin{figure*}
\includegraphics[width=0.85\textwidth]{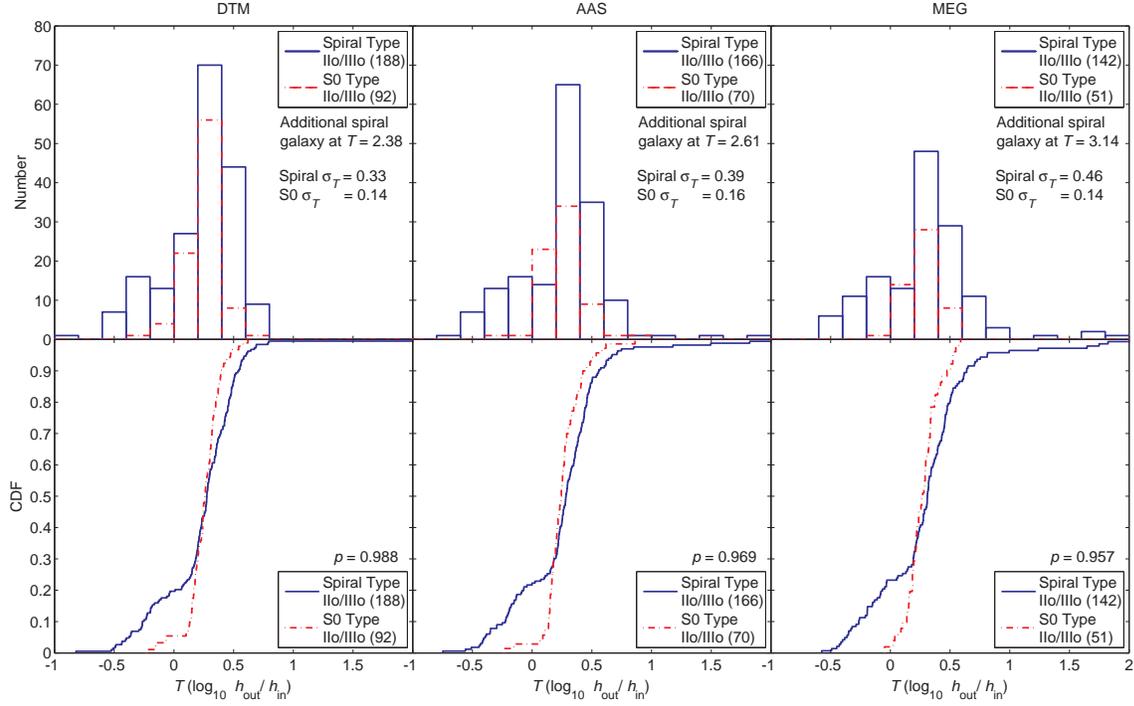}
\centering
\caption{\label{SpS0: Type II/III analysis} Comparing break strength $T$
(${\rm log}_{10}\,h_{\rm out}/{h}_{\rm in}$) distributions for spiral and S0 galaxies. Top row: break
strength $T$ distributions for spiral (blue line) and S0 (red dashed line) Type~{$\rm II_o/III_o$} galaxies
as classified by DTM (left-hand column), AAS (centre column) and MEG (right-hand column). Bottom row: the
corresponding break strength $T$ CDFs showing the probability $p$ that compared samples are {\em not} drawn
from the same continuous $T$ distributions in the bottom right of each plot. Respective sample sizes are
shown in the legends. Random errors in $T$ are $\sim\pm0.1$. Systematic errors in $T$ due to the sky
subtraction error are also $\sim\pm0.1$. Contamination of the cluster sample by the field is $<25$ per
cent. We find a significant ($>3\sigma$) difference between the $T$ CDFs for spiral and S0 galaxies.}
\end{figure*}

Fig.~\ref{SpS0: Type II/III analysis} shows a comparison of the break strength $T$ distributions for our
Type~{$\rm II_o/III_o$} spiral/S0 galaxies. These comparisons are independent of the galaxy environment (our
field and cluster samples are combined) and allow for an assessment of whether the break strength $T$ of the
stellar disc is affected by the Hubble-type morphology. Similar parallel analyses and statistical tests are
also carried out as in previous tests and the results of the K--S tests are presented in
Table~\ref{SpS0: K-S results tbl}.

In all cases (all assessor samples and sky versions), we observe a {\em significant difference} between the
$T$ distributions for spiral and S0 Type~{$\rm II_o/III_o$} galaxies. The probability $p_{\rm(spiral/S0)}$ is
above the $3\sigma$ level in each case. Additionally, the break strength $T$ distribution of our S0s has a
much smaller variance (S0 $\sigma_T\sim0.15$) compared to that of our spiral galaxies (spiral
$\sigma_T\sim0.4$). These results suggest that both Type~{$\rm II_o$} and Type~{$\rm III_o$} features in S0s
are weaker (smaller $|T|$) than in spiral galaxies.

\begin{figure*}
\includegraphics[width=0.85\textwidth]{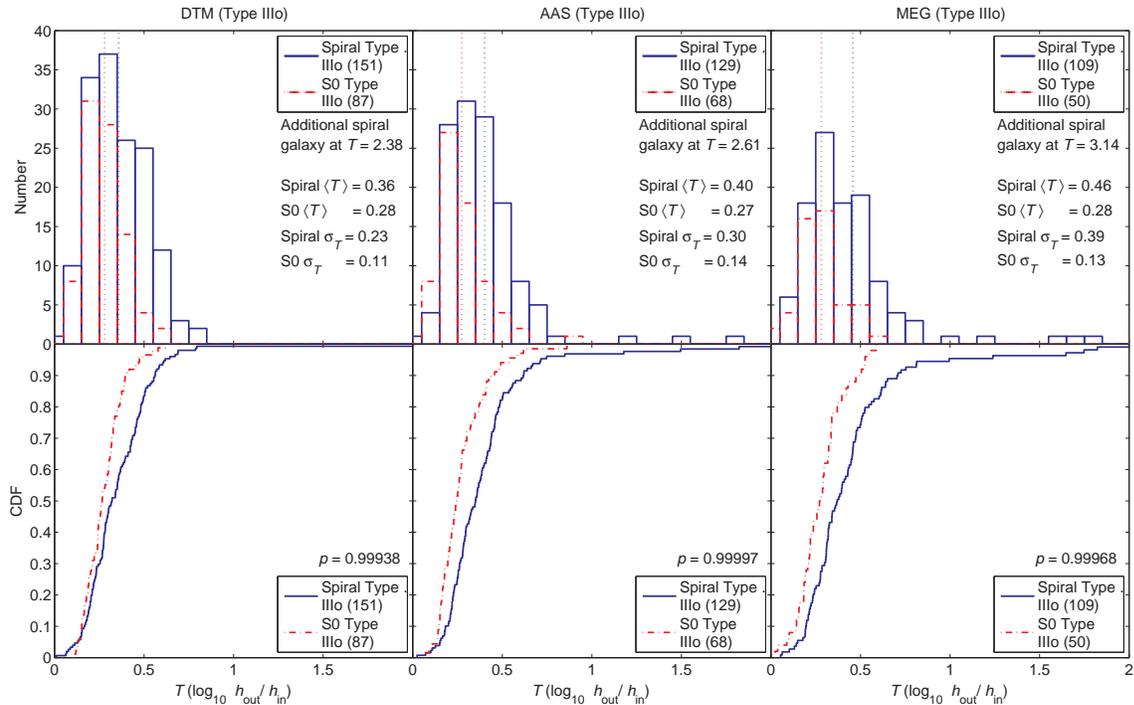}
\centering
\caption{\label{SpS0: Type III analysis} Comparing break strength $T$
(${\rm log}_{10}\,h_{\rm out}/{h}_{\rm in}$) distributions for Type~{$\rm III_o$} spiral and S0 galaxies.
Figure the same as Fig.~\ref{SpS0: Type II/III analysis} but for Type~{$\rm III_o$} galaxies only. We find a
significant ($>3\sigma$) difference between the $T$ CDFs for spiral and S0 galaxies. The mean $T$ for spirals
(blue dotted line) and S0s (red dotted line) are also shown for reference. Respective sample sizes are shown
in the legend.}
\end{figure*}

However, it is possible that the high significance of this result could be driven by the lack of
Type~II profiles in our S0s. Therefore, we repeat our analysis using the break strength $T$ distributions for
just our Type~{$\rm III_o$} galaxies (see Fig.~\ref{SpS0: Type III analysis}). The results of the K--S tests
are presented in \mbox{Table~\ref{SpS0: K-S results tbl}}. In all cases (all assessor samples and sky
versions), we still observe a significant difference between the $T$ distributions for our spiral and S0
Type~{$\rm III_o$} galaxies. The probability $p_{\rm(spiral/S0)}$ is above the $3\sigma$ level in each case.
Also, the break strength $T$ distribution of our Type~{$\rm III_o$} S0s has a much smaller variance and mean
(S0s: $\sigma_T\sim0.13$, $\langle{T}\rangle\sim0.28$) compared to that of our spiral Type~{$\rm III_o$}
galaxies (spirals: $\sigma_T\sim0.3$, $\langle{T}\rangle\sim0.4$).

Therefore, we conclude that there is some evidence to suggest that the break strength $T$ of our
Type~{$\rm II_o/III_o$} galaxies is dependent on the galaxy morphology, with $\mu(r)$ breaks in S0s being
generally weaker (smaller $|T|$) than those of spiral galaxies. This result is also robust to the error in the
sky subtraction and the subjective nature of the profile classifications. Additionally, since the S0 $T$
distribution is much tighter than that of spiral galaxies the potential contamination of our
Type~{$\rm III_o$} S0s by non-genuine stellar disc antitruncations (i.e.\ extended bulge components; see
Section~\ref{Antitruncated surface brightness profiles: bulge or disc related?}) has no effect on this
conclusion.

We propose that this result is consistent with current theories on the origin of Type~II/III profiles. For
classical truncations (Type~II-CT), the $\mu(r)$ break is thought to be related to a radial variation in
the age of the stellar population, with the outer disc being populated by old stars
\citep{Debattista_etal:2006, Bakos_etal:2008, Martinez-Serrano_etal:2009}. In this scenario, once the galaxy
has depleted its gas supply, star formation will cease in the `break region' causing it to gradually fade due to
the ageing of the stellar population. This would lead to a weakening of the Type~II break strength $T$ (see
Section~\ref{The absence of S0 Type II profiles}). For bar-related truncations (Type~II-OLR), the
depletion/removal of galactic gas should weaken the OLR effect and may also result in a weakening of the
Type~II break strength $T$ \citep{Erwin_etal:2012}. For Type~III galaxies, further radial mixing related to
the suspected minor-merger history could be responsible for the weakening of the Type~III feature in S0
galaxies.

\vspace{-0.3cm}
\section{Antitruncated surface brightness profiles: bulge or disc related?}

\label{Antitruncated surface brightness profiles: bulge or disc related?}

Antitruncated (Type~III) surface brightness $\mu(r)$ profiles have a broken exponential with a shallower
region beyond the break radius $r_{\rm brk}$ (i.e.\ up-bending break, see Fig.~\ref{Profile types}). However,
the excess light at large radii is not necessarily related to an outer exponential disc and could also be
associated with an extended spheroidal bulge or halo. This idea was first postulated by
\cite{Erwin_etal:2005}, who suggest that Type~III profiles can be separated into two distinct sub-classes
depending on whether the outer profile $r > r_{\rm brk}$ is dominated by a stellar disc (Type~III-d) or a
spheroidal component (Type~III-s).

In \cite{Erwin_etal:2005}, they propose that antitruncations with a smooth gradual transition and outer
isophotes that are progressively rounder than that of the main disc, suggest an inclined disc embedded in a
more spheroidal outer region such as an extended bulge or halo (i.e.\ Type~III-s). Using this `ellipse'
method, previous works \citep{Erwin_etal:2005, Erwin_etal:2008, Gutierrez_etal:2011} have found that $\sim40$
per cent of their Type~III profiles are Type~III-s. However, the ellipse method is limited for face-on discs
and cases where the outer/inner disc may have different orientations and axis ratios. In these cases, an
alternative method would be to use bulge--disc (B--D) decomposition \citep[e.g.][]{Allen_etal:2006} to
determine the contribution of the two main structural components (bulge and disc) to the galaxy's light
distribution and should provide more robust results.

In this section, we explore the nature of Type~III $\mu(r)$ profiles in S0 galaxies, using bulge--disc (B--D)
decomposition in order to determine the maximum possible contribution of bulge light in their outer regions
($r>r_{\rm brk}$). Using these analyses, we determine the maximum fraction of Type~III S0s for which the
excess light at large radii could be caused or affected by the spheroidal component (i.e.\ Type~III-s). This
study complements our previous work, \cite{Maltby_etal:2012b}, which presents an analogous study using
Type~III {\em spiral} galaxies. We review the findings of \cite{Maltby_etal:2012b} in
Section~\ref{Antitruncations in spiral galaxies}, before presenting the results for our S0s in
Section~\ref{Antitruncations in S0 galaxies}.

\subsection{Antitruncations in spiral galaxies}

\label{Antitruncations in spiral galaxies}

\cite{Maltby_etal:2012b} explore the nature of Type III $\mu(r)$ profiles in spiral galaxies and determine
the {\em maximum} fraction of Type~III spirals for which the excess light at large radii could be caused or
affected by the spheroidal component (i.e.\ Type~III-s). They achieve this by comparing azimuthally-averaged
radial $\mu(r)$ profiles with analytical \mbox{B--D} decompositions (de Vaucouleurs, $r^{1/4}$ bulge plus
single exponential disc) for $78$ Type~III spirals, and determine the {\em maximum} possible contribution of
bulge light in the outer regions ($r > r_{\rm brk}$) of each galaxy. B--D decomposition was performed on all
$327$ spirals defined in \cite{Maltby_etal:2012a} (see Section~\ref{SpS0: galaxy samples}) and their Type~III
spirals were drawn from this parent sample\footnote{Note: in \cite{Maltby_etal:2012b}, B--D decomposition
failed for two spiral galaxies, hence a total sample of $325$ spirals was used.}.

Using these comparisons, \cite{Maltby_etal:2012b} find that for the majority of their Type~III spirals
($\sim70$ per cent), the excess light beyond the break radius $r_{\rm brk}$ is clearly related to an outer
shallow disc (Type~III-d). In addition, they also find a further $\sim15$ per cent which are also
\mbox{Type~III-d}, but where the contamination by extended bulge light affects the measured properties of
the outer, disc-dominated region (e.g.\ $\mu_{\rm brk}$, outer scalelength). For the remaining Type~III
spirals ($\sim15$~per~cent), the excess light at $r > r_{\rm brk}$ could potentially be attributed to the
bulge profile (Type~III-s).

In considering these results it is important to note that the methodology adopted by \cite{Maltby_etal:2012b}
has two main drawbacks: i)~in a two-component B--D decomposition, an outer antitruncated disc, bar feature
or outer halo could cause the bulge profile to be constrained; and ii) in many cases the bulge component may
be not be de Vaucouleurs in nature (i.e. less concentrated -- pseudo bulge). These issues lead to an
over-estimation of bulge light in the outer regions of some galaxies. This naturally enhances the fraction of
Type III-s profiles in \cite{Maltby_etal:2012b}, which therefore represents an upper limit to the fraction of
genuine Type~III-s profiles in their Type III spirals. Adding more degrees of freedom to their galaxy model
(more components; S{\'e}rsic bulge profile) would address this issue. However, degeneracy issues would affect
the reliability of their \mbox{B--D} decompositions and could lead to many Type~III-s profiles being
misclassified. Consequently, in assessing the potential impact of the bulge component on the outer regions of
the galaxy, it becomes desirable to consider the {\em maximum} possible effect (i.e.\ a de Vaucouleurs,
$r^{1/4}$ bulge). Taking this into consideration, \cite{Maltby_etal:2012b} conclude that in the vast majority
of cases Type~III profiles in spiral galaxies are indeed a true disc phenomenon.

\subsection{Antitruncations in S0 galaxies}

\label{Antitruncations in S0 galaxies}

We explore the nature of Type~III $\mu(r)$ profiles in our S0s by using an analogous method to that used for
spiral galaxies in \cite{Maltby_etal:2012b} (see Section~\ref{Antitruncations in spiral galaxies}). We
therefore determine the {\em maximum} fraction of Type~III S0s for which the excess light at large radii
could be caused or affected by the spheroidal component.

For this study, we use the low-axis-ratio sample of $173$ S0s defined in Section~\ref{Galaxy inclination}
and used throughout Section~\ref{The structure of galactic discs in spiral and S0 galaxies} (hereafter
refered to as the total S0 sample). However, we remove five of these galaxies for which B--D decomposition
fails (probably due to contamination from a nearby star or companion galaxy; $N_{\rm tot} = 168$). We also
use the disc profile classifications from Section~\ref{Results} to obtain a robust S0 sub-sample of $39$
Type~III $\mu(r)$ profiles (i.e.\ where classifications from the three independent assessors were in
agreement). We use both this Type~III sub-sample and the total S0 sample in this study.

\subsubsection{Bulge--disc decompositions}

\label{Bulge-disc decompositions}

\begin{figure*}
\centering
\includegraphics[width=0.195\textwidth]{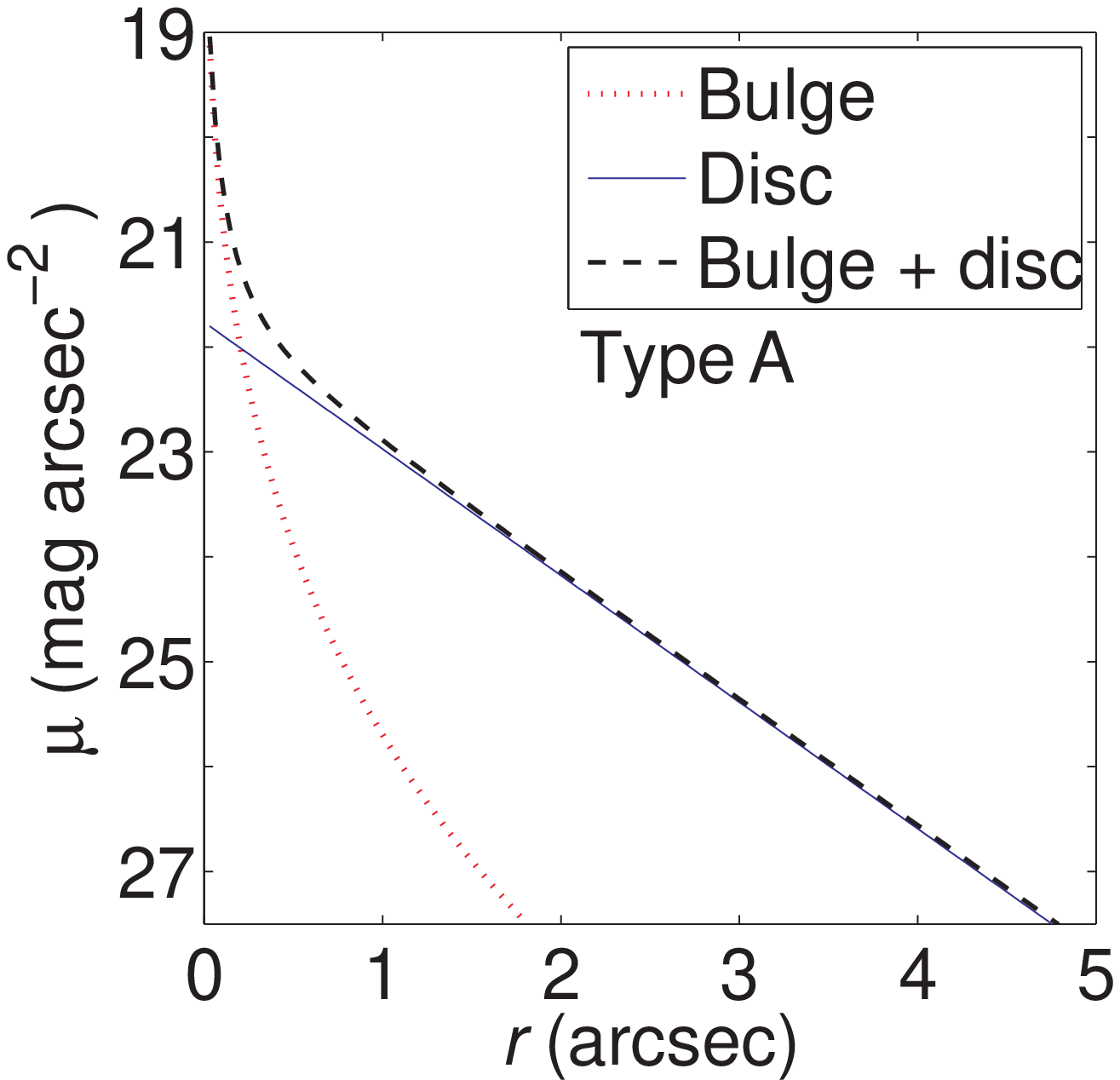}
\includegraphics[width=0.195\textwidth]{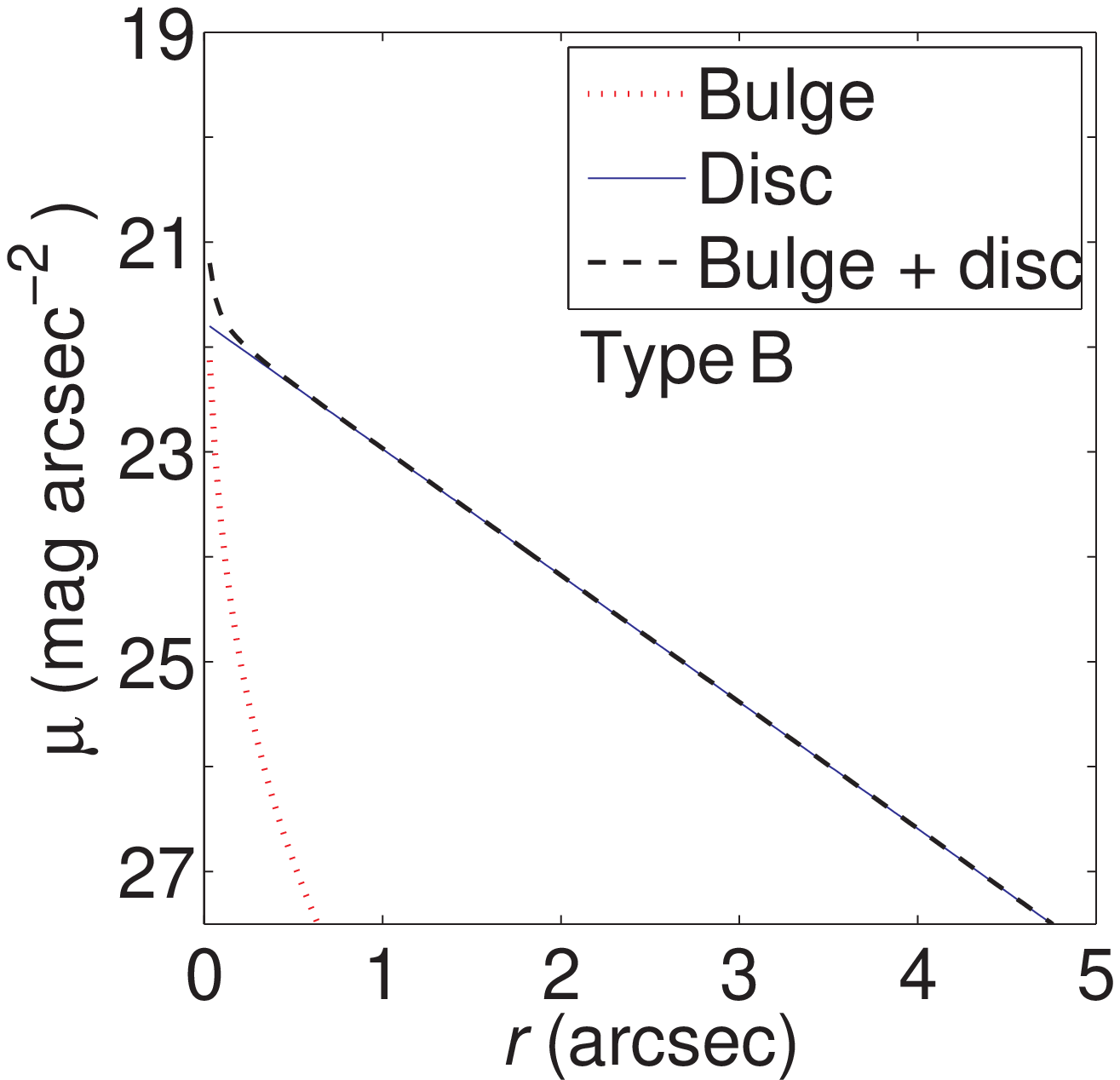}
\includegraphics[width=0.195\textwidth]{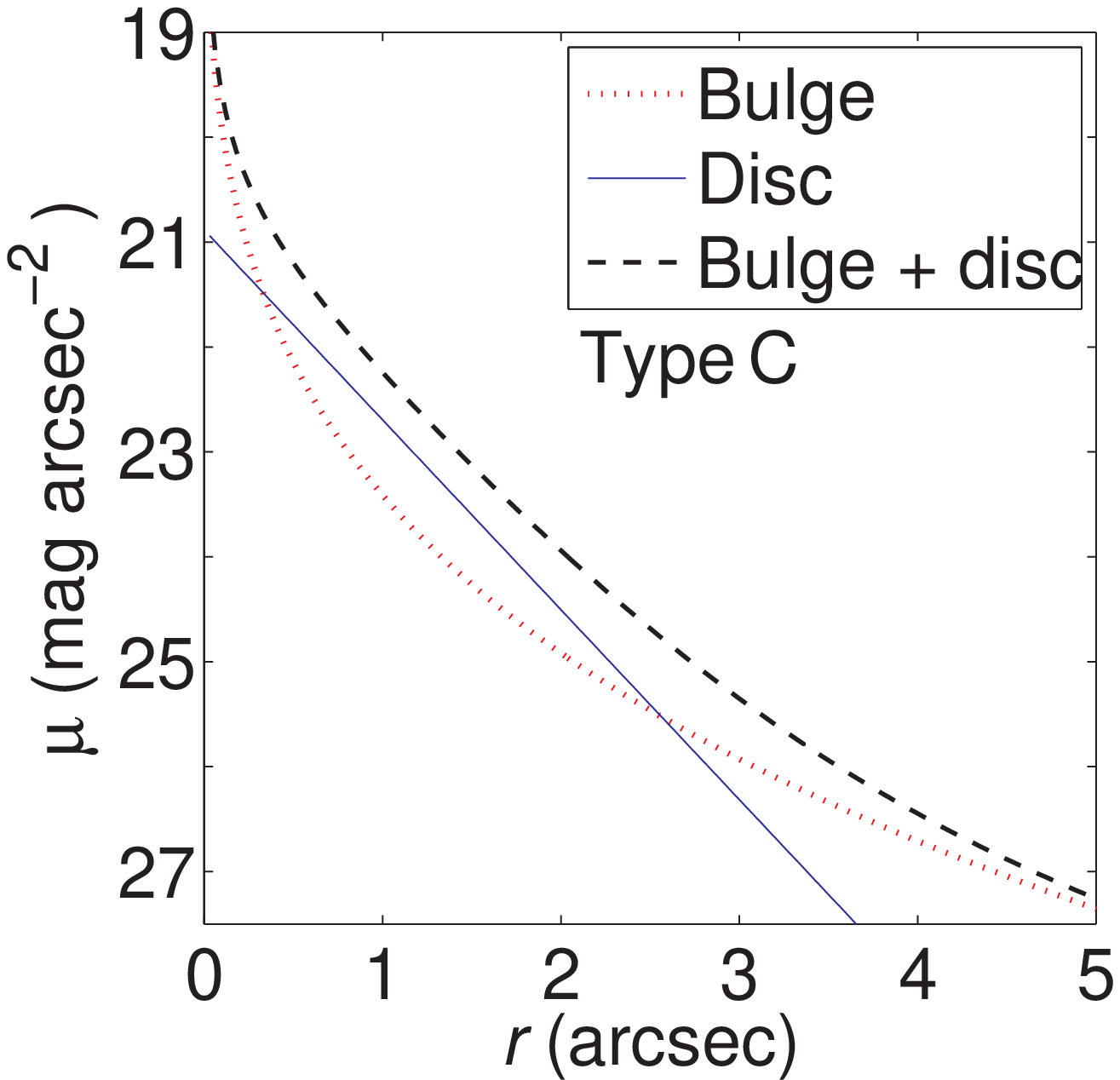}
\includegraphics[width=0.195\textwidth]{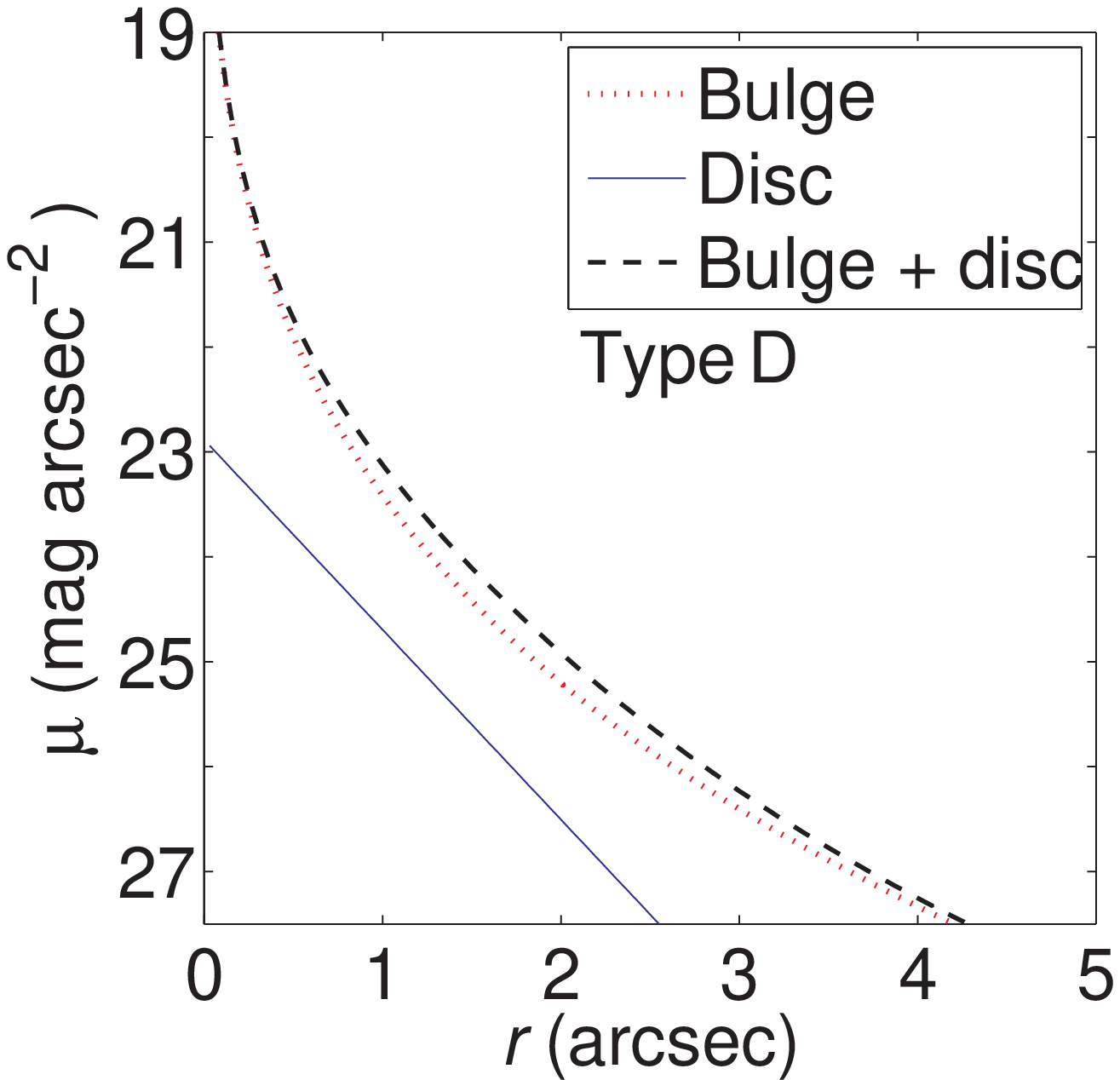}
\includegraphics[width=0.195\textwidth]{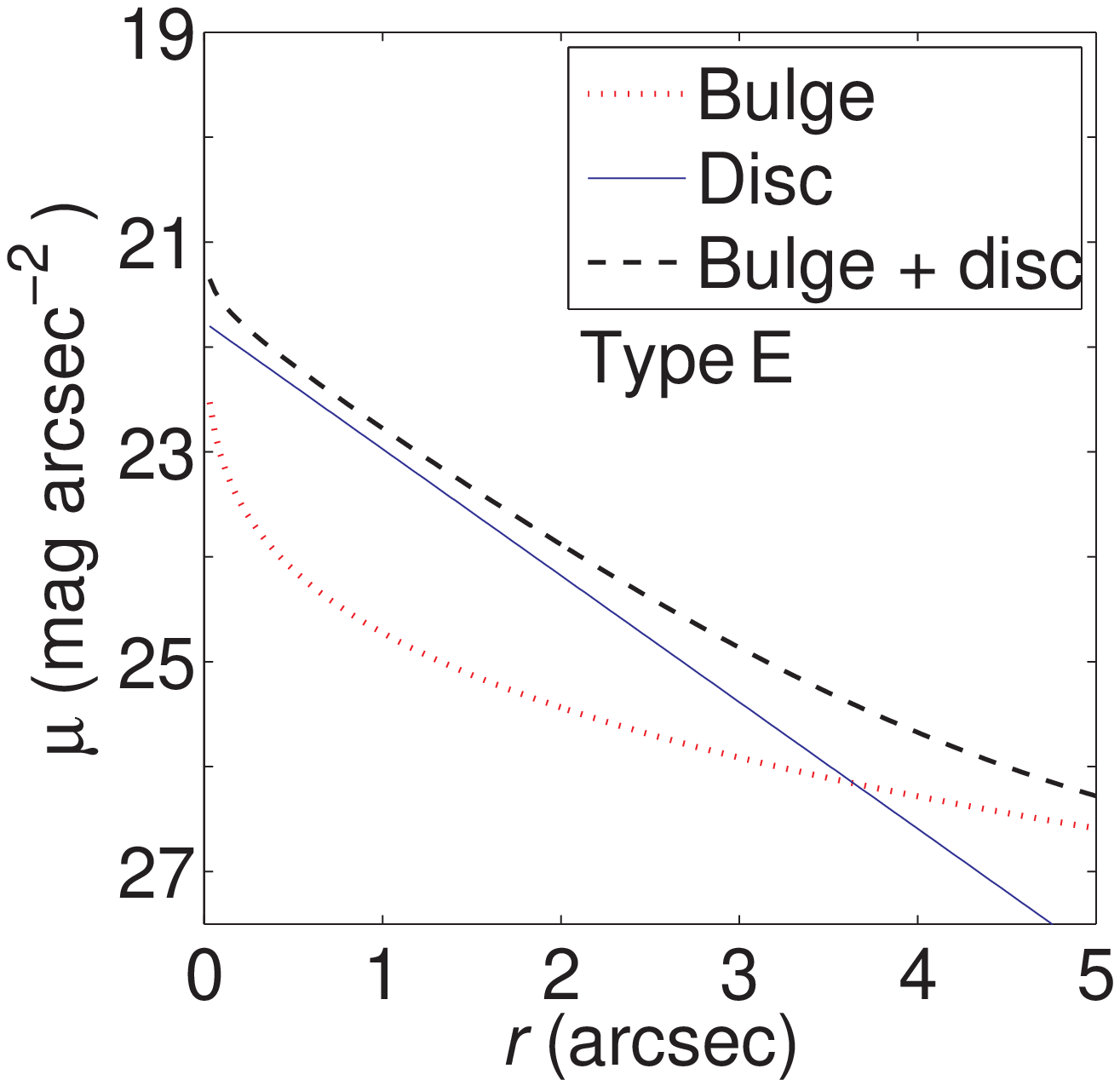}
\caption{\label{B-D profile types} B--D profile types. Left to right: Type~A -- `classical' system; Type~B
-- disc-dominated system; Type~C -- bulge-dominated at small/large radii but disc-dominated at intermediate
radii; Type~D -- bulge-dominated system; Type~E -- probable `constrained' outer bulge (caused by an outer
shallow disc).}
\end{figure*}

For each S0 galaxy in our total sample, we perform a two-dimensional \mbox{B--D} decomposition based on a
two-component galaxy model comprising a de Vaucouleurs ($r^{1/4}$) bulge and a single exponential disc.
Decompositions were carried out on the STAGES $V$-band imaging using the {\sc galfit} code
\citep{Peng_etal:2002} and the method of \cite{Hoyos_etal:2012} adapted to perform two-component fits.
Several measurable properties are produced for each galaxy including position [$x$,$y$], effective radii,
total magnitudes, axial ratios, position angles for the bulge and disc components and a sky-level estimation.

In this work, it is important to note that our B--D decompositions are not intended to yield the actual
bulge components of our S0 galaxies. Instead the intention is to obtain the {\em maximum} possible
contribution a bulge profile can give to the light in the outer regions of a galaxy. This is the motivation
behind adopting a de Vaucouleurs ($r^{1/4}$) profile for our bulge components instead of a free S{\'e}rsic
profile (see Section~\ref{Antitruncations in spiral galaxies}, for further details). However, we acknowledge
that many of our S0s will have less concentrated bulge profiles (i.e.\ pseudo bulges) and that our B--D
decompositions will (by design) over-estimate the bulge light in the outer regions of some galaxies.

B--D decomposition can be sensitive to the initial conditions used to search the B--D parameter space
(e.g.\ initial estimate for bulge-to-disc ratio $B/D$). Therefore, we perform two runs of the B--D
decomposition with different initial conditions taken from the two extremes: one run starting from a
bulge-dominated system ($B/D = 9$) and the other run starting from a disc-dominated system ($B/D = 1/9$).
Comparison of these runs (hereafter Run~$1$ and Run~$2$, respectively) allows for an assessment of the
uniqueness/stability of B--D decomposition on a galaxy--galaxy basis.

In the vast majority of cases ($\sim90$ per cent) the results were effectively the same, $\sim85$ per cent
being exactly the same and $\sim5$ per cent showing only minor differences. In only a few cases ($<5$ per
cent) were the decompositions catastrophically unstable with Run~$1$/$2$ yielding both bulge- and
disc-dominated systems. The remaining cases ($\sim5$ per cent), showed moderate instabilities great enough
to affect the assessment of bulge light in the outer regions of the galaxy. The unstable solutions are
mainly driven by differences in the sky level determined during the decomposition. However, the overall
conclusions of this study are not affected by these unstable solutions. The stability fractions quoted are
the same for both the total sample and the Type~III sub-sample.

\subsubsection{B--D profile types}

B--D decompositions using a de Vaucouleurs ($r^{1/4}$) bulge plus an exponential disc can be classified
into four distinct profile types \citep[e.g.][]{Maltby_etal:2012b}, see Fig.~\ref{B-D profile types}.

\begin{enumerate}
\renewcommand{\theenumi}{(\arabic{enumi})}
\item {\em Type~A:} `classical' system. The bulge profile dominates in the central regions, while the disc
profile dominates at larger radii. The bulge/disc profiles cross only once.
\item {\em Type~B:} disc-dominated system. The disc profile dominates at all radii, with a weak contribution
from the bulge profile in centre. The bulge/disc profiles never cross.
\item {\em Type~C:} the bulge profile dominates at small/large radii, but the disc profile dominates at
intermediate radii.
\item {\em Type~D:} bulge-dominated system. The bulge profile dominates at all radii with a weak underlying
disc component. The bulge/disc profiles never cross\footnote{Note: capital letters are used in our B--D
profile types to avoid confusion with the other classification schemes used in this paper.}.
\end{enumerate}

In addition to these profile types, a further class (hereafter Type~E) is also observed where the disc
profile dominates in the central regions, but the bulge profile dominates at larger radii. In such cases,
it is probable that an outer \mbox{antitruncated} disc has incorrectly affected the bulge profile fit.
Consequently, for Type~E profiles B--D decomposition is not a true representation of the galaxy at large
radii and in reality these galaxies probably have Type~B compositions (or similar). Analogous constraints
may also occur in some Type~D profiles.

\begin{figure}
\centering
\includegraphics[width=0.47\textwidth]{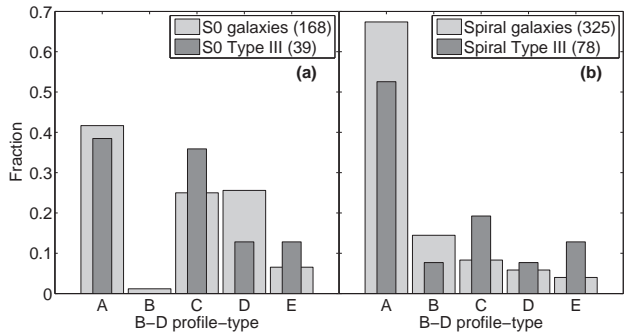}
\caption {\label{B-D profile type distributions} The distribution of B--D profile types in disc galaxies.
Left-hand panel: the distribution for S0 galaxies (from Run~$1$). Distributions are presented for the total
S0 sample (light grey) and the Type~III S0 sub-sample (dark grey). Right-hand panel: analogous distributions
for the spiral galaxies presented in \protect\cite{Maltby_etal:2012b}. Respective sample sizes are shown in
the legends.}
\end{figure}

Fig.~\ref{B-D profile type distributions}(a) shows the distribution of B--D profile types for both the total
S0 sample and Type~III sub-sample. Comparing these distributions, we find that the fraction of Type~C/E
profiles is greater in the Type~III sub-sample. This is expected from the nature of Type~C/E profiles
(i.e.~excess light at large radii, see Fig.~\ref{B-D profile types}). A similar trend is also observed in the
distribution of B--D profile types for {\em spiral} galaxies in \cite{Maltby_etal:2012b} [see Fig.~\ref{B-D
profile type distributions}(b)]. However, a comparison of the B--D profile type distributions between spiral
and S0 galaxies shows some distinct differences. The fraction of disc-dominated B--D profile types
(Type~A/B) is much larger in spiral galaxies compared with S0s. Conversely, the fraction of bulge-dominated
B--D profile types (Type~C/D) is much larger in S0s compared with spiral galaxies. Therefore, these
distributions indicate that on average S0s have a higher B/D than spiral galaxies, an observation that has
also been reported previously by several authors \citep[e.g.][]{Simien_deVaucouleurs:1986,
Laurikainen_etal:2010}.

\begin{figure}
\centering
\includegraphics[width=0.47\textwidth]{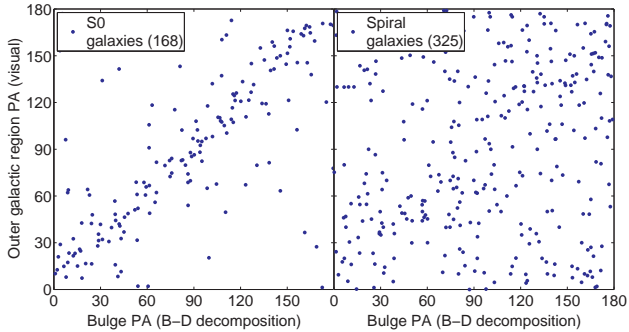}
\caption{\label{PA comparisons} A comparison of the bulge $\rm PA$ (from B--D decomposition) with a visual
estimate for the $\rm PA$ of the outer galactic regions (i.e.~outer stellar disc, see
Section~\ref{Profile fitting}) for both our S0 galaxies (left-hand panel) and the spiral galaxies from
\protect\cite{Maltby_etal:2012b} (right-hand panel). Respective sample sizes are shown in the legends.}
\end{figure}

We expand on this result by comparing the bulge $\rm PA$ (from B--D decomposition) with a visual estimate
for the $\rm PA$ of the outer galactic region (i.e.\ $\rm PA$ of the outer stellar disc used in our
fixed-ellipse fits, see Section~\ref{Profile fitting}). Fig.~\ref{PA comparisons} shows these comparisons
for both our sample of S0s and the spiral galaxies from \cite{Maltby_etal:2012a,Maltby_etal:2012b}. For
spiral galaxies, there is no obvious correlation between the $\rm PA$ of the bulge and the outer galactic
region. This indicates that our visual estimates are not related to the inner bulge and are indeed probing
an outer disc component. Additionally, since galactic discs are essentially axisymmetric (a consequence of
rotational motion), the misalignment between the bulge and the disc also suggests a triaxial
(non-axisymmetric) nature for the bulges of our spiral galaxies \citep[see e.g.][]{Bertola_etal:1991}. In
contrast, for S0 galaxies there is a significant correlation between the $\rm PA$ of the bulge and the outer
galactic region. This alignment between the bulge and the disc component implies a bulge that is axisymmetric
(oblate) in nature for many of our S0 galaxies. However, this alignment also suggests the possibility that
for some of our S0s, the visual estimates are actually probing the outer regions of the bulge component (and
not the stellar disc). This result is consistent with the observation that S0s tend to have a higher $B/D$
than spiral galaxies. Considering the high fraction of bulge-dominated B--D profiles (Type~C/D) in our S0s,
this is not surprising (see Fig.~\ref{B-D profile type distributions}). However, more importantly, this
result clearly indicates that light from the spheroidal (bulge) component potentially contributes a
significant amount of light to the outer regions of some S0 galaxies. Consequently, bulge light may account
for more Type~III profiles in S0s than observed for spiral galaxies in \cite{Maltby_etal:2012b}. We explore
this issue in further detail in the following section.

\subsubsection{Measured surface brightness $\mu(r)$ profiles}

\label{Measured surface brightness mu(r) profiles}

\vspace{0.1cm}

For each S0 galaxy, we also obtain new azimuthally-averaged radial $\mu(r)$ profiles from the STAGES
{\em HST}/ACS $V$-band imaging. We achieve this using a similar methodology to that described in
Section~\ref{Profile fitting}, but with two minor differences:

\vspace{0.05cm}

\begin{enumerate}

\item in the fixed-parameter fits (fixed centre, ellipticity $e$ and position angle $\rm PA$), we use the
galaxy centre determined from our B--D decomposition. As in previous sections, the $e$ and $\rm PA$ used are
for the outer stellar disc (see Section~\ref{Profile fitting});

\item the necessary sky subtraction is performed using the sky-level estimates generated during B--D
decomposition. Note: these sky values sometimes differ slightly from those of \cite{Gray_etal:2009} used
throughout Sections~\ref{Profile fitting}--\ref{The structure of galactic discs in spiral and
S0 galaxies}.

\end{enumerate}

\vspace{0.05cm}

Analogous fixed-parameter fits (using the same $e$ and $\rm PA$, i.e.\ the same isophotes) are also carried
out on the disc-residual images (ACS image minus bulge-only model) resulting in a measured $\mu$ profile for
the disc component~$\mu_{\rm disc}(r)$. We also obtain azimuthally-averaged radial $\mu$ profiles for the
decomposed B--D model using the same fixed-parameter ellipses (isophotes) as in the other profiles. This
results in separate analytical radial $\mu$ profiles for both the bulge- and disc-model along the semimajor
axis of the elliptical isophotes (i.e.\ outer stellar disc).

\begin{figure}
\centering
\includegraphics[width=0.47\textwidth]{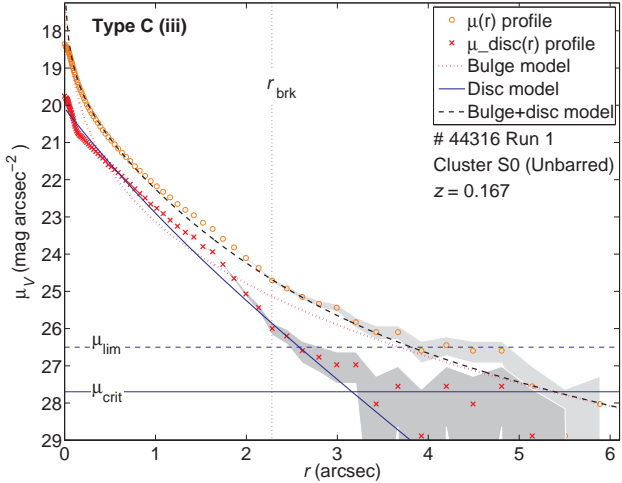}
\caption{\label{S0 Type C} A common example of a bulge profile causing an anti-truncation in an S0 $\mu(r)$
profile (break radius $r_{\rm brk}$). The bulge (red dotted line), disc (blue line), and bulge $+$ disc
(black dashed line) profiles from B--D decomposition are overplotted on the measured $\mu(r)$ profile (red
circles). The disc-residual $\mu_{\rm disc}(r)$ profile (measured $\mu$ profile minus bulge-only model, red
crosses) shows no antitruncation. Errors in $\mu(r)$/$\mu_{\rm disc}(r)$ are for an oversubtraction and
an undersubtraction of the sky by $1\sigma$. The $\mu_{\rm crit}$/$\mu_{\rm lim}$ levels represent
$+1\sigma$/$+3\sigma$ above the sky, respectively.}
\end{figure}

For our Type~III S0s, we compare these measured $\mu(r)$ profiles with the model $\mu$ profiles from B--D
decomposition in order to determine the {\em maximum} contribution of bulge light in the outer regions of the
galaxy ($r > r_{\rm brk}$)\footnote{Note: due to our B--D decomposition using a de Vaucouleurs, $r^{1/4}$
bulge, the model bulge profile often over-estimates the contribution of bulge light in the outer regions of
the $\mu(r)$ profile.}. Using an analogous scheme to that presented in \cite{Maltby_etal:2012b}, we find that
bulge light in the outer profile ($r > r_{\rm brk}$) either had:

\vspace{0.05cm}

\begin{enumerate}

\item{\em little or no contribution} ($\sim45$ per cent): for all Type~A/B profiles the bulge contributes
virtually no light at $r > r_{\rm brk}$ and in some Type~C/E profiles the contribution is negligible.
This can be determined by inspection of the measured disc-residual profile $\mu_{\rm disc}(r)$ and
assessing if the properties of the outer profile/break ($r_{\rm brk}$, $\mu_{\rm brk}$, scalelength)
have been affected with respect to the sky-subtraction error. For our Type~III S0s, these cases are entirely
comprised of Type~A profiles. No Type~B profiles are observed in our Type~III S0 galaxies.

\item{\em minor contribution} ($\sim10$ per cent): the majority of these cases are Type~C profiles (with one
exception which is Type~E). The amount contributed is enough to affect the outer profile causing
$\mu_{\rm brk}$ and the outer scalelength to be different in the disc-residual profile $\mu_{\rm disc}(r)$.
However, the antitruncation remains present.

\item{\em major contribution} ($\sim45$ per cent): the bulge contributes the majority of the light at
$r > r_{\rm brk}$. For these cases, one quarter are Type~D, one quarter are Type~E, and the remaining half
are Type~C (where the \mbox{antitruncation} can be entirely accounted for by bulge light, see
Fig.~\ref{S0 Type C} for one such an example).

\end{enumerate}

\vspace{0.05cm}

These results suggest that for only about a half of Type~III S0s, the excess light beyond the break radius
$r_{\rm brk}$ is related to an outer shallow disc \mbox{(Type~III-d)}. For the remaining cases, the excess
light at $r > r_{\rm brk}$ could be attributed to a bulge profile (Type~III-s). More importantly, for these
latter cases a considerable fraction ($\sim20$ per cent of the Type~III S0 sub-sample) exhibit profiles where
the bulge profile extends beyond a dominant disc (Type~C) and causes an antitruncation in the $\mu(r)$
profile.

However, it is important to note that our adopted methodology does have two main drawbacks: i) in a
two-component B-D decomposition an outer antitruncated disc, bar feature or outer halo could cause the bulge
profile to be contrained; and ii) in many cases the bulge component may not be de Vaucouleurs in nature
(i.e.\ less concentrated - pseudo bulge). These issues lead to an over-estimation of bulge light in the outer
regions of some galaxies (see Section~\ref{Antitruncations in spiral galaxies} and
\ref{Bulge-disc decompositions}, for the justification of this methodology). This naturally enhances the
fraction of Type~III-s profiles, which therefore represents an upper limit to the fraction of genuine
Type~III-s profiles in our Type III S0s. Due to this fact, this method is unsuitable for removing Type~III-s
profiles from our Type~III analysis in Sections~\ref{Results} and
\ref{The structure of galactic discs in spiral and S0 galaxies}.

With respect to these results, we conclude that bulge light is an important component in the $\mu(r)$
profiles of some S0 galaxies at large radii and that antitruncations in S0s are frequently caused by either
the bulge or disc component. This is in sharp contrast to the conclusion of our analogous study
for spiral galaxies, \cite{Maltby_etal:2012b}, where in the vast majority of cases Type~III profiles appear
to be a true disc phenomenon.

\subsection{Implications for the formation of S0 galaxies}

\label{Implications for the formation of S0 galaxies}

Various mechanisms have been suggested to explain the transformation of spiral galaxies into S0s. However,
these mechanisms generally require some type of interaction that causes a quenching of star formation and the
subsequent passive evolution of the spiral galaxy into an S0 \citep[see e.g.][]{Aragon-Salamanca_etal:2006}.
The nature of this interaction could be directly related to the gaseous component, e.g.\ tidal/ram-pressure
stripping of the inter-stellar medium \citep{Gunn_Gott:1972,Faber:1973} or the removal of the outer gas halo
\cite[starvation;][]{Larson_etal:1980}. Alternatively, the interaction could be related to minor mergers
which trigger starbursts that deplete the gas supply throughout the disc \citep[e.g.][]{Mihos_Hernquist:1994}. 

Regardless of the precise mechanism, the transition of a spiral galaxy into an S0 is expected to have a
distinct effect on the $B/D$. For example, the suppression of star formation in the stellar disc (e.g.\ by
gas stripping) would cause the disc component to gradually fade as the stellar population ages and the $B/D$
to increase. However, the degree to which the disc fades in this scenario is currently uncertain. Another
consideration is that many suggested formation mechanisms lead to gas concentration in the very centre of the
S0 galaxy during its transformation phase. This has recently been supported by observations suggesting a
central starburst (i.e.\ bulge growth) occurred in many S0s during the process of star formation being
quenched in the outer disc \citep[e.g.][]{Bedregal_etal:2011, Johnston_etal:2012, Johnston_etal:2014}. This
process would also lead to an increase in the $B/D$ as the bulge becomes more luminous.

The comparison of our B--D decompositions for Type~III S0s with those for the Type~III spiral galaxies from
\cite{Maltby_etal:2012b} should build on these ideas and provide some further insight into the potential
processes by which spiral galaxies transform into S0s. In the following, we highlight two key observations
from our results:

\vspace{0.35cm}

(i) {\em S0s generally have a higher $B/D$ than spiral galaxies:} our B--D decompositions indicate that the
fraction of bulge-dominated $\mu(r)$ profiles is larger in S0s than spiral galaxies, and that spiral
galaxies have mainly disc-dominated $\mu(r)$ profiles. This observation is in agreement with various similar
observations by previous works \citep[e.g.][]{Simien_deVaucouleurs:1986, Laurikainen_etal:2010}. With respect
to S0 formation theories, this result is consistent with both a fading stellar disc and a central starburst
(i.e.\ bulge growth). Consequently, it appears a galaxy undergoing a spiral $\rightarrow$ S0 transformation
should naturally evolve into a more bulge-dominated system.
\vspace{0.2cm}

(ii) {\em Bulge light can account for more Type~III profiles in S0s than spiral galaxies:} our results
suggest that an extended bulge component can account for Type~III features in as many as $\sim45$ per cent of
Type~III S0s, but in only $\sim15$ per cent of Type~III spirals \citep{Maltby_etal:2012b}. This result can
also be understood by the concept of a fading stellar disc as spirals transform into S0s. As the stellar disc
fades, the $B/D$ increases and consequently the tail end of the bulge profile may eventually dominate over
the disc at large radii (i.e.~as in Type~C/D profiles; see Fig.~\ref{B-D profile types}). This process would
naturally lead to an increase in the fraction of Type~III-s profiles in S0 galaxies. Currently, it is
uncertain as to whether the disc could fade to the required degree in order for this to occur. However, this
result is difficult to understand using only the concept of a central starburst (i.e.~bulge growth) since the
outer regions of the bulge should not be affected in this scenario. Consequently, it appears some disc fading
may be required to in order to explain this result.

\vspace{0.2cm}

We note that the fading stellar disc hypothesis is also consistent with the other structural comparisons
presented in Section~\ref{The structure of galactic discs in spiral and S0 galaxies} for spiral and S0
galaxies. A fading stellar disc would preserve the disc scalelength. Therefore, the scalelength of Type~I
profiles would be expected to be independent of morphology (see Fig.~\ref{SpS0: Type I analysis}). For
Type~II profiles, the suppression of star formation would cause a fading (ageing) of the `break region' with
respect to the inner/outer discs, leading to a convergence of $M/L$ across the $\mu(r)$ break. This would
cause the Type~II feature to get weaker and may even disappear. Consequently, the observation that Type~II
profiles are weaker/rarer in S0s compared to spiral galaxies is also consistent with this scenario (see
Section~\ref{SpS0: profile type} and Fig.~\ref{SpS0: Type II/III analysis}). We therefore conclude that in
addition a central starbust (i.e.\ bulge growth), a fading stellar disc seems to be an inherent process in
the morphological transformation of spiral galaxies into S0s.

\section[]{Conclusions}

\label{Conclusions}

We present an analysis of $V$-band radial surface brightness profiles $\mu(r)$ for S0 galaxies from the field
and cluster environment using {\em HST}/ACS imaging and data from the STAGES survey. Using a large,
mass-limited \mbox{($M_* > 10^9\rm\,M_\odot$)}, visually classified sample of $\sim280$ field and cluster
S0s, we assess the effect of the galaxy environment on the shape of S0 $\mu(r)$ profiles and the structure of
S0 stellar discs. We also compare the structure of our S0 stellar discs with those for spiral galaxies from
our previous works in order to provide insight into the potential evolutionary mechanisms by which spiral
galaxies evolve into S0s.

\subsection{Environmental analyses: S0 galaxies}

We classify our S0s according to $\mu$ break features in their $\mu(r)$ profiles and find that the frequency
of profile types (Type~I, II and III) is approximately the same in both the field and cluster environments.
For both field and cluster S0s, $\sim25$ per cent have a simple exponential profile (Type~I), $<5$ per cent
exhibit a down-bending break (truncation, Type~II) and $\sim50$ per cent exhibit an up-bending break
(antitruncation, Type~III). For the remaining S0 galaxies ($\sim20$ per cent), no discernible exponential
component was observed (i.e.\ general curvature in the $\mu$ profile, Type~c). These profile fractions are
robust to the subjective nature of the profile classifications, agreeing for classifications performed by
three independent assessors. We also find that limiting our analysis to only low-axis ratio systems
($q > 0.5$) has no significant effect on our profile fractions. These results imply that the shape of S0
$\mu(r)$ profiles is not dependent on the galaxy environment.

The distinct lack of truncations (Type~II profiles) in both our field and cluster S0s is of particular
interest. In previous works on the disc structure of {\em spiral} galaxies \citep[e.g.][]
{Pohlen_Trujillo:2006,Erwin_etal:2008,Gutierrez_etal:2011}, Type~II profiles are very common with the
distribution of profile types I:II:III being approximately $20$:$50$:$30\pm10$ per cent. Therefore, it seems
whatever mechanism transforms spiral galaxies into S0s may erase these truncations from their $\mu(r)$
profiles. We shall return to this result in Section~\ref{Conclusion: structural analyses}. This result is in
partial agreement with a similar result reported recently by \cite{Erwin_etal:2012}, who find no Type~II S0s
in the cluster environment but a Type~II S0 fraction of $\sim30$ per cent in the field. Therefore, our
Type~II S0 fractions are in perfect agreement with \cite{Erwin_etal:2012} for the cluster environment, but
differ significantly for the field. The origin of this disagreement is uncertain but may be related to the
lower fraction of barred S0s (and hence bar-related truncations) in our field sample.

For S0s with a pure exponential disc (Type~I), we find no evidence to indicate that the disc scalelength $h$
is dependent on the galaxy environment. Additionally, for S0s with an antitruncated disc (Type~III) we find
no evidence to suggest any environmental dependence on either the break surface brightness $\mu_{\rm brk}$ or
the break strength $T$ (outer-to-inner scalelength ratio, ${\rm log}_{10}\,h_{\rm out}/{h}_{\rm in}$). These
results have been shown to be robust to the error in the sky subtraction and the subjective nature of the
profile classifications. We also find that limiting our analysis to only low-axis-ratio systems ($q > 0.5$)
has no effect on these results. Therefore, we conclude that there is no evidence to suggest that the stellar
distribution in the stellar disc of S0 galaxies is directly affected by the galaxy environment.

These results are consistent with our analogous work, \cite{Maltby_etal:2012a}, which reaches the same
conclusion but for a sample of {\em spiral} galaxies from STAGES. Our results are also consistent with other
studies carried out on the effect of the galaxy environment on disc features in the STAGES survey. For
example, \cite{Marinova_etal:2009} find that the optical fraction of bars among disc galaxies show no
evidence for any strong variation between the field and the A901/2 clusters, suggesting the mass
redistribution associated with bar formation within galactic discs is not a strong function of environment
from the general field to the intermediate densities of the A901/2 clusters. 

However, our results are for one survey field (STAGES), and one multicluster complex of intermediate galaxy
density at low redshift ($z\sim0.167$). Therefore, it is important to investigate whether we see the same
trends observed in the STAGES A901/2 field in other survey fields across a wide range of redshift and cluster
mass. Extending these studies to higher redshifts is of key importance. In the relatively local Universe,
structural evolution in a galaxy's stellar distribution may already have ceased in both the field and cluster
environments, even if the environment is the principal driver. However, in the more distant Universe
structural changes may still be occurring in both the field and cluster environments and at different rates.
Probing denser, more massive cluster environments (e.g.\ the Coma cluster) is also important because some
environmental drivers may only be significant in very high-density environments (i.e.\ rich cluster cores).
Ultimately, the comparison of high-redshift studies with those from the local Universe across a wide range of
environments will allow for a complete assessment of whether or not the galaxy environment has any direct
effect on a galaxy's stellar distribution.

\subsection{Structural analyses: implications for S0 formation}

\label{Conclusion: structural analyses}

We complement our environmental studies by comparing the structural analyses of our S0s with those for
spiral galaxies from our previous works. These comparisons provide some insight into the potential
evolutionary paths by which spiral galaxies transform into S0s. Two structural comparisons were made:

\vspace{0.2cm}

(i) {\em The structure of galactic discs.} We compare the disc structure of our S0s with the spiral galaxies
from our analogous work, \cite{Maltby_etal:2012a}. For spiral/S0 galaxies with a pure exponential disc
(Type~I), we find no evidence to suggest that the disc scalelength $h$ is dependent on the galaxy morphology.
For spiral/S0 galaxies with an antitruncated disc (Type~III), we also find no evidence to suggest that the
break surface brightness $\mu_{\rm brk}$ is related to the galaxy morphology. However, we do find some
evidence (significance $>3\sigma$) that the break strength $T$ of spiral/S0 galaxies is somehow related to
the galaxy morphology, with $T$ for both Type~II and III profiles being generally smaller (weaker) in S0s
compared to spiral galaxies.

In order to understand this result, we need to consider the current theory for the formation of stellar disc
truncations. For classical truncations (Type~II-CT), current theories suggest that their formation is via a
radial star-formation threshold and the outward scattering of inner disc stars to regions beyond this
threshold \citep[i.e.\ break radius;][]{Debattista_etal:2006, Bakos_etal:2008, Martinez-Serrano_etal:2009}.
Consequently, the outer disc should be populated by old stars as these are the ones that have had enough
time to make the disc migration. In this scenario, the truncation (Type~II feature) is related to a radial
change in the age of the stellar population throughout the disc. Assuming this formation scenario, and an
inside-out growth for the inner disc [i.e.\ negative age(r) gradient], the suppression of star formation in
the galaxy (e.g.\ via gas stripping) would cause the age of the stellar population in the `break region' to
increase and the $M/L$ across the $\mu(r)$ break to converge. Consequently, the $\mu$ break will get weaker
and may even disappear. For bar-related truncations (Type~II-OLR), the $\mu$ break is expected to be related
to a resonance phenomenon and therefore the above scenario does not hold. However, \cite{Erwin_etal:2012}
suggest that the depletion/removal of gas from a barred galaxy would cause a weakening of the resonance
effect and may weaken or remove the Type~II-OLR break from the $\mu(r)$ profile. Considering these theories,
the absence/weakening of Type~II profiles in our S0s may actually be the natural consequence of the
termination of star formation in the stellar disc as spiral galaxies transform into S0s. For Type~III
profiles, the observed weakening may be the consequence of radial mixing throughout the disc related to the
suspected minor-merger history of the Type~III system.

In order to fully test these hypotheses, we require high-quality colour profiles, or better still stellar
age profiles, on a large sample of spiral/S0 galaxies covering a wide range of stellar masses. Only then will
we be able to test whether the weakening of the $\mu(r)$ break in S0 galaxies is related to a weakening of
their age profile gradient.

%

\vspace{0.2cm}

(ii) {\em The nature of antitruncated stellar light profiles.} We also explore the nature of
antitruncated (Type~III) stellar light profiles in S0 galaxies and assess the effect of a `classical',
\cite{deVaucouleurs:1948} bulge on the outer regions of their $\mu(r)$ profiles. In Type~III $\mu(r)$
profiles (up-bending breaks), the excess light beyond the break radius $r_{\rm brk}$ can either be related to
an outer exponential disc (Type~III-d) or an extended spheroidal component (Type~III-s). Using analytical
B--D decomposition (de Vaucouleurs, $r^{1/4}$ bulge plus single exponential disc) on a sample of $39$
Type~III S0s, we assess the {\em maximum} fraction of Type~III S0s for which the excess light at large radii
($r > r_{\rm brk}$) could be caused or affected by the spheroidal component.

Our results indicate that for only about a half of Type~III S0s, the antitruncation is related to an outer
shallow disc \mbox{(Type~III-d)}. For the remaining cases, the excess light at $r > r_{\rm brk}$ can be
accounted for by the bulge profile (Type~III-s). More importantly, for these latter cases there are many
S0s ($\sim20$ per cent of the Type~III sub-sample) that exhibit an antitruncated $\mu(r)$ profile caused by
the bulge profile extending beyond a dominant disc. However, with respect to these results, it is important
to note that our adopted methodology (the use of a de Vaucouleurs bulge in our B-D decomposition) leads to an
over-estimation of the bulge light in the outer regions of the $\mu(r)$ profile. This naturally enhances the
fraction of Type~III-s profiles, which therefore represents an upper limit to the fraction of genuine
Type~III-s  profiles in our Type~III S0s. Therefore, we conclude that bulge light is an important component
in the $\mu(r)$ profiles of many S0s at large radii and that \mbox{antitruncations} in S0 galaxies are
commonly caused by either the bulge or disc component. This result is in sharp contrast to the result of
\cite{Maltby_etal:2012b} for {\em spiral galaxies} where in the vast majority of cases ($\sim85$ per cent),
Type~III profiles are a true disc phenomenon (i.e. Type~III-d).

We propose that these results are consistent with the hypothesis that spiral galaxies transform into S0s by
the termination of star formation. The suppression of star formation in the stellar disc (e.g.\ by gas
stripping), would cause the disc component to gradually fade and the $B/D$ to increase. Consequently, any
galaxy undergoing a spiral $\rightarrow$ S0 transformation should naturally evolve into a more
bulge-dominated system. As the stellar disc fades the tail end of the bulge profile may eventually dominate
over the stellar disc at large radii. Consequently, if a fading stellar disc is inherent to the morphological
transformation of spirals into S0s, this would naturally lead to an increase in the fraction of
antitruncations caused by the bulge component (Type~III-s) in S0 galaxies. Alternative processes inherent to
the morphological transformation of spirals into S0s, such as a central starburst \citep{Bedregal_etal:2011,
Johnston_etal:2012, Johnston_etal:2014}, could also influence the bulge profile. However, the emergence
of the bulge profile over the disc in the {\em outer} regions implies some degree of disc fading is required.
Unfortunately, at present it is uncertain as to whether the disc could fade to the required degree in order
for this to occur and stellar population synthesis modelling would be required to study this further.

\vspace{0.3cm}

Taken together, the results of our environmental studies suggest that the galaxy environment has little
direct effect on the structure of a galaxy's stellar distribution (at least from the general field to the
intermediate densities probed by the STAGES survey). Consequently, our results imply that environmental
processes directly affecting the structure of the stellar distribution, i.e.\ galaxy-galaxy or galaxy-cluster
gravitational interactions (e.g.\ mergers and harassment), are not driving the observed morphology--density
relation. The results of our morphological comparisons are also consistent with this finding, implying that
a fading stellar disc is a likely process inherent to spiral $\rightarrow$ S0 transformations. This result is
also supported by the environmental comparisons of \cite{Bosch_etal:2013b}, using rotational gas kinematics.
\cite{Bosch_etal:2013b} find that in the cluster environment, disc galaxies with smooth morphology (i.e.\ S0s)
exhibit greater kinematic disturbances in the gas disc than disc galaxies with greater morphological asymmetry
(i.e.\ spirals). This suggests that a subtle cluster-related gas process (e.g.\ ram-pressure stripping) is
directly affecting the gaseous disc of cluster galaxies. Such processes could cause star formation to be
quenched in the stellar disc and bring about disc fading. Consequently, we conclude that more subtle processes
acting on the gaseous component of a galaxy (e.g.\ ram-pressure stripping) are more likely to play an important
role in the origin of the morphology--density relation and the transformation of spiral galaxies into S0s.

\section[]{Acknowledgements}

We thank the anonymous referee for their detailed and insightful comments on the original version of this
manuscript, which helped to improve it considerably. The support for STAGES was provided by NASA through
GO-10395 from STScI operated by AURA under NAS5-26555. DTM was supported by STFC. MEG was supported by an
STFC Advanced Fellowship. AB acknowledges the funding of the Austrian Science Foundation FWF (projects
P19300-N16 and P23946-N16).


\bibliographystyle{mn2e} \bibliography{DTM_bibtex} \bsp

\begin{thebibliography}{}

\bibitem[\protect\citeauthoryear{{Allen}, {Driver}, {Graham}, {Cameron},
  {Liske} \& {de Propris}}{{Allen} et~al.}{2006}]{Allen_etal:2006}
{Allen} P.~D.,  {Driver} S.~P.,  {Graham} A.~W.,  {Cameron} E.,  {Liske} J.,
  {de Propris} R.,  2006, \mnras, 371, 2

\bibitem[\protect\citeauthoryear{{Arag{\'o}n-Salamanca}, {Bedregal} \&
  {Merrifield}}{{Arag{\'o}n-Salamanca}
  et~al.}{2006}]{Aragon-Salamanca_etal:2006}
{Arag{\'o}n-Salamanca} A.,  {Bedregal} A.~G.,    {Merrifield} M.~R.,  2006,
  \aap, 458, 101

\bibitem[\protect\citeauthoryear{{Azzollini}, {Trujillo} \&
  {Beckman}}{{Azzollini} et~al.}{2008}]{Azzollini_etal:2008}
{Azzollini} R.,  {Trujillo} I.,    {Beckman} J.~E.,  2008, \apj, 684, 1026

\bibitem[\protect\citeauthoryear{{Bakos}, {Trujillo} \& {Pohlen}}{{Bakos}
  et~al.}{2008}]{Bakos_etal:2008}
{Bakos} J.,  {Trujillo} I.,    {Pohlen} M.,  2008, \apjl, 683, L103

\bibitem[\protect\citeauthoryear{{Barden}, {H{\"a}u{\ss}ler}, {Peng},
  {McIntosh} \& {Guo}}{{Barden} et~al.}{2012}]{Barden_etal:2012}
{Barden} M.,  {H{\"a}u{\ss}ler} B.,  {Peng} C.~Y.,  {McIntosh} D.~H.,    {Guo}
  Y.,  2012, \mnras, 422, 449

\bibitem[\protect\citeauthoryear{{Bedregal}, {Cardiel}, {Arag{\'o}n-Salamanca}
  \& {Merrifield}}{{Bedregal} et~al.}{2011}]{Bedregal_etal:2011}
{Bedregal} A.~G.,  {Cardiel} N.,  {Arag{\'o}n-Salamanca} A.,    {Merrifield}
  M.~R.,  2011, \mnras, 415, 2063

\bibitem[\protect\citeauthoryear{{Bertola}, {Vietri} \& {Zeilinger}}{{Bertola}
  et~al.}{1991}]{Bertola_etal:1991}
{Bertola} F.,  {Vietri} M.,    {Zeilinger} W.~W.,  1991, \apjl, 374, L13

\bibitem[\protect\citeauthoryear{{Bland-Hawthorn}, {Vlaji{\'c}}, {Freeman} \&
  {Draine}}{{Bland-Hawthorn} et~al.}{2005}]{BlandHawthorn_etal:2005}
{Bland-Hawthorn} J.,  {Vlaji{\'c}} M.,  {Freeman} K.~C.,    {Draine} B.~T.,
  2005, \apj, 629, 239

\bibitem[\protect\citeauthoryear{{Borch} et~al.,}{{Borch}
  et~al.}{2006}]{Borch_etal:2006}
{Borch} A.,  et~al., 2006, \aap, 453, 869

\bibitem[\protect\citeauthoryear{{B{\"o}sch} et~al.,}{{B{\"o}sch}
  et~al.}{2013a}]{Bosch_etal:2013a}
{B{\"o}sch} B.,  et~al., 2013a, \aap, 549, A142

\bibitem[\protect\citeauthoryear{{B{\"o}sch} et~al.,}{{B{\"o}sch}
  et~al.}{2013b}]{Bosch_etal:2013b}
{B{\"o}sch} B.,  et~al., 2013b, \aap, 554, A97

\bibitem[\protect\citeauthoryear{{Bournaud}, {Elmegreen} \&
  {Elmegreen}}{{Bournaud} et~al.}{2007}]{Bournaud_etal:2007}
{Bournaud} F.,  {Elmegreen} B.~G.,    {Elmegreen} D.~M.,  2007, \apj, 670, 237

\bibitem[\protect\citeauthoryear{{Brown}, {Cai} \& {DasGupta}}{{Brown}
  et~al.}{2001}]{Brown_etal:2001}
{Brown} L.~D.,  {Cai} T.~T.,    {DasGupta} A.,  2001, Stat. Sci., 16, 101

\bibitem[\protect\citeauthoryear{{Buitrago}, {Trujillo}, {Conselice},
  {Bouwens}, {Dickinson} \& {Yan}}{{Buitrago}
  et~al.}{2008}]{Buitrago_etal:2008}
{Buitrago} F.,  {Trujillo} I.,  {Conselice} C.~J.,  {Bouwens} R.~J.,
  {Dickinson} M.,    {Yan} H.,  2008, \apjl, 687, L61

\bibitem[\protect\citeauthoryear{{de Vaucouleurs}}{{de
  Vaucouleurs}}{1948}]{deVaucouleurs:1948}
{de Vaucouleurs} G.,  1948, Annales d'Astrophysique, 11, 247

\bibitem[\protect\citeauthoryear{{de Vaucouleurs}}{{de
  Vaucouleurs}}{1959}]{deVaucouleurs:1959b}
{de Vaucouleurs} G.,  1959, Handbuch der Physik, 53, 311

\bibitem[\protect\citeauthoryear{{Debattista}, {Mayer}, {Carollo}, {Moore},
  {Wadsley} \& {Quinn}}{{Debattista} et~al.}{2006}]{Debattista_etal:2006}
{Debattista} V.~P.,  {Mayer} L.,  {Carollo} C.~M.,  {Moore} B.,  {Wadsley} J.,
    {Quinn} T.,  2006, \apj, 645, 209

\bibitem[\protect\citeauthoryear{{Dressler}}{{Dressler}}{1980}]{Dressler:1980}
{Dressler} A.,  1980, \apj, 236, 351

\bibitem[\protect\citeauthoryear{{Elmegreen} \& {Parravano}}{{Elmegreen} \&
  {Parravano}}{1994}]{Elmegreen_Parravano:1994}
{Elmegreen} B.~G.,  {Parravano} A.,  1994, \apjl, 435, L121

\bibitem[\protect\citeauthoryear{{Erwin}, {Beckman} \& {Pohlen}}{{Erwin}
  et~al.}{2005}]{Erwin_etal:2005}
{Erwin} P.,  {Beckman} J.~E.,    {Pohlen} M.,  2005, \apjl, 626, L81

\bibitem[\protect\citeauthoryear{{Erwin}, {Guti{\'e}rrez} \& {Beckman}}{{Erwin}
  et~al.}{2012}]{Erwin_etal:2012}
{Erwin} P.,  {Guti{\'e}rrez} L.,    {Beckman} J.~E.,  2012, \apjl, 744, L11

\bibitem[\protect\citeauthoryear{{Erwin}, {Pohlen} \& {Beckman}}{{Erwin}
  et~al.}{2008}]{Erwin_etal:2008}
{Erwin} P.,  {Pohlen} M.,    {Beckman} J.~E.,  2008, \aj, 135, 20

\bibitem[\protect\citeauthoryear{{Faber}}{{Faber}}{1973}]{Faber:1973}
{Faber} S.~M.,  1973, \apj, 179, 423

\bibitem[\protect\citeauthoryear{{Ferguson}, {Irwin}, {Chapman}, {Ibata},
  {Lewis} \& {Tanvir}}{{Ferguson} et~al.}{2007}]{Ferguson_etal:2007}
{Ferguson} A.,  {Irwin} M.,  {Chapman} S.,  {Ibata} R.,  {Lewis} G.,
  {Tanvir} N.,  2007, in de Jong, R.~S., ed., Island Universes: Structure and
  Evolution of Disk Galaxies.
Springer, Dordrecht, p.~239

\bibitem[\protect\citeauthoryear{{Freeman}}{{Freeman}}{1970}]{Freeman:1970}
{Freeman} K.~C.,  1970, \apj, 160, 811

\bibitem[\protect\citeauthoryear{{Gray} et~al.,}{{Gray}
  et~al.}{2009}]{Gray_etal:2009}
{Gray} M.~E.,  et~al., 2009, \mnras, 393, 1275

\bibitem[\protect\citeauthoryear{{Gunn} \& {Gott}}{{Gunn} \&
  {Gott}}{1972}]{Gunn_Gott:1972}
{Gunn} J.~E.,  {Gott} J.~R.~I.,  1972, \apj, 176, 1

\bibitem[\protect\citeauthoryear{{Guti{\'e}rrez}, {Erwin}, {Aladro} \&
  {Beckman}}{{Guti{\'e}rrez} et~al.}{2011}]{Gutierrez_etal:2011}
{Guti{\'e}rrez} L.,  {Erwin} P.,  {Aladro} R.,    {Beckman} J.~E.,  2011, \aj,
  142, 145

\bibitem[\protect\citeauthoryear{{Heiderman} et~al.,}{{Heiderman}
  et~al.}{2009}]{Heiderman_etal:2009}
{Heiderman} A.,  et~al., 2009, \apj, 705, 1433

\bibitem[\protect\citeauthoryear{{Hoyos} et~al.,}{{Hoyos}
  et~al.}{2012}]{Hoyos_etal:2012}
{Hoyos} C.,  et~al., 2012, \mnras, 419, 2703

\bibitem[\protect\citeauthoryear{{Ibata}, {Chapman}, {Ferguson}, {Lewis},
  {Irwin} \& {Tanvir}}{{Ibata} et~al.}{2005}]{Ibata_etal:2005}
{Ibata} R.,  {Chapman} S.,  {Ferguson} A.~M.~N.,  {Lewis} G.,  {Irwin} M.,
  {Tanvir} N.,  2005, \apj, 634, 287

\bibitem[\protect\citeauthoryear{{Icke}}{{Icke}}{1985}]{Icke:1985}
{Icke} V.,  1985, \aap, 144, 115

\bibitem[\protect\citeauthoryear{{Johnston}, {Arag{\'o}n-Salamanca} \&
  {Merrifield}}{{Johnston} et~al.}{2014}]{Johnston_etal:2014}
{Johnston} E.~J.,  {Arag{\'o}n-Salamanca} A.,    {Merrifield} M.~R.,  2014,
  \mnras, 441, 333

\bibitem[\protect\citeauthoryear{{Johnston}, {Arag{\'o}n-Salamanca},
  {Merrifield} \& {Bedregal}}{{Johnston} et~al.}{2012}]{Johnston_etal:2012}
{Johnston} E.~J.,  {Arag{\'o}n-Salamanca} A.,  {Merrifield} M.~R.,
  {Bedregal} A.~G.,  2012, \mnras, 422, 2590

\bibitem[\protect\citeauthoryear{{Kennicutt}
  Jr.}{{Kennicutt}}{1989}]{Kennicutt:1989}
{Kennicutt} Jr. R.~C.,  1989, \apj, 344, 685

\bibitem[\protect\citeauthoryear{{Kron}}{{Kron}}{1980}]{Kron:1980}
{Kron} R.~G.,  1980, \apjs, 43, 305

\bibitem[\protect\citeauthoryear{{Larson}, {Tinsley} \& {Caldwell}}{{Larson}
  et~al.}{1980}]{Larson_etal:1980}
{Larson} R.~B.,  {Tinsley} B.~M.,    {Caldwell} C.~N.,  1980, \apj, 237, 692

\bibitem[\protect\citeauthoryear{{Laurikainen}, {Salo}, {Buta}, {Knapen} \&
  {Comer{\'o}n}}{{Laurikainen} et~al.}{2010}]{Laurikainen_etal:2010}
{Laurikainen} E.,  {Salo} H.,  {Buta} R.,  {Knapen} J.~H.,    {Comer{\'o}n} S.,
   2010, \mnras, 405, 1089

\bibitem[\protect\citeauthoryear{{Maltby} et~al.,}{{Maltby}
  et~al.}{2010}]{Maltby_etal:2010}
{Maltby} D.~T.,  et~al., 2010, \mnras, 402, 282

\bibitem[\protect\citeauthoryear{{Maltby} et~al.,}{{Maltby}
  et~al.}{2012a}]{Maltby_etal:2012a}
{Maltby} D.~T.,  et~al., 2012a, \mnras, 419, 669

\bibitem[\protect\citeauthoryear{{Maltby}, {Hoyos}, {Gray},
  {Arag{\'o}n-Salamanca} \& {Wolf}}{{Maltby} et~al.}{2012b}]{Maltby_etal:2012b}
{Maltby} D.~T.,  {Hoyos} C.,  {Gray} M.~E.,  {Arag{\'o}n-Salamanca} A.,
  {Wolf} C.,  2012b, \mnras, 420, 2475

\bibitem[\protect\citeauthoryear{{Marinova} et~al.,}{{Marinova}
  et~al.}{2009}]{Marinova_etal:2009}
{Marinova} I.,  et~al., 2009, \apj, 698, 1639

\bibitem[\protect\citeauthoryear{{Mart{\'{\i}}nez-Serrano}, {Serna},
  {Dom{\'e}nech-Moral} \&
  {Dom{\'{\i}}nguez-Tenreiro}}{{Mart{\'{\i}}nez-Serrano}
  et~al.}{2009}]{Martinez-Serrano_etal:2009}
{Mart{\'{\i}}nez-Serrano} F.~J.,  {Serna} A.,  {Dom{\'e}nech-Moral} M.,
  {Dom{\'{\i}}nguez-Tenreiro} R.,  2009, \apjl, 705, L133

\bibitem[\protect\citeauthoryear{{Mihos} \& {Hernquist}}{{Mihos} \&
  {Hernquist}}{1994}]{Mihos_Hernquist:1994}
{Mihos} J.~C.,  {Hernquist} L.,  1994, \apjl, 425, L13

\bibitem[\protect\citeauthoryear{{Moore}, {Katz}, {Lake}, {Dressler} \&
  {Oemler}}{{Moore} et~al.}{1996}]{Moore_etal:1996}
{Moore} B.,  {Katz} N.,  {Lake} G.,  {Dressler} A.,    {Oemler} A.,  1996,
  \nat, 379, 613

\bibitem[\protect\citeauthoryear{{Peng}, {Ho}, {Impey} \& {Rix}}{{Peng}
  et~al.}{2002}]{Peng_etal:2002}
{Peng} C.~Y.,  {Ho} L.~C.,  {Impey} C.~D.,    {Rix} H.-W.,  2002, \aj, 124, 266

\bibitem[\protect\citeauthoryear{{P{\'e}rez}}{{P{\'e}rez}}{2004}]{Perez:2004}
{P{\'e}rez} I.,  2004, \aap, 427, L17

\bibitem[\protect\citeauthoryear{{Pohlen}, {Dettmar}, {L{\"u}tticke} \&
  {Aronica}}{{Pohlen} et~al.}{2002}]{Pohlen_etal:2002}
{Pohlen} M.,  {Dettmar} R.,  {L{\"u}tticke} R.,    {Aronica} G.,  2002, \aap,
  392, 807

\bibitem[\protect\citeauthoryear{{Pohlen} \& {Trujillo}}{{Pohlen} \&
  {Trujillo}}{2006}]{Pohlen_Trujillo:2006}
{Pohlen} M.,  {Trujillo} I.,  2006, \aap, 454, 759

\bibitem[\protect\citeauthoryear{{Pohlen}, {Zaroubi}, {Peletier} \&
  {Dettmar}}{{Pohlen} et~al.}{2007}]{Pohlen_etal:2007}
{Pohlen} M.,  {Zaroubi} S.,  {Peletier} R.~F.,    {Dettmar} R.,  2007, \mnras,
  378, 594

\bibitem[\protect\citeauthoryear{{Ro{\v s}kar}, {Debattista}, {Quinn},
  {Stinson} \& {Wadsley}}{{Ro{\v s}kar} et~al.}{2008b}]{Roskar_etal:2008b}
{Ro{\v s}kar} R.,  {Debattista} V.~P.,  {Quinn} T.~R.,  {Stinson} G.~S.,
  {Wadsley} J.,  2008b, \apjl, 684, L79

\bibitem[\protect\citeauthoryear{{Ro{\v s}kar}, {Debattista}, {Stinson},
  {Quinn}, {Kaufmann} \& {Wadsley}}{{Ro{\v s}kar}
  et~al.}{2008a}]{Roskar_etal:2008a}
{Ro{\v s}kar} R.,  {Debattista} V.~P.,  {Stinson} G.~S.,  {Quinn} T.~R.,
  {Kaufmann} T.,    {Wadsley} J.,  2008a, \apjl, 675, L65

\bibitem[\protect\citeauthoryear{{Schaye}}{{Schaye}}{2004}]{Schaye:2004}
{Schaye} J.,  2004, \apj, 609, 667

\bibitem[\protect\citeauthoryear{{S{\'e}rsic}}{{S{\'e}rsic}}{1968}]{Sersic:196%
8}
{S{\'e}rsic} J.~L.,  1968, {Atlas de galaxias australes}.
Cordoba, Argentina: Observatorio Astronomico, 1968

\bibitem[\protect\citeauthoryear{{Sil'Chenko}, {Chilingarian}, {Sotnikova} \&
  {Afanasiev}}{{Sil'Chenko} et~al.}{2011}]{SilChenko_etal:2011}
{Sil'Chenko} O.~K.,  {Chilingarian} I.~V.,  {Sotnikova} N.~Y.,    {Afanasiev}
  V.~L.,  2011, \mnras, 414, 3645

\bibitem[\protect\citeauthoryear{{Simien} \& {de Vaucouleurs}}{{Simien} \& {de
  Vaucouleurs}}{1986}]{Simien_deVaucouleurs:1986}
{Simien} F.,  {de Vaucouleurs} G.,  1986, \apj, 302, 564

\bibitem[\protect\citeauthoryear{{Trujillo} \& {Pohlen}}{{Trujillo} \&
  {Pohlen}}{2005}]{Trujillo_Pohlen:2005}
{Trujillo} I.,  {Pohlen} M.,  2005, \apjl, 630, L17

\bibitem[\protect\citeauthoryear{{van der Kruit}}{{van der
  Kruit}}{1979}]{vanderKruit:1979}
{van der Kruit} P.~C.,  1979, \aaps, 38, 15

\bibitem[\protect\citeauthoryear{{Weinmann}, {van den Bosch}, {Yang} \&
  {Mo}}{{Weinmann} et~al.}{2006}]{Weinmann_etal:2006}
{Weinmann} S.~M.,  {van den Bosch} F.~C.,  {Yang} X.,    {Mo} H.~J.,  2006,
  \mnras, 366, 2

\bibitem[\protect\citeauthoryear{{Wilson}}{{Wilson}}{1927}]{Wilson:1927}
{Wilson} E.~B.,  1927, J. Am. Stat. Assoc., 22, 209

\bibitem[\protect\citeauthoryear{{Wolf} et~al.,}{{Wolf}
  et~al.}{2004}]{Wolf_etal:2004}
{Wolf} C.,  et~al., 2004, \aap, 421, 913

\bibitem[\protect\citeauthoryear{{Wolf} et~al.,}{{Wolf}
  et~al.}{2009}]{Wolf_etal:2009}
{Wolf} C.,  et~al., 2009, \mnras, 393, 1302

\bibitem[\protect\citeauthoryear{{Wolf}, {Hildebrandt}, {Taylor} \&
  {Meisenheimer}}{{Wolf} et~al.}{2008}]{Wolf_etal:2008}
{Wolf} C.,  {Hildebrandt} H.,  {Taylor} E.~N.,    {Meisenheimer} K.,  2008,
  \aap, 492, 933

\bibitem[\protect\citeauthoryear{{Wolf}, {Meisenheimer}, {Rix}, {Borch}, {Dye}
  \& {Kleinheinrich}}{{Wolf} et~al.}{2003}]{Wolf_etal:2003}
{Wolf} C.,  {Meisenheimer} K.,  {Rix} H.-W.,  {Borch} A.,  {Dye} S.,
  {Kleinheinrich} M.,  2003, \aap, 401, 73

\bibitem[\protect\citeauthoryear{{Younger}, {Cox}, {Seth} \&
  {Hernquist}}{{Younger} et~al.}{2007}]{Younger_etal:2007}
{Younger} J.~D.,  {Cox} T.~J.,  {Seth} A.~C.,    {Hernquist} L.,  2007, \apj,
  670, 269

\end{thebibliography}

\label{lastpage}

\end{document}